\documentclass[sensors,article,submit,pdftex,moreauthors]{Definitions/mdpi} 
\preto{\abstractkeywords}{\nolinenumbers}
\def\Fig#1{Fig.~\ref{fig:#1}}

\def\Tab#1{Table~\ref{tab:#1}}

\def\Sec#1{Section~\ref{sec:#1}}

\firstpage{1} 
\pubvolume{1}
\issuenum{1}
\articlenumber{0}
\pubyear{2023}
\copyrightyear{2023}
\datereceived{ } 
\daterevised{ } 
\dateaccepted{ } 
\datepublished{ } 
\hreflink{https://doi.org/} 

\Title{Human Stress Assessment: A Comprehensive Review of Methods Using Wearable Sensors and Non-wearable Techniques}

\TitleCitation{Human Stress Assessment: A Comprehensive Review of Methods Using Wearable Sensors and Non-wearable Techniques}

\Author{Aamir Arsalan $^{1}$\orcidA{}, Muhammad Majid $^{2}$*\orcidB{}, Imran Fareed Nizami $^{3}$, Waleed Manzoor $^{4}$, Syed Muhammad Anwar $^{5, }$$^{6}$, and Jihyoung Ryu $^{7}$\orcidC{}}

\AuthorCitation{Arsalan, A.; Majid, M.; Nizami, I. F.; Manzoor, W.; Anwar, S.M.; Ryu, J.}

\address{%
$^{1}$ \quad Department of Software Engineering, Fatima Jinnah Women University, Rawalpindi, Pakistan.; aamir.arsalan@fjwu.edu.pk\\
$^{2}$ \quad Department of Computer Engineering, University of Engineering and Technology, Taxila, 47050 Pakistan.; m.majid@uettaxila.edu.pk\\
$^{3}$ \quad Department of Electrical Engineering, Bahria University, Islamabad, 44000 Pakistan.; 
imnizami.buic@bahria.edu.pk\\
$^{4}$ \quad Department of Computer Engineering, Bahria University, Islamabad, 44000 Pakistan.\\
$^{5}$ \quad Sheikh Zayed Institute for Pediatric Surgical Innovation, Children’s National Hospital, Washington, DC.; sanwar@childrensnational.org \\ 
$^{6}$ \quad School of Medicine and Health Sciences, George Washington University, Washington, DC.\\
$^{7}$ \quad Electronics and Telecommunications Research Institute (ETRI), Gwangju 61012, South Korea.; 
jihyoung@etri.re.kr\\
}

\corres{Correspondence: Muhammad Majid (m.majid@uettaxila.edu.pk); 
}

\abstract{This paper presents a comprehensive review of methods covering significant subjective and objective human stress detection techniques available in the literature. The methods for measuring human stress responses could include subjective questionnaires (developed by psychologists) and objective markers observed using data from wearable and non-wearable sensors. In particular, wearable sensor-based methods commonly use data from electroencephalography, electrocardiogram, galvanic skin response, electromyography, electrodermal activity, heart rate, heart rate variability, and photoplethysmography both individually and in multimodal fusion strategies. Whereas, methods based on non-wearable sensors include strategies such as analyzing pupil dilation and speech, smartphone data, eye movement, body posture, and thermal imaging. Whenever a stressful situation is encountered by an individual, physiological, physical, or behavioral change is induced which help in coping with the challenge at hand. A wide range of studies has attempted to establish a relationship between these stressful situations and the response of human beings by using different kinds of psychological, physiological, physical, and behavioral measures. Inspired by the lack of availability of a definitive verdict about the relationship of human stress with these different kinds of markers, a detailed survey about human stress detection methods is conducted in this paper. In particular, we explore how stress detection methods can benefit from artificial intelligence utilizing relevant data from various sources. This review will prove to be a reference document that would provide guidelines for future research enabling effective detection of human stress conditions.}


\begin{document}




\section{Introduction}
\label{sec:intro}
Human stress research is significant since it plays a vital role in social, physiological, and psychological health. Stress research has a wide range of applications that include 
stress monitoring during the daily routine, stress assessment for improving health and work productivity, and preventing the onset of serious diseases. This research domain is beneficial for both individuals and society. Stressful conditions manifest in the form of adverse effects on an individual's working abilities and health. This makes them vulnerable to various kinds of diseases and weakens the recovery process for the human body from various clinical conditions~\citep{subhani2017machine}. Long-term exposure to heightened stress could cause symptoms of depression. Besides this, the strong connection between depression and stress could boost anxiety and mood disorders~\citep{dempsey2018stress}. Depression affects almost 350 million people worldwide and the situation seems to be worse in developing countries~\citep{world2015depression}. There are multiple causes of stress in human life that could also lead to mental disorders. These factors include but are not limited to, internal conflicts, political unrest, economic instability, rising poverty, crime rates, and natural disasters, etc. Hence, stressful conditions could severely affect human life in day-to-day activities and could trigger mental and clinical conditions. Therefore, methods that are efficient in stress detection are needed for the time. Recently, the global change in lifestyle owing to COVID-19 is also believed to infest various mental conditions. We are forced to a change in our daily life and have minimal social interaction, which is bound to affect our mental health. Experts believe that this could result in a mental pandemic if not handled properly \cite{spoorthy2020mental} \cite{ransing2020can}. Hence, monitoring and detecting stress at the right time has become a need of time.  

A large segment of the human population is living in a constant state of stress without even knowing its serious consequences. Most individuals are unaware of their current stress levels, 
and ironically one of the reasons for heart attacks and strokes is a high level of human stress. There are varied causes for stress such as poor income, joblessness, higher crime rates, natural disasters, and many others. According to a survey by the American Psychological Association in 2017, 62\% of the individuals have stress due to financial problems, 62\% due to problems at the workplace, 57\% due to political unrest in the country, and 51\% due to violence and crime in the society~\citep{american2017stress}. Despite such serious consequences, the definition of human stress in medicine is sometimes vague. Hence, priority should be given to focusing on stress quantification with precise numerical indexes. In the near future, it will not be enough to tell patients (dealing with stress) that it is all in their heads since advancements in diagnostic tools would aid in quantifying stress levels more precisely. Human stress is a difficult phenomenon to explain because every individual perceives it differently. There is a common perception in society that stress is a bad thing, but this is not always true as good stress also exists. For instance, an individual might have increased productivity under a stressful or challenging situation. In recent years, there has been an effort to develop strategies for stress assessment in a variety of daily life activities. For instance, in \cite{can2020personal}, authors have proposed a stress detection scheme based on heart activity, skin conductance, accelerometer, and skin temperature of the subject. The stress measurement protocol is composed of baseline, lecture, exam, and recovery sessions. Maximum accuracy of 94.52\% is achieved for three class stress classification using a random forest classifier. A study focused on cognitive training and stress detection in older adults suffering from mild cognitive impairment (MCI) while they were participating in a cognitive and motor rehabilitation session \cite{delmastro2020cognitive}.   
An acute multi-level stress classification framework using photoplethysmography signals in response to the mental arithmetic task was presented in \cite{zubair2020multilevel}.
A stress and anxiety detection mechanism using physiological signals for the academic environment was proposed in \cite{rodriguez2020towards}. 
A study to evaluate the physical and psychological stress in firefighters using the heart rate variability parameter was presented in \cite{pluntke2019evaluation}.  
Detection of human stress using ECG data with car driving and mental arithmetic task as a stimulus and deep neural network classifier was performed in a study conducted in \cite{cho2019ambulatory} with an achieved stress classification accuracy of 90.19\%. Regression analysis for the measurement of perceived stress using rest state EEG signal was presented in \cite{gillani2021prediction}. A review study about human stress assessment using physiological signals was presented in \cite{gedam2021review}. One of the major shortcomings of this review is that it only focuses on physiological measures of stress and does not enlighten other common methods of stress measurement like physical and psychological measures. Moreover, discussion about publicly existing datasets of human stress is not available.

Generally, stress is defined as a response of the human body to different kinds of situations like a threat or a challenge \citep{muthukumar2010cadmium}. There are two main systems of the human body i.e., the autonomic nervous system (ANS) and hypothalamic-pituitary-adrenal (HPA) axis, which responds to stress~\citep{ulrich2009neural}. When a stressor is encountered by a person, it activates neurons present in the hypothalamus. It releases a hormone called corticotropin-releasing hormone (CRH), which consequently causes the release of another hormone (adrenocorticotropin hormone (ACTH)) from the pituitary gland. ACTH travels in the blood and affects the adrenal glands, which in turn triggers the release of stress hormones including cortisol, epinephrine, and norepinephrine~\citep{kajantie2006effects}. Cortisol is released in response to stress and it helps an individual to cope with an immediate threat.

In terms of treatment and cure, stress is categorized into three main types i.e., acute stress, episodic stress, and chronic stress based on the symptoms and duration ~\citep{werner1993risk}. Acute stress (also termed instantaneous stress) originates from a specific event that is novel or unpredictable for an individual. For instance a public speaking task, a nearly missed road accident, and an interview. These stressors are not prone to affect an individual's health, rather are good for human health. Since such events provide a chance for the human body to practice and develop the fight response to any stressful situation in the future~\citep{segerstrom2004psychological}. Whereas severe stressors, which persist for a longer duration, can lead to serious health disorders. Episodic stress occurs when a person faces multiple acute stressors over a shorter period. Episodic stress is commonly faced by individuals that take on more responsibilities than they can easily manage at a given time. Individuals facing episodic stress are often in a hurry and have disorganized personalities. Individuals who have a pessimistic approach toward daily life routine tasks tend to have episodic stress. Unlike acute stress, episodic stress has negative effects on individual health. People facing episodic stress have low confidence in their abilities and they assume that they will never be able to come out of a stressful situation~\citep{sincero2012three}. Chronic stress (also called long-term stress) originates due to a variety of reasons like an unsatisfactory job, tense family life, and financial crises~\citep{hiriyappa2013stress}. Unlike acute stressors which can be negative as well as positive, chronic stress is always negative. Chronic stress affects the personality of an individual and can be the cause of many serious diseases which include heart attacks, cancer, and lung diseases~\citep{salleh2008life}.

For the assessment of human stress, subjective and objective measures have been used~\citep{gross2016standard,fohr2015subjective}. For subjective stress assessment, two different ways are used including standard stress measurement questionnaires designed by field experts and conducting sessions with psychologists~\citep{gurung2013health}. Whereas, objective measures of stress further include physiological and physical measures~\citep{onorati2013reconstruction,arsalan2019classification}. In physical measures,  visible changes in the human body are observed such as facial expressions~\citep{deschenes2015facial}, eye blinking rate~\citep{gowrisankaran2012asthenopia}, and dilation of the pupil~\citep{schulte2011handbook}. Whereas, for physiological measures, sensors are placed on the human body to measure internal changes. Towards this, various biomarkers have been employed including heart rate variability (HRV), heart rate (HR)~\citep{subahni2012association}, electrodermal activity~\citep{liapis2015recognizing}, respiration~\citep{wielgosz2016long}, and cortisol~\citep{liew2015classifying}.

Further, in recent years the application of machine learning for developing artificially intelligent systems has gained pace. One of the driving factors for this scientific advancement has been the success of deep learning algorithms. Machine learning is poised to significantly change and improve how healthcare systems work. The improvement in computational power will allow the development of embedded systems that are AI-enabled in healthcare. Human stress detection could also benefit from these advancements. Machine learning can be deployed for both offline and online stress assessments. Some challenges that need to be handled include dealing with unpaired data, assigning reliable labels, and developing algorithms that reliably work with limited data and are explainable.

Some studies focus on reviewing the current state of affairs related to human stress detection. For instance, a review on human stress detection using bio-signals is presented in~\citep{giannakakis2019review}. However, a discussion about the psychological, physical, and behavioral measures of human stress is found lacking. Further, publicly available databases for human stress measurement were also not explored. In another study, objective, subjective, physical, and behavioral measures for stress detection, as well as publicly available data used for human stress, are discussed. Another application-specific human stress measurement survey focusing on driver stress level is presented in~\citep{rastgoo2018critical}. Physical and physiological measures of human stress for driver stress detection are explored in detail. The limitation of this survey is that it only discusses a specific application i.e., driver stress level, and is not generic. Similarly, a review of methods developed for human stress measurement at the workplace is discussed in~\citep{carneiro2017new}. The limitation of this survey is that it only discusses the stress measurement methods about a specific application i.e., workplace environment and there is also no discussion about the publicly available existing databases for human stress assessment. Human stress measurement survey using smartphones and wearable sensors is presented in~\citep{can2019stress}. The paper presents the in-lab and out-of-laboratory environment stress measurement studies. The major limitation of the presented survey was the lack of discussion of existing publicly available datasets for human stress detection and the presentation of a limited amount of literature as compared to other available stress assessment surveys. A survey of devices available in the market was presented in~\citep{thapliyal2017stress} without any information on the studies using those devices. In summary, there is a need for a comprehensive presentation of the available human stress measurement methods. To the best of our knowledge, the current review addresses most of the shortcomings of existing human stress assessment survey papers by thoroughly investigating all the subjective and objective measures of human stress. In particular, our major contributions include,

\begin{enumerate}
  \item Subjective measures, which include psychological questionnaires, are explored in detail for completeness.
  \item Objective measures of stress comprising of data acquired from wearable and non-wearable sensors are elaborated.
  \item Publicly available human stress measurement databases and commonly used stimuli for inducing stress are also discussed in detail.
  \item Future research directions in the domain of automated human stress detection using artificial intelligence are identified.
\end{enumerate}

The organization of this review paper is as follows. \Sec{intro} presents an introduction to available stress measurement techniques categorization and a discussion about the existing stress measurement reviews and their limitations. \Sec{sads} presents a review of the commonly used stressors adopted in stress measurement studies for inducing stress in the participants followed by a brief discussion about publicly available stress detection databases. Subjective stress measurement techniques commonly used in literature are explored in \Sec{ssa} followed by objective stress measurement techniques, its general framework, and its associated literature is explored in \Sec{osa}. \Sec{msa} presents the multimodal stress detection schemes available in the literature followed by a discussion about the limitations of the existing schemes and future directions in \Sec{fd} and conclusion in \Sec{conc}.

\section{Stress Detection Datasets and Stressors}
\label{sec:sads}
The section is subdivided into two parts: first, we discuss some commonly used stressors for inducing stress in humans, secondly, we summarize publicly available datasets for human stress detection.

\subsection{Stress Inducers: Stressors}
Human stress measurement methods presented in the literature use a wide variety of stressors, which could include a public speaking task, an interview, an arithmetic task stressor, and many others. Stress is measured in response to these stressors by using different physiological and psychological techniques. Herein, we will review the most commonly used stressors for inducing stress in the participants and their related literature.

\subsubsection{Stroop Color Word Test (SWT)}
SWT is a neuropsychological test which has been developed by J.R. Stroop in 1935 and it has been widely adopted for experimental as well as clinical purposes. SWT was composed of three different tasks~\citep{stroop1935stroop}, where the first task consists of names of all colors written in black, in the second task the names of the colors and the color of the written text are different, whereas, in the third task, there are squares of different colors. During the test, a participant should answer the color of the word and not the word itself. In another version of SWT, three tasks were named as neutral (introductory session), congruent or non-conflict task, and non-congruent or conflict task. In the introductory session, all color names are written in black. In the congruent session, all color names are written in the same color as the color name. Whereas, in the non-congruent session, the name of the color is written in a different color from the color name. SWT has undergone a wide range of changes since its inception in 1935. The alterations include an increase or decrease in the task duration, the addition of more colors to the experimental tasks, and the selection of one or more non-congruent colors among the number of congruent colors. Stroop color-word test has been widely used in brain imaging and human attention measurement studies~\citep{pujol2001effect} and for the measurement and identification of human stress~\citep{pehlivanouglu2005computer,tulen1989characterization,svetlak2010electrodermal,renaud1997stress,zhai2006stress,lundberg1994psychophysiological,alonso2015stress,ren2012affective,kurniawan2013stress,giannakakis2017stress,giakoumis2012using,karthikeyan2012descriptive,krantz2004consistency}.

\subsubsection{Mental Arithmetic Task (MAT)}
MAT is one of the most commonly used stimuli for inducing stress~\citep{lundberg1994psychophysiological,ushiyama1991physiologic,tomaka1994effects,seraganian1997effect,ring2002secretory,hassellund2010long,linden1991arithmetic}. Mental arithmetic task is a mechanism to increase the mental workload by performing a series of arithmetic operations with a varying range of difficulty. This stimulus is easy to implement and does not require any special instrument. Another variant of the mental arithmetic task is Montreal Imaging Stress Task (MIST)~\citep{dedovic2005montreal}, which is a computer-based stress-inducing protocol mainly consisting of mental arithmetic problems and has been used as a stressor in several studies~\citep{setz2009discriminating,minguillon2016stress,al2015mental,al2016mental}

\subsubsection{Cold Pressor Test (CPT)}
The CPT is another stimulus that is commonly used for inducing stress in stress measurement experiments. CPT was first introduced by Hines and Brown in 1932~\citep{hines1932standard}. In particular, CPT involves immersion of the human hand or limb in cold water for a duration of 2 to 3 minutes. During this experiment, the subject feels uncomfortable and it is painful to adapt to a particular temperature for quite some time. The CPT protocol is widely used in laboratory experiments because of its ease of use. CPT triggers the activation of the sympathetic nervous system which increases blood pressure, heart rate, and skin conductance of the human body~\citep{lovallo1975cold}. A rise in cortisol level is also observed during CPT~\citep{al2002adrenocortical,bullinger1984endocrine}. Various versions of CPT have been used in different experiments which include immersion of both hands~\citep{suter2007cold} or both feet in hot or cold water~\citep{previnaire2012severity}. In~\citep{frings2013stress}, bilateral foot immersion was used to elicit stress response increasing salivary cortisol concentration and heart rate. In~\citep{hassellund2010long}, the author conducted a study in which the right hand of the subject was immersed completely in cold water for a time duration of one minute. In another study ~\citep{shi2010personalized}, the participant was asked to keep their hand in ice water until they started to feel discomfort.

\subsubsection{Social Evaluative Tasks}
Psycho-social stress is a type of human stress which occurs when an individual has to face people or a group of people as in public speaking task. When a socially threatening situation occurs, two mechanisms of the human body are affected, which include the autonomic nervous system and the neuroendocrine system. Hypothalamus activates both these systems to monitor the environmental demand (i.e., stress) as well as the internal state of the subject~\citep{bitsika2014hpa}. Based on these two mechanisms, physiological as well as behavioral response is activated to generate a fight-or-flight response~\citep{taylor2000biobehavioral}. The physiological system of a human being is affected and has an immediate impact with exposure to a social stressor~\citep{dickerson2004acute}. Exposure to social stressors has been the cause of many diseases including depression~\citep{mcewen2005glucocorticoids}, cardiovascular diseases~\citep{kemp2012depression}, and immune dysfunction~\citep{glaser2005stress}. Obesity, anxiety, and psychosocial stress have also been found interlinked with each other~\citep{pittig2013heart}. Hence, curing social stress is important, towards which exposure therapies have been developed to treat anxiety. Real-life social evaluative situations generate psychosocial stress~\citep{wolpe2013systematic}. 

Instead of real-life events exposure, virtual reality has also been used as a stressor~\citep{parsons2008affective}. Virtual reality exposure therapy (VRET) is an intermediate phase between thoughts and real-life events. Virtual reality is useful for a person who has difficulty imagining fearful tasks. VRET has also the advantage that if the stimuli become too threatening for the patient, the therapist has the control to stop the stimuli. VRET is a very effective method of treating social anxiety and based on this VRET patients learn methods to face such a threatening situation in real life~\citep{bordnick2012feasibility}. The public speaking task as a social stressor has been a focus in very few studies. Existing literature either focuses on the real audience~\citep{kudielka2009we} or a virtual audience~\citep{slater2006experimental,felnhofer2014afraid}. A complete study based on different physiological measures to find the impact of social stress in the real world as well as a controlled environment is still pending.

Existing literature has shown that a virtual audience has been able to induce stress based on a virtual public speaking task based on heart rate and self-reported anxiety measure~\citep{slater2006experimental,felnhofer2014afraid,pertaub2002experiment}. Moreover, literature exists on the comparison of stress based on gender. It is shown in~\citep{kudielka2009we} that when men and women are both subjected to real-life stressors, no significant difference based on gender was found. HPA has also been found to have no changes between male and female participants~\citep{kelly2008sex}. It has been established in the literature that women have a decreased happiness level after facing social stressors. A study presented  in~\citep{hemmeter2005modification} shows that men have higher cortisol concentration than females when facing a virtual stressor. Social rejection shows higher cortisol levels in women as compared to men in a study given in~\citep{stroud2002sex}. In~\citep{kothgassner2016salivary} author examined the stress response of a public speaking task in front of a real audience, a virtual audience, and in an empty lecture hall. Gender difference in stress response was also evaluated and heart rate, heart rate variability, and saliva cortisol were used as a parameter of measurement.

\subsubsection{Music}
The effect of music on human stress has also been the subject of various studies. In~\citep{escher1993music}, authors experimented with cortisol changes and found that positive cortisol changes occur when the subject was asked to listen to music before and during stressful medical treatment. In~\citep{suda2008emotional}, the author has demonstrated the effect of music on suppressing stress with an increase of cortisol level whereas, on the other hand in another study, the authors have demonstrated the fact that for a non-music condition, the cortisol levels decreased after the stressor period~\citep{khalfa2003effects}. Another parameter of research is the effect of music on the SNS system of an individual. Different experiments have been conducted in this regard to establish the effect of music on the SNS system. In~\citep{bartlett1996physiological}, a decrease in SNS activity was observed in response to music. But few other studies contradict these findings. An investigation into the fact that whether human stress is relieved due to music is reported in~\citep{allen2001normalization}. It was concluded that the level of relaxation and the ability to cope with challenges is increased with a decrease in the perceived level of stress of an individual. A decrease in anxiety in response to listening to music is a consistent finding of many studies~\citep{knight2001relaxing}. Few studies exist that have reported no reduction in anxiety in response to music~\citep{evans2002effectiveness}.

\subsubsection{International Affective Picture System (IAPS)}
IAPS is a collection of photos that have been widely used to elicit an emotional response either positive or negative in the viewers~\citep{lang1997international}. IAPS is a set of photos that have been evaluated on a 9-scale rating of valance and arousal. IAPS has been used as a very effective tool to induce stress in stress recognition experiments~\citep{baltaci2016stress,liao2005real,giannakakis2017stress,nhan2009classifying,khalilzadeh2010qualitative}. The database is developed by the National Institute of Mental Health Center for Emotion and Attention at the University of Florida and is composed of 956 images that have been categorized into pleasant (to elicit positive feelings), non-pleasant (to elicit negative feelings) and neutral images. The database consists of a normative rating which is developed on three dimensions i.e., valance, arousal, and dominance, representing the average rating of the emotion induced by each picture. This rating helps the researchers using IAPS in their research to select an appropriate set of images for inducing relevant emotions. The establishment of this type of average rate is termed standardization by psychologists. The standard rating of IAPS was obtained from 100 students composed of 50 males and 50 females having US-American origin. Normative rating of IAPS is also obtained from non-US participants of other origins i.e., Hungarian~\citep{deak2010hungarian}, German~\citep{gruhn2008age}, Portuguese~\citep{lasaitis2008brazilian}, Indian~\citep{lohani2013cross}, and Spanish~\citep{dufey2011adding}. Various kind of physiological modalities which include fMRI~\citep{caria2010volitional}, EEG~\citep{hajcak2009brain}, magnetoencephalography~\citep{styliadis2015distinct}, skin conductance~\citep{d2010early}, heart rate~\citep{bradley2001emotion}, and electromyography~\citep{baglioni2010psychophysiological} have been used along with IAPS stimulus.

\subsubsection{Trier Social Stress Test (TSST)}
TSST is a psychological stress-inducing protocol in the laboratory environment and was developed by Clemens Kirschbaum in 1993~\citep{kirschbaum1993trier}. TSST consists of two parts which include an anticipation period of 10 minutes and a test period of 10 minutes in which the subject has to deliver a speech and perform a mental arithmetic task in front of an audience. TSST has been used in a variety of stress measurement studies for inducing stress~\citep{kurniawan2013stress,engert2014exploring,vinkers2013effect,nater2005human}.

\subsection{Publicly Available Datasets for Human Stress Detection}

Only a few human stress assessment datasets have been curated by the research community and are publicly available for further research. In this section, we present details of publicly available data for this task using physiological signals. A human stress measurement data (\url{https://physionet.org/content/drivedb/1.0.0/}) to measure the driver stress using physiological signals of electrocardiogram, electromyogram, skin conductance, and respiration is presented in~\citep{healey2005detecting}. The physiological signals from 24 drivers were acquired during three different phases i.e., the rest condition, highway driving, and city driving. The three conditions (rest, highway, city) under which the data is acquired were mapped onto three stress levels i.e., low stressed, medium stressed, and highly stressed, respectively. One of the major limitations of this database is that the sampling rate of all the acquired physiological sensors is low e.g., the electromyogram signal is recorded at a sampling rate of 15.5 Hz. 

Another dataset to measure driver workload (\url{http://www.hcilab.org/automotive/}) using physiological signals of heart rate, skin conductance, and body temperature and GPS, acceleration, and brightness level data obtained from the smartphone are presented in~\citep{schneegass2013data}. The data from 10 drivers (7 males and 3 females) is acquired while driving on the pre-defined route of 23.6 km on five different road types i.e., 30 km/h zone, 50 km/h zone, highway, freeway, and tunnel. Moreover, the labels depicting different levels of workload i.e., no workload to maximum workload are also provided in the database. The dataset can be used for the assessment of different levels of workload based on physiological signals. 

Another publicly available human stress measurement dataset (\url{https://physionet.org/content/noneeg/1.0.0/}) using bio-signals of electrodermal activity, temperature, acceleration, heart rate, and arterial oxygen level is presented in~\citep{birjandtalab2016non}. It consists of data from 20 participants consisting of 16 males and 4 females. Data acquisition is performed under four different conditions i.e., relaxed state, physical stress, cognitive stress, and emotional stress. Relaxation condition is achieved by asking the participants to listen to a soothing music track. Physical stress is induced by making the participants jog on a treadmill at 3 miles/hour. Cognitive stress is elicited by asking the participants to count backward from 2485 in a step of seven. Lastly, emotional stress is evoked by watching a video clip from a movie. 

Another dataset (\url{https://osf.io/c42cn/wiki/home/}) to measure the driver's behavior under different kinds of emotional, cognitive, and startling stressors which are the major cause of accidents is presented in~\citep{taamneh2017multimodal}. The dataset was acquired by involving 68 drivers, who drove under four different conditions i.e., no distraction, emotional distraction, cognitive distraction, and sensorimotor distraction in a controlled environment in a driving simulator. Modalities used for acquiring the driver response include heart rate, respiration rate, facial expressions, gaze, and electrodermal activity from the palm of the subject. Different types of subjective questionnaires were used to measure the cognitive state, personality type, and task load of the subject. 

\textbf{WE}arable \textbf{S}tress and \textbf{A}ffect \textbf{D}ataset (WESAD) (\url{https://ubicomp.eti.uni-siegen.de/home/datasets/}) is a publicly available data consisting of physiological and motion data of subjects for both emotion and stress stimuli~\citep{schmidt2018introducing}. Here 15 participants were involved in the experiment which includes 12 male and 3 female participants and the data were acquired in the laboratory setting. Data for each subject were recorded in three different conditions i.e., baseline recording done by performing a reading task, funny condition achieved by watching a set of funny videos, and stressed condition achieved by exposure to trier social stressor test. Sensor modalities used in the data acquisition include electrocardiography, electrodermal activity, electromyography, blood volume pulse, respiration, body temperature, and three-axis acceleration. 

A multimodal \textbf{S}mart reasoning system for \textbf{WELL}-being at work and at home \textbf{K}nowledge \textbf{W}ork (SWELL-KW) dataset (\url{http://cs.ru.nl/~skoldijk/SWELL-KW/Dataset.html}) for research on human stress and user modeling is developed in~\citep{koldijk2014swell}. Data were curated from 25 participants performing a variety of knowledge work tasks, which included report writing, preparing presentations, checking emails, and information search. Stress was induced by telling the participants that they have to present one of the prepared presentations to get the full experiment participation fee. Data for each participant were recorded for a duration of three hours, which was subdivided into three one-hour blocks. Each block started with an eight-minute relaxation period block and then after that, the participant was assigned to the tasks on which he/she has to work. The participants had to write two reports and prepare one presentation in each block of the experiment. Stress was induced by showing a countdown timer flashing the remaining time for task completion. 

A dataset (\url{https://catalog.ldc.upenn.edu/LDC99S78}) to measure the effect of human stress on speech signals was presented in~\citep{steeneken1999speech}. Three different databases named Speech Under Stress Conditions (SUSC), Speech Under Simulated and Actual Stress (SUSAS), and DERA License Plate (DLP) datasets were developed in this research work to develop robust speech processing algorithms for the identification of human stress in the speech signals. Towards this, 32 speakers constituting 19 male and 13 female participants with an age bracket ranging from 22 to 76 years participated in the experiment to record 16,000 voice samples. The speech signals were sampled using a 16-bit analog-to-digital converter at a sampling rate of 8 kHz. 

Another dataset (\url{https://www.sensornetworkslab.com/clas}) for \textbf{C}ognitive \textbf{L}oad, \textbf{A}ffect and \textbf{S}tress Recognition (CLAS) was presented in~\citep{markova2019clas}. The database consists of the physiological recording of ECG, PPG, and EDA and motion data of accelerometer from 62 healthy participants (45 men and 17 women) with ages ranging from 20-50 years. The data was acquired while performing three interactive and two perspective tasks.  The interactive tasks include mathematical problems, logic tasks, and Stroop color-word tests, whereas perspective tasks are composed of images and audio-video stimuli. All the physiological signals were acquired at a sampling rate of 256 Hz with a resolution of 16 bits per sample. 

Another publicly available dataset for the assessment of social stress in humans using physiological signals of blood volume pulse and electrodermal activity is presented in \cite{meziatisabour2021ubfc}. Moreover, video recording was done to measure the remote photoplethysmography and facial features. A total of 68 undergraduate students from the psychology department participated in the experiment. The Competitive State Anxiety Inventory (CSAI) questionnaire was used to measure the three dimensions of self-reported anxiety which include cognitive anxiety, somatic anxiety, and self-confidence. Trier Social Stress Test has been used as a stimulus during which the physiological signals and the video recording is performed.  

A perceived human stress measurement dataset (\url{https://sites.google.com/site/simplrgp/resources}) using EEG signal was presented in~\citep{arsalan2019classification}. The database consists of EEG recordings from 28 participants (13 male and 15 females) with ages ranging from 18 to 40 years, in three different phases of the experiment i.e., pre-activity, during activity, and post-activity. EEG recording was performed while the participant was delivering a presentation on an unknown topic for a time duration of five minutes. Subjective scores from the perceived stress scale questionnaire were also recorded. 

\section{Subjective Stress Assessment}
\label{sec:ssa}
Subjective measures for human stress assessment have been traditionally used for many decades. While here our objective is to review methods that use data from wearable and non-wearable sensors for automated stress detection using artificial intelligence. However, subjective measures are explored herein for completeness. Further, such assessments have been used to benchmark machine learning-based methods. Towards this, there exists a wide range of questionnaires developed by psychologists for measuring different types of stress. These measures are based on the questionnaire being filled out by the subject. Psychological questionnaires are being used by the researchers to validate the objective measures obtained from the sensors. Perceived stress scale (PSS) questionnaire~\citep{cohen1983global} is commonly used by psychologists to measure chronic stress. Acute stress disorder (ASD) scale questionnaire~\citep{bryant2000acute} is developed by psychologists to measure acute stress. Some of the other questionnaires used by the psychologists are relative stress scale (RSS)~\citep{ulstein2007relative}, daily stress inventory (DSI)~\citep{brantley1987daily}, brief symptom inventory~\citep{derogatis1993brief} and trier inventory for the assessment of chronic stress (TICS)~\citep{schulz1999trier}. A brief review of the commonly used questionnaires for stress assessment is given below.

\subsection{Acute Stress Disorder (ASD)}
ASD is a subjective self-reporting questionnaire inventory that is used to quantify acute stress disorder and post-traumatic stress disorder. ASD is a self-reporting version of the Acute Stress Disorder Interview (ASDI) questionnaire. ASD was developed with three aims i.e., (a) identification of ASD, (b) a self-report version of ASDI, and (c) a measure of post-traumatic stress disorders (PSTD). It is a 19-item questionnaire and is compliant with the Diagnostic and Statistical Manual of Mental Disorders criteria. The scale has been successfully used to measure the acute stress order among a wide range of subjects~\citep{bryant2000acute}. The 19 questions of the ASD questionnaire are composed of 5 dissociatives, 4 reexperiencing, 4 avoidance, and 6 arousal items. The questions for the ASD questionnaire are rated on a five-point Likert scale, where 1 means that a condition did not occur at all and 5 means the particular situation occurred very strongly. The minimum score of the questionnaire can be 19 and the maximum score of 85. A study to analyze the factor structure of acute stress disorder in the earthquake victims of the Chinese population is conducted in~\citep{wang2010factor}. The study was conducted on a sample of 353 samples consisting of 180 men and 173 women with a mean age of 29.36 and a standard deviation of 11.45. The study concluded that a four-factor model consisting of dissociation, reexperiencing, avoidance, and arousal is consistent with the conceptualization of ASD. A wide range of studies has been conducted to establish a correlation between PSTD and ASD. The studies report that around three-quarters of the survivors of trauma patients who show symptoms of ASD ultimately develop PSTD~\citep{harvey1998relationship,harvey1999two,harvey2000two,brewin1999acute}. A study conducted for motor vehicle/industrial accidents in~\citep{harvey1998relationship} found a 3-factor model consisting of acute post-traumatic stress reactions, dissociative symptoms, and dissociative amnesia. The study was conducted on 99 participants consisting of 65 men and 34 women with a mean age of 31.59 and a standard deviation of 11.28.

\subsection{Brief Symptom Inventory (BSI)}
BSI is a questionnaire developed by a psychologist to measure psychological distress and psychiatric disorders in people~\citep{derogatis1993brief}. The data collected from the questionnaire can be used to diagnose and treat the patients. BSI is a 53-item questionnaire with each question being answered on a five-point scale of 1 to 5. The 53 items of BSI consist of questions of nine symptoms dimensions including Somatization, Obsession-Compulsion, Interpersonal Sensitivity, Depression, Anxiety, Hostility, Phobic Anxiety, Paranoid Ideation, and Psychoticism, and three indices of distress i.e.,  Global Severity Index, Positive Symptom Distress Index, and Positive Symptom Total. The time required by the subject to complete the questionnaire is approximately 8 to 12 minutes. The respondent to the questionnaire answers the questions on a scale from 0 (condition never occurs) to 5 (a condition that occurs very frequently). The minimum score of the questionnaire can be 53 whereas a maximum score of 265 can be recorded. The somatization dimensions are calculated from items 2, 7, 23, 29, 30, 33, and 37, obsession-compulsion dimension is obtained from items 5, 15, 26, 27, 32, and 36, interpersonal sensitivity is measured from items 20, 21, 22, and 42, depression dimension is evaluated from items 9, 16, 17, 18, 35, and 50, anxiety is obtained from items 1, 12, 19, 38, 45, and 49, hostility dimension is calculated from items 6, 13, 40, 41, and 46, phobic anxiety is taken from items 8, 28, 31, 43, and 47, paranoid ideation is evaluated from items 4, 10, 24, 48, and 51, and psychoticism is measured from items 3, 14, 34, 44, and 53 of the questionnaire. Items 11, 25, 39, and 52 did not contribute to the calculation of any dimension but they are recorded because of their clinical importance. Global Severity Index is calculated by the sum of the items of all the nine dimensions as well as the four items which were not included for the calculation of any dimension and then dividing the sum by the total number of items a particular person answered. The positive Symptom Total is calculated by counting the number of items whose responses are non-zero. Positive Symptom Distress Index is obtained by dividing the sum of the non-zero response items by the positive symptom total. BSI has been used for examining the relationship among psycho-social family risk factors, parental psychological distress, and quality of life in pediatric cancer survivors in a study conducted in~\citep{racine2018quality}. The study reports that families having a low level of distress have a lesser impact on the quality of life of pediatric cancer survivors. The relationship between psycho-pathological symptoms and technological addictions has also been studied. In a study conducted among 126 university students, the nine dimensions obtained from the BSI questionnaire were found to be significantly correlated to internet addiction~\citep{adalier2012relationship}. A significant association was found between anxiety level and internet addiction in adolescents in a study conducted in~\citep{stavropoulos2017longitudinal}.

\subsection{Relative Stress Scale (RSS)}
RSS is a commonly used subjective marker to measure the psychiatric disorders among the individuals who act as caretakers of dementia patients. It is a 15-item questionnaire and is a reliable measure of stress disorders among the carers of dementia patients. The items of the questionnaire are scored on a scale of 0 to 5, where 0 means a particular event never occurred and 5 means a particular event occurs very frequently. The minimum score of the questionnaire is 0 whereas the maximum score is 60. Age gender, education, occupation, and the relationship of the carer with the patient are also recorded in RSS. The carer of the patients was also asked to specify their routine and estimated time they used to take care of and assist the patient in a week. The RSS questionnaire covers many different aspects of the carer burden like a subjective emotional response (emotional distress), the negative feeling associated with the behavior of patients (negative feelings), and restrictions in the patient carer's social life (social distress). Question items 1, 2, 3, 4, 5, and 6 of the RSS questionnaire measures emotional distress, items 7, 8, 9, 10, 11, and 13 measure social distress, and items 12, 14, and 15 measure the negative feelings of the patient carers. Emotional distress in carers is directly proportional to the amount of time spent per week with the patient and more specifically the emotional distress is higher in female carers supporting the fact that female carers are more emotional in their approach~\citep{fitting1986caregivers} whereas their counterpart males are more task or goal oriented~\citep{corcoran1992gender}. Social distress is also higher in carers who spend more time with the patients and patients with high social distress need help and a break from caring for dementia patients. The negative feelings in the carer are associated with the patient's age, i.e., the younger the patient, the more negative feelings that occur in the patient's carers. RSS has been widely adopted in Norway for clinical purposes and research to measure the carer burden~\citep{braekhus1999social,thommessen2002psychosocial}. RSS has been used in literature for the validation of the distress scale of the Neuropsychiatric Inventory~\citep{kaufer1998assessing}. In~\citep{greene1982measuring}, RSS has been used to provide a useful basis for discussion with carers of dementia patients.

\subsection{Daily Stress Inventory (DSI)}
DSI is another measure developed to provide research scientists and doctors with information about the psychiatric issues of patients after they have gone through some stressful events. DSI is specifically designed for the measurement of small stressful events that need to be measured on daily basis. DSI possesses useful and unique qualities for the measurement of stressful events. DSI is a 58-item questionnaire that allows the participant to indicate the events which occurred in the last 24 hours. After indicating the events that occurred, the subject rated the stressfulness of those events on a Likert-type scale from 1 to 7. 1 refers to the events which occurred but they were not stressful whereas a score of 7 means that a particular event caused panic to the subject. At the end of the DSI inventory, two blank items were provided to let the subject report those events which were not included in the 58 items. However, the scores of the blank items were not counted toward the calculation of stress scores. The minimum score of the DSI inventory can be 58 whereas a maximum score of 406 can be obtained. Three different scores are computed for each individual, (i) the number of events that are reported by the subject to have occurred, (ii) the sum of the total score of all these events, and (iii) the average of the scores of these events. DSI inventory has been used frequently in research studies and has shown good validity ad reliability~\citep{maslach1997evaluating}. DSI was aimed at daily monitoring over a course of seven to ten days to measure the changes in daily faced stressors and to observe the relationship of these stressors to physical and psychological symptoms~\citep{brantley1993daily}. This daily monitoring leads to a better association between these small stressors and the illness of the subject. A study conducted in~\citep{goreczny1988daily} monitored 24 patients with asthma and chronic obstructive pulmonary disease. Daily stress score was recorded via DSI inventory and respiratory symptoms was recorded for a time of 21 days. The study depicted the fact that on a highly stressful day, asthma symptoms in patients worsened. DSI has been used for correlation with other medical conditions like headache~\citep{mosley1991time,waggoner1986investigation}, Crohn’s disease~\citep{garrett1991relation} and diabetes~\citep{goetsch1990stress}.

\subsection{Perceived Stress Scale (PSS)}
PSS is a questionnaire developed to measure the chronic stress of an individual. The questionnaire assesses the extent to which an individual has been stressed in the last thirty days. PSS is a 10-item questionnaire scored on a scale of 0 to 4 where 0 means that a particular situation never occurred whereas 4 means a situation occurred very frequently. The score from all the items of the questionnaire is summed up to get the PSS questionnaire score. The minimum and maximum score that can be obtained from the PSS questionnaire is 0 and 40, respectively. PSS's final score is obtained by reversing the four-item of the questionnaire which include items 4, 5, 7, and 8, and using the other items of the questionnaire as it is. PSS has been used in a wide range of studies for the assessment of chronic stress among individuals. 

\subsection{Trier Inventory for the Assessment of Chronic Stress (TICS)}
The TICS is a standardized questionnaire for assessing nine interrelated factors of chronic psychosocial stress and is a very reliable and effective tool. The nine factors which are addressed by TICS include Work Overload (e.g., "I have too many tasks to perform."), Social Overload (e.g., "I must frequently care for the well-being of others."), Pressure to Perform (e.g., "I have tasks to fulfill that pressure me to prove myself."), Work Discontent (e.g., "Times when none of my tasks seem meaningful to me."), Excessive Demands at Work (e.g., "Although I try, I do not fulfill my duties as I should."), Lack of Social Recognition (e.g., "Although I do my best, my work is not appreciated."), Social Tensions (e.g. "I have unnecessary conflicts with others."), Social Isolation (e.g.,  "Times when I have too little contact with other people."), and Chronic Worrying (e.g., "Times when I worry a lot and cannot stop). TICS is a 57-item questionnaire that is rated on a 5-point scale from 0 to 4 based on whether the participant experienced a particular situation in the last 3 months or not. On the 5-point scale, 0 means a situation never occurred, 1 means a situation very rarely occurs, 2 means a situation sometimes occurs, 3 means a particular situation often occurs, and 4 means a particular situation occurs very frequently. The total score of the TICS questionnaire can range from 0 to 228. In a study conducted in~\citep{sturmbauer2019stress}, a correlation between the Stress and Adversity Inventory (STRAIN) with TICS and PSS was examined. It was found that STRAIN is more correlated to TICS as compared to PSS. A correlation between the TICS score and central serous chorioretinopathy (CSC) named syndrome in young and middle-aged adults was established in~\citep{buehl2012trier}. The study found that people with CSC syndrome have higher TICS scores as compared to individuals with no CSC syndrome.

Subjective measures of human stress have been widely used in the literature but there exist some limitations and shortcomings in these methods. One of the major shortcomings of these subject measures is that these questionnaires are being responded to by the subject himself and if the subject answers the items of the questionnaire in a biased manner then the score obtained for stress measurement is unreliable and incorrect. Secondly, to answer the questionnaires, the subject has to be literate and able to properly read the items of the questionnaire. Thirdly, the questionnaires for stress measurement are not available in all the languages thus creating a bottleneck and hence cannot be used by individuals whose first language is not the one in which the questionnaire has been developed. Keeping in view these limitations, using only subjective measures is not a reliable indicator of stress, thus objective measures of stress are essential for the development of better stress measurement protocols.

\section{Objective Stress Detection}
\label{sec:osa}
\begin{figure*}
          \begin{center}
          \begin{tabular}{c}
          \includegraphics[width=\linewidth]{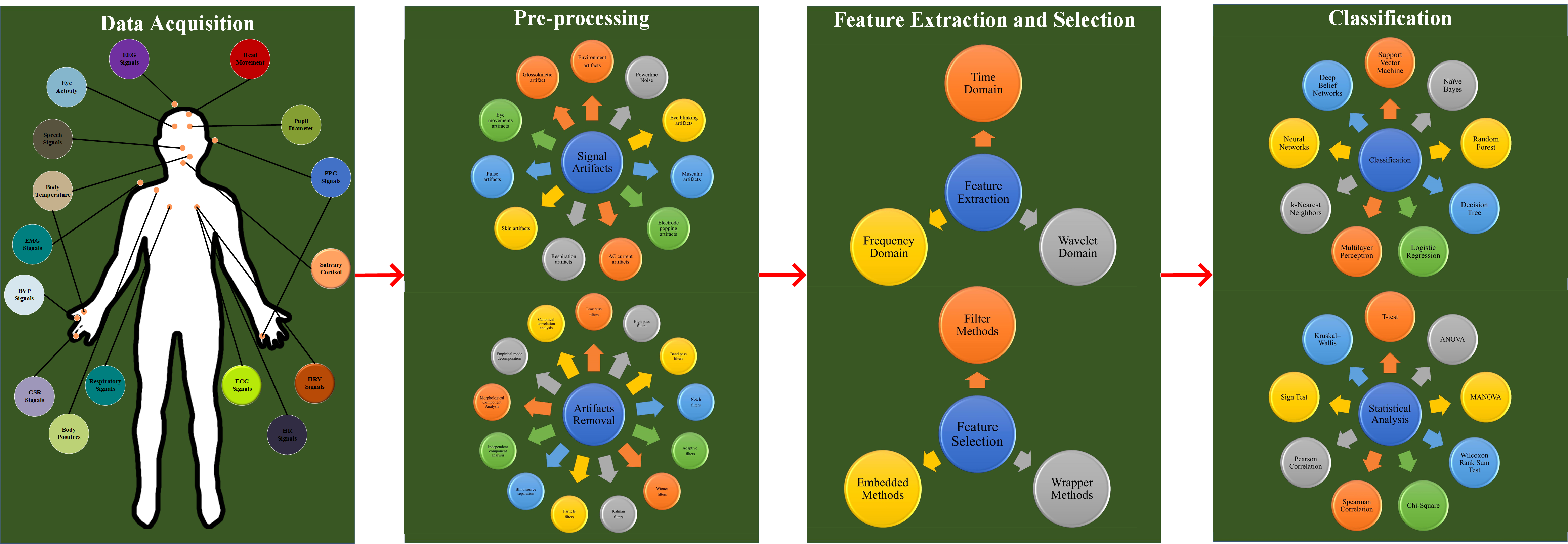}
          \end{tabular}
          \end{center}
          \caption
          { \label{fig:fig1a}
          { General machine learning framework for objective stress detection.}}
          \end{figure*}

Objective measures of stress include physiological and physical measures. Physiological measures of stress need sensors to be connected to the human body at some specified location e.g., EEG, ECG, and EDA whereas, in the case of physical sensors the measurement can be done at a distance from the subject without the need of any physical contact. Objective measures of stress are free from human intervention and hence cannot be biased like the subjective questionnaire and it is the major benefit of objective measures over the subjective assessment of stress. Moreover, the studies which use objective measures of stress also validate their finding using subjective questionnaires~\citep{healey2005detecting}. The data acquisition protocols in case of objective measures of stress are time-consuming and complicated and hence recording data for a large population sample is difficult. The limited capacity of the existing stress modeling protocol and the lack of a large data sample make it necessary to include the conventional subjective stress measurement methods for the validation of objective measures. It is because of these factors, subjective measures are still regarded as an efficient measure of stress~\citep{ulstein2007high,weidner1989hostility}. In this section, we will discuss a general framework for human stress assessment and review all the literature available for human stress measurement using objective methods. The general machine learning framework of human stress detection includes data acquisition and annotation, pre-processing, feature extraction and selection, and classification steps that are shown in \Fig{fig1a}. Each of these steps plays a vital role in accurate human stress detection and is discussed below.

\noindent\textbf{Data Acquisition and Annotation} is one of the most important steps in the human stress detection framework. The quality of the acquired data is of utmost importance for the robust analysis of human stress and to draw a reliable conclusion. Before the start of data acquisition, a good experimental design following the standard protocols is needed. In stress measurement studies, there exist two types of experimental protocols which include (i) measuring stress induced by an external stimulus also called acute or instantaneous stress, and (ii) measuring perceived or chronic stress without using any external stimulus. Before data acquisition, the stress-inducing protocol which will be used needs to be defined. Another important factor that needs to be considered for data acquisition is whether the data need to be acquired in laboratory settings or out-of-laboratory environment. The number of participants in the experiment is also important because if the number of participants is small, the findings of the study could not be generalized and there is also a chance that data acquired from some participants might get corrupted on the other hand, acquiring data from a large number of participants is a time-consuming and cumbersome process to follow. The physical and physiological sensors for whom data is to be recorded should be selected before the start of data acquisition. In addition to the above-mentioned parameters, another important factor that needs to be considered in data acquisition is data annotation. Data annotation is the process of assigning each training example of the data to a particular class depending on some criteria. Commonly used criteria for data annotation in stress measurement studies include the use of subjective questionnaire~\citep{asif2019human,arsalan2019classification} and evaluation by psychologists~\citep{saeed2020eeg}. This variation in the labeling technique also poses challenges to the comparison of the available techniques with each other.

\noindent\textbf{Pre-processing} is the second step in the stress detection pipeline and plays an important role in the whole process. Signals acquired by using wearable and non-wearable sensors during the data acquisition phase are affected by different kinds of noises which include power line~\citep{lin2016removal}, eye blinking artifacts~\citep{shoker2005artifact}, muscular artifacts~\citep{chen2014preliminary}, and posture or physical activity~\citep{alamudun2012removal}. Noise removal techniques for pre-processing different kinds of modalities have been developed in the literature. Accelerometer sensor has been used in stress detection studies and the noise which affects accelerometer data is composed of high-frequency components and can be removed by using low-pass filtering. Authors have applied low pass filtering to remove high-frequency artifacts from the accelerometer signal in their stress assessment studies conducted in~\citep{mozos2017stress,gjoreski2017monitoring}. Two of the main steps for pre-processing an ECG signal are to identify the R-peaks and RR interval. Algorithms like Pan and Tompkin’s algorithm~\citep{pan1985real} have been developed to identify the R-peaks of the ECG signal. Moreover, to identify the valid RR-intervals algorithms have also been proposed~\citep{hovsepian2015cstress}. Pre-processing of PPG signals has been explored in the literature. PPG signals are affected by low-frequency noise which can be mitigated by the use of high-frequency filtering~\citep{elgendi2012analysis}. Meaningful information in an EDA signal is normally contained in low-frequency components and the noise is the high-frequency component of the signal which can be removed by passing the EDA signal through a low-pass filter. Another important pre-processing task performed with the EDA signals is its segmentation into a slow-varying baseline conductivity known as skin conductance level (SCL) and a high-frequency component called skin conductance response (SCR). Authors in~\citep{choi2011development} have proposed a technique to separate the SCL and SCR components of the EDA signal. Techniques for pre-processing an EMG have been proposed in the literature. A two-step noise removal technique for EMG signals is proposed in~\citep{wijsman2010trapezius}. In the first step, band-pass filtering is applied to the EMG signal to limit the signal from 20 to 450 Hz. In the second step, power line interference is negated by applying notch filters at frequencies of 50, 100, 150, 200, 250, and 350 Hz. Another important contamination source for EMG signals is ECG signals i.e., cardiac artifacts. Different algorithms to remove cardiac noise from EMG signals have been compared in a study conducted in~\citep{willigenburg2012removing}.

\noindent\textbf{Feature Extraction and Selection} are critical for an efficient machine learning model. Feature extraction corresponds to the process of extraction of meaningful features from the acquired data. Meaningful features are the extracted set of features that are descriptive i.e., the features have discriminating values for instances from different classes. The extracted features constitute a feature vector which is fed as input to the classification stage. The extracted features can be categorized differently e.g., time or frequency or wavelet domain features, linear features vs non-linear features, and unimodal vs multimodal features. The computational complexity of the extracted set of features can range from simple statistical features e.g., mean, median, minimum, and maximum to complex features based on certain modalities. A different set of features are extracted from each sensor for human stress recognition. Some of the commonly used features for accelerometer sensors in stress recognition studies include mean, standard deviation, variance, maximum, absolute value, signal magnitude area, root mean squared, energy, differential entropy, discrete Fourier transform, peak magnitude frequency, peak power, and zero crossing~\citep{garcia2015automatic,can2019continuous,sano2013stress}. List of some of the features extracted from ECG and the PPG signals in stress measurement studies include the mean and standard deviation of the R-R interval, root mean square difference of the consecutive R-R interval, heart rate, heart rate variability, mean R peak amplitude, mean standard deviation, skewness, kurtosis, percentile, geometric and harmonic mean, low-frequency power, high-frequency power, power ratio, crest time, and instantaneous pulse ratio ~\citep{bong2012analysis,ahn2019novel,mohino2015assessment,cho2019instant,charlton2018assessing}. Common features extracted from the EEG signals in human stress measurement studies include~divisional asymmetry, rational asymmetry, mean power, power spectral density, alpha asymmetry index, normalized band power, relative power, coherence, and amplitude asymmetry~\citep{arsalan2019classification,ahn2019novel,asif2019human}. Looking at EDA-based human stress measurement studies, statistical features of mean, standard deviation, mean of the absolute values, root mean square, the proportion of negative samples, the slope of the EDA level, mean EDA peak rate, and height, minimum and maximum have been commonly used~\citep{giakoumis2012using,setz2009discriminating}. Feature selection is defined as a process that aims at selecting the subset of features that have the highest discriminative power and yield the highest classification accuracy from among the extracted set of features. Different features selection algorithms have been used in stress classification studies which include genetic algorithm~\citep{shon2018emotional}, t-test~\citep{saeed2020eeg}, minimum redundancy maximum relevance (mRMR)~\citep{subhani2017mrmr}, principal component analysis (PCA)~\citep{deng2012evaluating}, particle swarm optimization (PSO)~\citep{yerigeri2019meta}, wrapper based feature selection~\citep{hasan2019hybrid}, Bhattacharya distance~\citep{subhani2017machine}, and independent component analysis (ICA)~\citep{palacios2019ica}.

\noindent\textbf{Classification} is the last step in the human stress detection framework and is an important part of the whole process. The classification process can be performed by either use of statistical measures (t-test or ANOVA) or by using machine learning techniques. For both types of techniques, the selected or extracted set of features is fed as input to the classification stage. The t-test is a type of inferential statistics test that is aimed at finding whether there is any significant difference between the means of two groups or not. The t-test is based on the assumption that the dependent variable of the data follows a normal distribution and we can identify the probability of a particular instance. T-test produces a p-value whose acceptable value is considered to be less than 0.05. A p-value of 0.01 means that the likelihood to get the difference in the two groups by chance is 1 out of 100 times. The t-test is applied to cases in which we need to find the difference between two groups where to find the difference between more than two groups ANOVA test is applied. The second type of method used for the classification of human stress in the literature includes machine learning techniques. A wide variety of algorithms depending on the situation have been employed in human stress recognition studies. Multilayer perceptron (MLP) is a type of feed-forward neural network which is composed of at least three layers i.e., input layer, hidden layer, and output layer. MLP has been used for binary as well as multi-class stress classification tasks in a wide range of human stress recognition studies~\citep{arsalan2019classification,arsalan2019classification_EMBC}. Another commonly used classification technique for human stress recognition is the Naive Bayes algorithm. The naive Bayes algorithm is a type of algorithm that is based on the Bayes probability theorem and the conditional probability rule. Some of the stress recognition studies which have used the Naive Bayes algorithm include~\citep{ahuja2019mental,saeed2017quantification}. Support vector machine has also been used in a sizable amount of human stress recognition studies. Support vector machine (SVM) is a supervised machine learning classifier and it works by defining a separating a hyperplane with the help of support vectors. Some of the human stress recognition studies involving the SVM classifier include~\citep{saeed2018selection,saeed2020eeg,vanitha2013hybrid,attallah2020effective}. k-nearest neighbors (kNN) is a type of supervised and non-linear machine learning algorithm used for classification tasks. k-nearest neighbors assign the new data point by calculating the distance of the point from the k-nearest neighbors and the data point is assigned to the class whose nearest neighbor has the lowest distance metric. The value of k can be any odd number i.e., 1, 3,5, etc. Higher the value of k, the more reliable results kNN produces. Some of the stress classification studies which have used kNN as a classifier include~\citep{rahman2015mental,karthikeyan2012study,shon2018emotional}. Some of the other classifiers used in stress recognition studies include logistic regression~\citep{asif2019human,vasavi2018regression}, deep belief networks~\citep{song2017development}, deep neural network~\citep{sardeshpande2019psychological,masood2019modeling}, and random forest~\citep{uddin2019synthesizing}.

\begin{figure*}
          \begin{center}
          \begin{tabular}{c}
          \includegraphics[width=\linewidth]{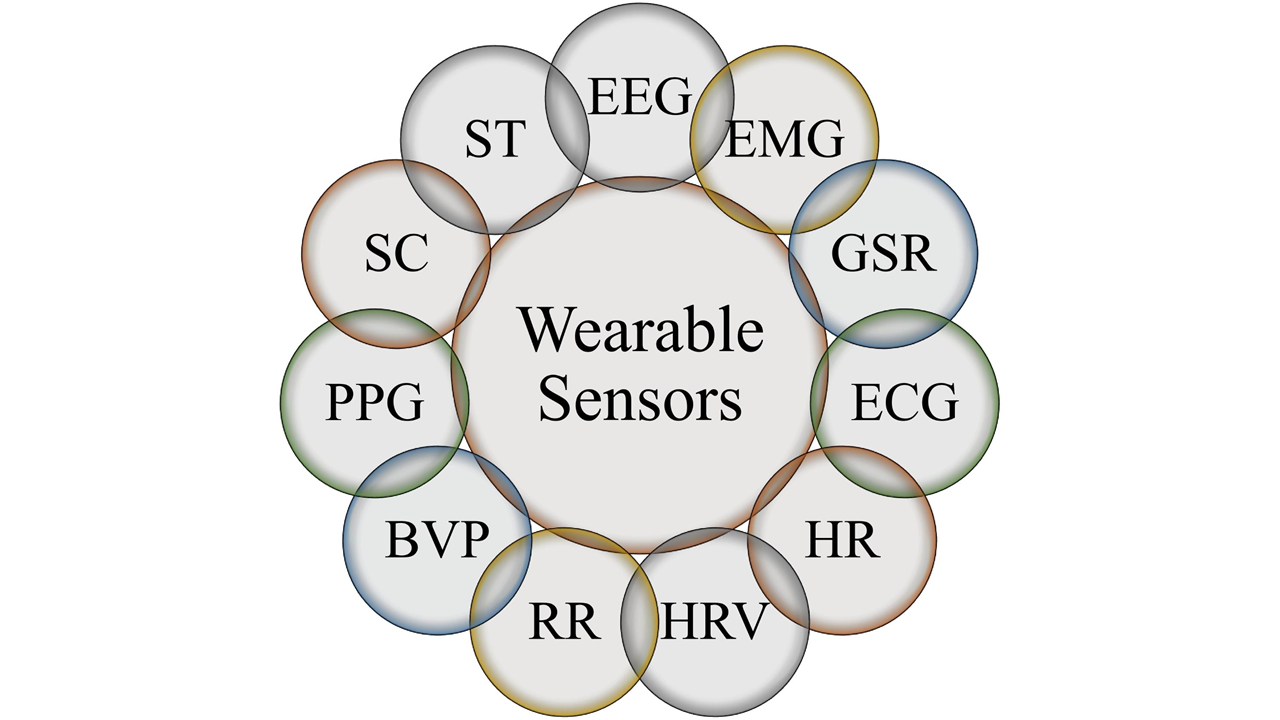}
          \end{tabular}
          \end{center}
          \caption
          { \label{fig:fig2a}
          { Categorization of objective measures of stress using wearable sensors.}}
          \end{figure*}

\begin{figure*}
          \begin{center}
          \begin{tabular}{c}
          \includegraphics[width=\linewidth]{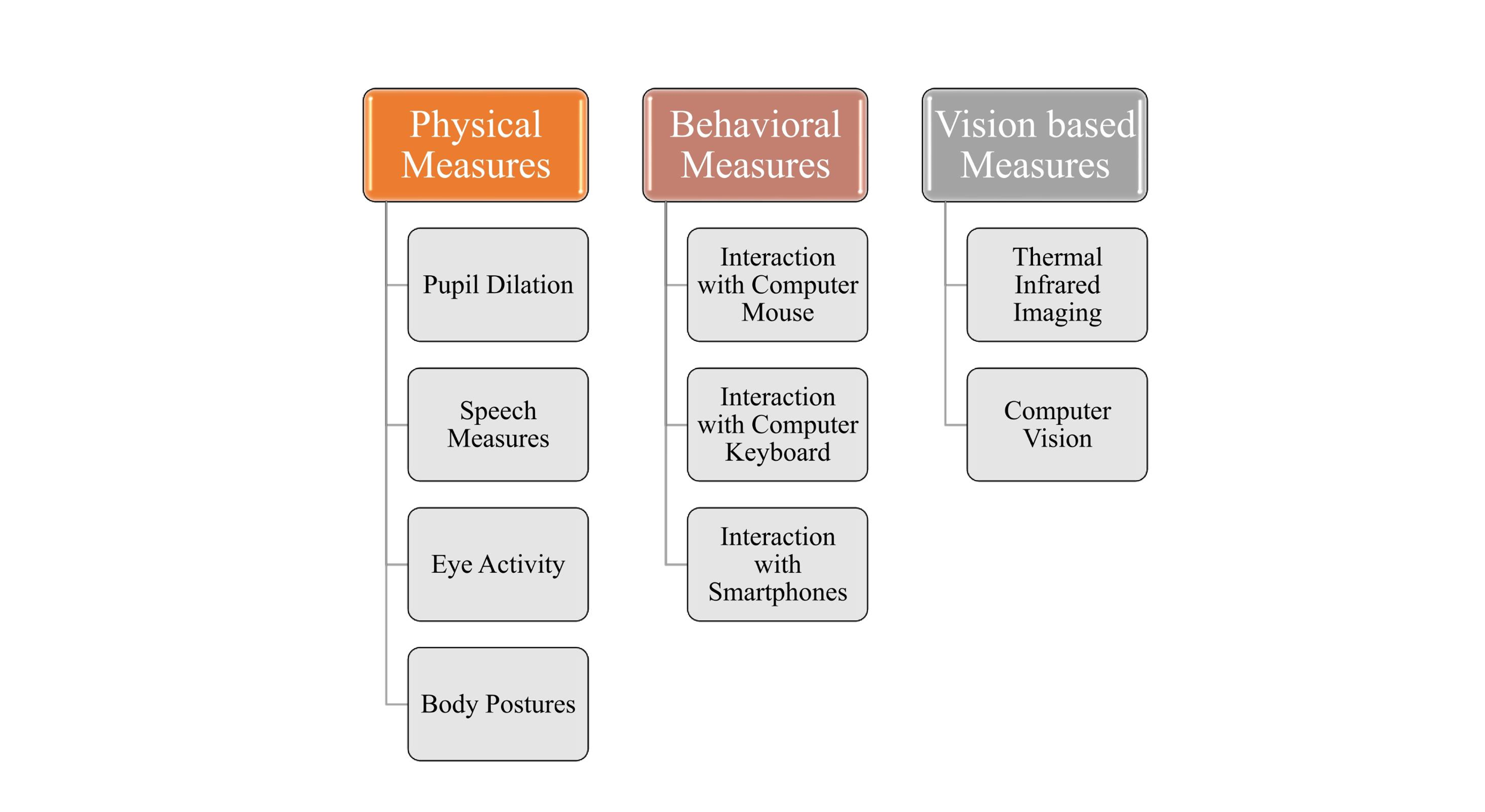}
          \end{tabular}
          \end{center}
          \caption
          { \label{fig:fig3a}
          { Categorization of objective measures of stress using non-wearable sensors.}}
          \end{figure*}

The objective measures of stress can be categorized into methods based on wearable sensors and non-wearable sensors as shown in \Fig{fig2a} and \Fig{fig3a} respectively. The literature corresponding to each of these categories is reviewed in the following subsections.

\subsection{Wearable Sensors based Human Stress Detection}

Wearable sensors need physical devices to be connected to the body of an individual to measure the stress response of the body. The autonomic nervous system (ANS) of a human being has two parts i.e., the sympathetic nervous system and the parasympathetic nervous system. When a person is under stress, changes in human ANS occur. Sympathetic nervous system (SNS) activity is increased whereas activity in the parasympathetic nervous system (PNS) decreases in stressful situations. Wearable sensors-based stress detection is important because it can overcome the limitation of wrong self-reporting by an individual~\citep{garcia1997science,northrup1997problem}. Wearable sensors used for human stress monitoring include electroencephalography (EEG), electromyography (EMG), galvanic skin response (GSR) or electrodermal activity (EDA), electrocardiography (ECG), heart rate (HR), skin temperature (ST), respiratory rate (RR), heart rate variability (HRV), blood volume pressure (BVP), photoplethysmography (PPG), and salivary cortisol (SC).

\subsubsection{Electroencephalography based Stress Detection}

Brain activity has a strong relationship with stress~\citep{dharmawan2007analysis}. For analysis of brain activity, functional magnetic resonance imaging (fMRI), positron emission tomography (PET), and EEG are commonly used. Out of all these methods, EEG is the most commonly used method due to its low cost and non-invasive nature. EEG field originated back in 1924 when the first time EEG recording was performed~\citep{berger1929elektroenkephalogramm}. EEG is a physiological measure used by the research community as well as physicians to record brain activity for the analysis and diagnosis of brain diseases and disorders~\citep{chandra2017role}. EEG signals acquisition can be performed using commercially available consumer-grade as well as medical-grade EEG devices. Medical-grade systems are quite expensive as well as sophisticated and are commonly used for patient monitoring in hospitals, whereas consumer-grade EEG headsets are less expensive but they are not as accurate when compared to medical-grade devices. Both types of systems can consist of dry as well as wet electrodes. Each type of electrode has its pros and cons and thus many factors contribute to the type of device, which is selected for data acquisition. Consumer-grade EEG headsets have several electrodes ranging from $1$ to $16$~\citep{sawangjai2019consumer}, whereas medical-grade EEG caps can have several electrodes ranging from $8$ to $256$ or even more~\citep{troy2012many}. EEG electrodes are small metal plates made of steel and have a silver coating to record brain activity. These electrodes are placed on the human scalp to record brain activity. International 10-20 electrode positioning system specifies the electrode positions in the EEG acquisition system~\citep{trans201210}. Each electrode has a specified name comprising of a letter and a number and a location on the human head, which is standardized. The letter in the name shows the area of the brain where the electrode is placed e.g., F for the frontal lobe and T for the temporal lobe. The right side of the head has even-numbered electrodes, whereas the left side has odd-numbered electrodes~\citep{oostenveld2001five}. EEG electrodes are connected to the data acquisition system in a wired or wireless manner. When there is a change in brain activity the voltage level at different electrodes varies, which corresponds to different diseases and disorders. Amplitude values of EEG signals are approximately around $100 \mu V$. EEG signal is composed of five different frequency bands. The behavior of each EEG frequency band is different in different situations. The frequency range of the EEG signal is from 1 Hz to 50 Hz. Based on the frequency ranges the descending order of brain waves is gamma, beta, alpha, theta, and delta.

\begin{enumerate}
\item \textbf{Gamma Band:} The brain activity that lies in the range of 30 - 45 Hz is usually regarded as a gamma wave or fast beta wave. The occurrence of this wave is rare and associated with brain diseases. The gamma wave is considered a good indicator of event-related synchronization (ERS) of the brain. Tongue movement, right and left index finger movement, and right toe movement have been demonstrated to relate to gamma waves. The association of gamma-band and human stress has been established in literature~\citep{minguillon2016stress}.\\

\item \textbf{Beta Band:} The electrical activity of the brain that lies in the range of 14 - 26 Hz is considered a beta wave. This rhythm is found in waking normal individuals and is associated with thinking, attention, focus, and a panic state. Beta activity is mainly originated in the frontal and central regions of the brain.  It occurs around the tumor regions of the brain. Among different neural oscillations, a higher level of beta waves acts as a marker denoting that a person is not in a calm state~\citep{sanei2013eeg}. The presence of stress has been shown to increase the spectral power in the EEG beta band~\citep{saeed2015psychological,hamid2015brainwaves}.\\

\item \textbf{Alpha Band:} Alpha waves (8 - 13 Hz) can be detected in all parts of the posterior lobes of the brain and commonly appears like a sine wave or a round-shaped signal. Relaxed alertness without attention is considered associated with alpha waves. The alpha wave is the most observable brain activity due to its prominence.  The alpha wave is claimed to be awaiting pattern by visual regions of the brain as in a closed eye state alpha wave is produced. Activities like opening the eyes, listening to unfamiliar sounds, anxiety, or mental attention can reduce or even eliminate the alpha waves. It has an amplitude that is normally less than 50 $\mu$V and is found over occipital regions. Its origin and significance are not known physiologically and require more research and experimentation. Stress has been shown to be associated with a fall in alpha waves~\citep{hoffmann2005brain}.\\

\item \textbf{Theta Band:} Theta waves (4 - 7.5 Hz) originate due to drowsiness and have been associated with creative inspiration and deep meditation. The arousal of an individual is determined by the theta wave. Pathological problems show larger groups of abnormal theta activity in waking adults. Variations in the theta activity are also used in human stress recognition-based studies~\citep{arsalan2019classification}.\\

\item \textbf{Delta Band:} Delta waves (0.5 - 4 Hz) are considered to reflect deep sleep. and the slowest brain waves. Newborn babies and very young children have strong delta wave activities. As the age of the individual increases, the amplitude and occurrence of delta waves are reduced. Delta waves are associated with a deep level of relaxation. These waves can be confused with muscular artifacts produced by the neck and jaw. Therefore, these artifacts need to be removed by applying simple signal processing methods to the EEG signals.
\end{enumerate}

Asymmetry analysis of the EEG signal is an established feature for the classification of different psychological states~\citep{gatzke2014role,giannakakis2015detection}. The asymmetry index of the EEG signal is the difference between the natural logarithm of the power of the right hemisphere from the left hemisphere of the brain. Commonly used locations for the estimation of the alpha asymmetry in stress-related studies are F3-F4~\citep{seo2008relation,lewis2007effect} because these locations are directly affected by the stressful events~\citep{qin2009acute}. However apart from the frontal part of the brain, stress-related studies involving lateral region i.e., F7-F8~\citep{lopez2012frontal}, anterior region i.e., Fp1-Fp2~\citep{peng2013method}, and posterior regions i.e., T5-T6~\citep{minguillon2016stress} of the brain have been reported in the literature. A large number of studies have a consensus over the fact that the alpha band activity in the right frontal part of the brain is dominating as compared to the left half of the brain under stressful condition~\citep{acharya2012application}. This phenomenon is present in a variety of stressful situations like students feeling stressed during examinations~\citep{giannakakis2015detection} when presented with a stimulus of sad/happy/horror movie~\citep{lopez2012frontal,tomarken1990resting} and even in case of chronic stress~\citep{peng2013method}.

The human stress response has been explored using power spectrum and relative power index in a significant number of studies~\citep{hosseini2010higher,khosrowabadi2011brain,sharma2014modeling,ko2009emotion,giannakaki2017emotional}. Alpha band activity is dominant in the relaxation phase when the cognitive demands are minimal whereas on the contrary it has been found that situations involving high strain or alertness, beta-band activity is found to be significant~\citep{hayashi2009beta}. Even though the findings of the stress-related studies are contradictory, still stressful conditions has found to decrease the alpha band activity~\citep{alonso2015stress,demerdzieva2011eeg,al2015mental,tran2007detecting,seo2010stress} and increase the beta band activity~\citep{katsis2011integrated}. Stress is found to be correlated to the beta wave in the temporal part of the brain~\citep{choi2015measurement}. Coherence among different brain regions is another important factor that varies with stress. Beta and theta band coherence is increased whereas the alpha band coherence is decreased in the anterior location of the brain hemisphere~\citep{alonso2015stress}. When the person is having a negative mood or depression the alpha and beta band activity is dominant~\citep{huiku2007assessment}. Alpha band activity of the prefrontal electrodes is reduced when the person is facing a stressful event~\citep{marshall2015effects}. The temporal lobe electrode shows a dominant activity of alpha-band when the stressful event occurs~\citep{choi2015measurement}. To classify the mental stress from the resting state EEG a method is proposed in~\citep{sharma2012objective}.

Different level of mental stress is measured in participants using single-channel EEG headset and mental workload and public speaking task as a stimulus in a study conducted in~\citep{secerbegovic2017mental}. Alpha and beta bands of the EEG signal are found to be statistically significant bands and a classification accuracy of 83.33\% for binary stress classification was reported. An EEG-based multiple-level human stress classification framework using 128 channels EEG cap and MIST as a stimulus was presented in~\citep{subhani2017machine}. The study reported an average classification accuracy of 94.6\% and 83.4\% for binary and multi-level stress classification respectively. Another study for acute stress classification using SWT as a stressor and EEG as a modality is presented in~\citep{hou2015eeg}. Two, three, and four levels of stress were measured in this study with an average classification accuracy of 85.71\%, 75.22\%, and 67.06\%, respectively. Another human stress classification study for two and three-class problems with the mental arithmetic task as a stimulus and Emotiv EPOC as an EEG data acquisition device is presented in~\citep{jun2016eeg}. Classification accuracy of 96\% and 75\% is achieved for two and three-level stress classes, respectively.  A study in~\citep{duru2013assessment} focuses on stress level detection of the surgeon during the most stressful phases of an operation via EEG. In~\citep{calibo2013cognitive} authors used the Stroop colored word test to elicit stress in the subjects. After applying preprocessing techniques to the EEG data, features were extracted for further analysis which was then classified using k nearest neighbor and logistic regression classifiers with an accuracy of 73.96\%. In~\citep{pomer2014methodology} authors proposed a methodology for analyzing the stress of military firefighters based on asymmetry levels of alpha waves. In~\citep{vijayaragavan2015eeg}, the authors developed an android application that reduces stress using music and yoga. EEG signals were recorded using the Neurosky headset and were preprocessed by using a high pass filter and proprietary algorithms of Neurosky. A survey of 100 users was conducted in which better relaxation waveforms were observed in the case of yoga by 67\% members and 29\% readings showed better results in the case of music and 4\% reported no results. Hand movement, heart rate variation, and EEG are used to analyze stress, and office syndrome was detected by an intelligent watch in~\citep{reanaree2016stress}.

In another study, the mind-ball game has been used to establish a correlation between human stress and the winning percentage of the game. The study concludes that the person with a lower stress level wins the game more often~\citep{lin2006quantifying}. An EEG and Hilbert Huang Transform-based stress measurement method with a support vector machine (SVM) as a classifier has been proposed in~\citep{vanitha2016real}. Another stress measurement scheme using EEG signals is proposed in~\citep{kalas2016stress}. A method of measuring mental stress is proposed based on the Montreal Imaging Stress Task (MIST) as a stimulus, power spectral density, and energy as a feature, and SVM as a classifier~\citep{al2015mental}. Another EEG-based mental stress classification scheme using mental arithmetic tasks as a stimulus to classify stress into three levels is proposed in~\citep{pandiyan2013mental}. A stress classification framework in response to Urdu and English music tracks is presented in~\citep{asif2019human}. Classification accuracy of 98.76\% and 95.06\% for two- and three classes is achieved using a logistic regression classifier. A driver stress measurement framework using an EEG signal has been presented in~\citep{halim2020identification}. In the proposed scheme data from 86 automobile drivers were used. EEG data were recorded to log the ongoing brain activity during driving to find a correlation between the brain activity with the emotional response of the driver. Three different classification algorithms which include SVM, neural network and RF were used to classify the driver's emotional state based on the labeling obtained by the self-reporting questionnaire. The SVM classifier was found to be the best among them with an achieved classification accuracy of 97.95\% for stress vs relaxed state. An EEG-based multilevel stress classification scheme in nurses and non-health professionals is presented in \cite{akella2021classifying}. An accuracy of 91\% is reported for four-level stress classification.  Another EEG-based stress measurement scheme using K-Means Clustering and Support Vector Machine classifier is presented in \cite{wen2022hybrid}. An accuracy of 98\% is achieved for three-level stress classification. An EEG-based stress recognition framework using Symmetric Convolutional and Adversarial Neural Networks is presented in \cite{fu2022symmetric}. The highest classification accuracy of 85.1\% is achieved for five-level stress classification. Another EEG-based study to analyze the brain activity of stressed and non-stressed individuals in response to high arousal images as a stimulus is presented in \cite{batool2022brain}. The study found no significant difference in the EEG data for the two groups.   

All of the EEG stress detection methods discussed above are for acute stress. EEG signals have also been used to assess and classify perceived stress. An EEG-based study to identify the appropriate phase of EEG recording for the classification of perceived stress is presented in~\citep{arsalan2019classification}. EEG data of the participants were recorded for a duration of three minutes in open-eye conditions before and after performing the public speaking activity. The study concluded that the pre-activity phase was better for perceived stress classification. Two-level and three-level stress classification were performed and the classification accuracy of 92.85\% and 64.28\% was achieved, respectively. A perceived stress classification study using closed-eye resting-state EEG data is presented in~\citep{saeed2015psychological}. The authors reported that there exists a relationship between the PSS questionnaire score and the EEG data of the subject. The beta band of the EEG signal was found to be directly proportional to the level of perceived stress i.e., individuals having high perceived stress have increased beta-band activity and vice versa. Another single-channel EEG headset-based stress quantification study using resting-state and closed-eye EEG data is presented in~\citep{saeed2017quantification}. Multiple linear regression analysis was performed and beta waves of the EEG signals were found to be the optimum frequency band for the prediction of the PSS questionnaire score of the subject with a confidence interval of 94\%. The correlation-based feature selection (CFS) method has been used in an EEG-based perceived human stress classification scheme proposed in~\citep{saeed2018selection}. The CFS method yielded the result that beta and low gamma frequency bands have the highest correlation with the PSS score of an individual. A significant difference in the energy spectral density of the alpha and beta bands of the EEG signals in the right and left hemispheres of the brain for stressed and non-stressed individuals is observed in a study conducted in~\citep{hamid2015brainwaves}. Alpha asymmetry has been found as a useful marker of the relationship between human stress and EEG signals in a study conducted in~\citep{sulaiman2011intelligent}. The left hemisphere of the brain has strong activity among individuals who have low chronic stress whereas, on the other hand, the right hemisphere of the brain has strong activation in subjects having moderate and high chronic stress. A study conducted in~\citep{hamid2010evaluation} found that the PSS questionnaire score and ratio of alpha and beta band of the EEG signal have a negative correlation among themselves. Therefore individuals with high PSS scores have a negative ratio whereas individuals with low PSS scores have been found to have a positive ratio. The correlation of the EEG temporal characteristics with the recorded EEG signals and the PSS questionnaire score is presented in~\citep{luijcks2015influence}. The study concluded the fact that the theta and delta wave of the EEG signal of the participants having high PSS questionnaire scores has an increased activation in the post-stimulus phase when compared to the pre-stimulus phase. Moreover, theta band activity in the frontal part of the brain was higher in the post-stimulus phase. Another perceived stress classification study using resting-state EEG data based on the PSS questionnaire as well as psychologist interview labeling is presented in~\citep{saeed2020eeg}. The study concluded that by using the psychologist interview labeling, a classification accuracy of 85.20\% was achieved. \Tab{tab1} presents a summary of human stress classification schemes using EEG signals.

\begin{table}[H] 
\caption{Summary of Human Stress Detection Studies using EEG signals.\label{tab:tab1}}
\scalebox{0.9}{
\newcolumntype{C}{>{\centering\arraybackslash}X}
\begin{tabularx}{\textwidth}{CCCCCCCC}
\toprule
\textbf{Method}	& \textbf{Type of Stress}	& \textbf{Number of Subjects (M/F)} & \textbf{Age} & \textbf{Stimulus}& \textbf{Features Domain}& \textbf{Classifier}& \textbf{Accuracy (Classes)}\\
\midrule
~\citep{halim2020identification} & Acute & 86 & -- & Driving & Time and Frequency & SVM, RF, NN & 97.95\% (2) \\
~\citep{subhani2017machine} & Acute &  22 & 19-25 & MIST & Frequency & LR, SVM, NB & 94.60\% (2) 83.40\% (multilevel) \\
~\citep{dharmawan2007analysis} & Acute & 20 (18/2) & 20-35 & Game & Frequency & DT & 79.08\%  \\
~\citep{secerbegovic2017mental} & Acute & 9 (6/3) & 19.3-22.7  & MAT, Computer Games & Time and Frequency & SVM & 86.66\% (3) \\
~\citep{hou2015eeg} & Acute & 9 & 21-28 &  SCWT & Frequency & SVM & 85.71\% (2) 75.22\% (3) 67.06\% (4) \\
~\citep{jun2016eeg} & Acute & 10 (9/1) & 20-35 & MAT, SCWT & Frequency & SVM & 75.00\% (3) 96.00\% (2)  88.00\% (2) \\
~\citep{calibo2013cognitive} & Acute & 18 & -- & SCWT & Frequency & LR, kNN & 73.96\% (2) \\
~\citep{hosseini2010higher} & Acute & 15 (15/0) & 20-24 & IAPS & Frequency & SVM & 82.00\% (2) \\
~\citep{giannakaki2017emotional} & Acute & 5 (5/0) & 22-38 & IAPS & Frequency & RF & 75.12\% (2) \\
~\citep{al2015mental} & Acute & 12 (12/0) & 20-24 & MIST & Wavelet & SVM & 94.00\% (L1), 85.00\% (L2), 80.00\% (L3) \\
~\citep{vanitha2016real} & Acute & 6 & -- & MQ & Frequency & hierarchical SVM & 89.07\% (2) \\
~\citep{asif2019human} & Acute & 27 (13/14) & 20-35 & Music Tracks & Frequency & SMO, LR & 98.76\% (2) 95.06\% (3) \\
~\citep{khosrowabadi2011brain} & Chronic & 26 (20/6) & 18-30 & University Exam & Frequency & kNN, SVM & 90.00\% (2) \\
~\citep{saeed2015psychological} & Chronic & 28 (18/10) & 22-33 & Baseline & Frequency & SVM & 71.42\% (2) \\
~\citep{saeed2017quantification} & Chronic & 28 (18/10) & 22-33 & Baseline & Frequency & NB & 71.42\% (2) \\
~\citep{saeed2018selection} & Chronic & 28 (18/10) & 22-33 & Baseline & Frequency & SVM & 78.57\% (2) \\
~\citep{saeed2020eeg} & Chronic & 33 (20/13) & 18-40 & Baseline & Frequency & SVM & 85.20\% (2) \\
~\citep{arsalan2019classification} & Chronic & 28 (13/15) & 18-40 & Baseline & Frequency & MLP & 92.85\% (2) 64.28\% (3) \\

\bottomrule
\end{tabularx}
}

   \noindent{\footnotesize{LR: Logistic Regression, SVM: Support Vector Machine, kNN: k- Nearest Neighbors, NB: Naive Bayes, SMO: Sequential minimal optimization, RF: Random Forest, MLP: Multilayer Perceptron, DT: Decision Tree, NN: Neural Networks, MIST: Montreal Imaging Stress Task, MAT: Mental Arithmetic Task, SCWT: Stroop color-word test, IAPS: International Affective Picture System, MQ: Mathematical questions}}
\end{table}

\subsubsection{Electromyography based Stress Detection}

EMG is a biomedical signal that deals with the electric current generated in the muscles of the human body during its contraction representing neuromuscular activities. EMG signal is recorded via a device called an electromyograph. EMG signals are recorded by placing the sensors near the muscle whose movement needs to be measured. The amplitude of the signal lies in the range of 1-10 mV and the frequency range of the EMG signal is 0-500 Hz with the dominant frequencies between 50-150 Hz~\citep{de2002surface}. EMG is a complicated signal being controlled by the human nervous system and is strongly dependent on the physiological and anatomical characteristics of the skeletal muscles. EMG signals become noisy while moving through different tissues of the human skin. Moreover, the EMG data acquisition device acquires signals from various motor units resulting in an overlap of other muscles' movement to the desired muscle movement.

Recently, the measurement of EMG signals using sophisticated equipment has gained a lot of interest in the field of biomedical engineering~\citep{farfan2010evaluation}. EMG has also been a focus of biomedical experts because of its clinical and diagnostic applications. Robotic arms and rehabilitation of patients have been potentially identified as key areas of applications for EMG signal recording. Motor Unit Action Potentials (MUAPs) (their shapes and firing rates) are useful for the treatment of a variety of neuromuscular disorders.

Advancement in the available signal processing techniques has made the design and development of state-of-the-art EMG detection and diagnosis techniques a practical possibility. A wide range of mathematical and artificial intelligence (AI) based techniques have gained attention~\citep{reaz2006techniques}. Mathematical models used for EMG signal analysis include wavelet transform, Wigner-Ville Distribution (WVD), Fourier transform, and higher-order statistics. On the other hand, AI-based techniques used include artificial neural networks, dynamic recurrent neural networks, and fuzzy logic. The recorded EMG signal faces two important challenges which include (i) the signal-to-noise ratio (i.e., the ratio of the energy of the EMG signal to the energy of the noise) and (ii) distortion in the recorded EMG signal (i.e., there should be no alteration in the contribution of each frequency component of the signal). A typical EMG recording is done in two phases i.e., the baseline recording and EMG recording in response to some stimulus, and then measured as a ratio of baseline and stimulus-response. Baseline recording is necessary because this level is different for every individual depending on a variety of factors~\citep{weyers2006electromyographic}. Facial EMG has been extensively used in the literature to record facial expressions in response to some kind of stimulus. This finding has been reported by  Ekman and Friesen in their study in~\citep{ekman1978technique}. 
  
The relationship between human stress and EMG signal has been discussed in a wide range of studies in the literature. A study to investigate the relationship between the changes in human stress level and muscular tension via EMG signal is presented in~\citep{karthikeyan2012emg}. Stroop color-word test was used as a stimulus and EMG data were acquired from the left trapezius muscle of the participants. Pre-processing of the acquired EMG data was performed using the wavelet de-noising technique and time-domain features were extracted from the data. kNN classifier was used and a classification accuracy of 90.70\% was achieved. A study to validate the stress-EMG paradigm using an unpredictable and uncontrollable stimulus is presented in~\citep{luijcks2014experimentally}. The stimulus given to the participants was an electro-shocker to give electric shocks of 10 milliseconds duration. The experiment performed includes a 3-minutes baseline recording, 3 minutes recording before the stimulus, and 2 minutes post-stimulus recording. EMG activity of the trapezius muscles was significantly higher than in the pre-activity phase when compared to the other two phases of the experiment. The study concluded that the presented stimulus is a reliable and valid test to identify the difference between stressed and non-stressed individuals. The activities of the human muscles like the trapezius are associated with stress~\citep{lundberg1994psychophysiological,wijsman2010trapezius,larsson1995effects}. In~\citep{lundberg1994psychophysiological}, the author has designed an experimental study to investigate the effect of mental stress and physical workload separately and in the combined form on the perceived human stress, physiological signals, and the muscular tension faced by an individual by using EMG signal. Stressor given to the subjects includes mental arithmetic task, Stroop color-word test, cold pressor test, standardized test contractions (TC), and a combination of SWT with TC. The results indicate that when compared to baseline recording, stressors induced an increase in blood pressure, heart rate, salivary cortisol, urinary catecholamines, and self-reported questionnaire score. Mental arithmetic tasks caused a significant amount of increase in the EMG activity, SWT when used alongside TC produced more pronounced changes in the EMG signal as compared to SWT alone. The study concluded that an increase in muscular tension was observed when facing only mental stress as well as facing mental stress along with physical workload. Human stress measurement using EMG signals recorded from the upper trapezius muscle is presented in a study conducted in~\citep{wijsman2010trapezius}. The authors have designed two new stress measurement tests for the experiment. Three different stressful situations which include a memory task, a logical puzzle, and a calculation task were presented to the subject and EMG signals of the upper trapezius muscle were recorded. The study revealed the fact that EMG activity of the upper trapezius muscle was higher when facing a stressor as compared to rest condition thus making EMG signal a good indicator of mental stress. Another study to correlate mental stress, blood flow, and the EMG signals recorded from the upper trapezius muscles using the Stroop color-word test is presented in~\citep{larsson1995effects}. The study concluded that there was a decrease in muscle blood flow and an increase in heart rate during the stressor phase. EMG activity of trapezius muscle is increased in response to stress during cold pressor and Stroop color-word test~\citep{krantz2004consistency}. Moreover, an increase in the blood pressure, heart rate, and urinary epinephrine and norepinephrine was observed when facing the stressor but no correlation could be found with the salivary cortisol measure. Another important finding of the study was that men have higher blood pressure and an increase in epinephrine as compared to women, whereas, on the other hand, women have an increased heart rate as compared to men. A positive correlation between the negative stress rating and EMG signal during work has been found in a study conducted in~\citep{rissen2000surface}. A study reported that the EMG of the trapezius muscle is increased under low or high mental workload during computer data entry work~\citep{schleifer2008mental}. Moreover, a decrease in the EMG-gap of the left, as well as right trapezius muscles, was greater during high mental workload as compared to low mental workload. Another study to measure the influence of EMG-based methods and human stress-based methods on the shoulder muscles forces is presented in~\citep{engelhardt2015comparison}. Another study to analyze the stress in the lower back of a subject while at work for different postures and positions using the EMG signal is presented in~\citep{tyagi2017stress}. The study founds an application in the area of chair design for a comfortable sitting posture of an individual at work. \Tab{tab2} presents a summary of human stress classification schemes using EMG signals.

\begin{table}[H] 
\caption{Summary of Human Stress Detection Studies using EMG signals.\label{tab:tab2}}
\scalebox{0.9}{
\newcolumntype{C}{>{\centering\arraybackslash}X}
\begin{tabularx}{\textwidth}{CCCCCCCC}
\toprule
\textbf{Method}	& \textbf{Type of Stress}	& \textbf{Number of Subjects (M/F)} & \textbf{Age} & \textbf{Stimulus}& \textbf{Features Domain}& \textbf{Classifier}& \textbf{Accuracy (Classes)}\\
\midrule
~\citep{karthikeyan2012emg} & Acute & 10 (0/10) & -- &  SCWT & Wavelet & kNN & 90.70\% (4) \\

\bottomrule
\end{tabularx}
}

    \noindent{\footnotesize{SCWT: Stroop color-word test, kNN: k- Nearest Neighbors}}
\end{table}

\subsubsection{GSR based Stress Detection}

The skin response of an individual is affected whenever we confront an emotional stimulus like listening to audio, watching a video, or an emotional real-life event. It is pertinent to mention that whatever the reason for the emotional arousal i.e., whether it is due to happiness or excitement or due to fear, anger, depression, or stress, in either case, the skin response of the person changes~\citep{farnsworthgsr}. The response of the human skin is not under human conscious control~\citep{udovivcic2017wearable} and is dependent on the changes in the sweating pattern of a subject and thus reflects the behavior of the sympathetic nervous system~\citep{wu2010analysis}. Another study supports the fact that some signals are generated from the sympathetic nervous system when a change in the skin conductance of a person occurs~\citep{lidberg1981sympathetic}. Sweat reaction occurs due to any emotional change that can be seen in the fingers and palm. The amount of salt in the human skin varies as a result of a sweat reaction thus causing a change in the electrical conductance of the skin~\citep{ayata2017emotion}. As time goes on the sweat glands of a person become more active resulting in a dis-balance of positive and negative ions and thus as a result affecting the flow of current through the skin~\citep{critchley2002electrodermal}.

GSR measurement locations are part of the body with a large number of sweat glands. There exist a variety of possible locations on the human body for the measurement of GSR. Common locations for measuring skin response include the fingers, shoulders, feet, and wrists of a person. According to the studies, the palm and fingers of the skin have the highest number of sweat glands and are used as a location for GSR recording in experiments. GSR activity is typically measured in “micro-Siemens ($\mu S$)” or “micro-Mho ($\mu M$)”. Sweating secretion is increased when a person faces some emotional stimuli, whether positive or negative and due to these measurable changes in skin conductance occurs.

One of the most widely used measures of skin activity is GSR, which is also called Electrodermal Activity (EDA) or Skin Conductance (SC). EDA is a physiological measure of the flow of electricity through human skin. Even a small amount of sweating, which is not visible to the naked eye on the surface of the human skin causes a change in its electrical conductivity. EDA can be divided into (i) Skin Conductance Level (SCL) which is the slowly changing part of the EDA, (ii) Skin Conductance Response (SCR) which corresponds to the peaks in the EDA due to some kind of stimulus, and (iii) Non-specific Skin Conductance Response (NS.SCR) which exists even without the existence of any external stimulus. The pattern of the skin response data is distinct according to the state of the person and is considered one of the reliable stress measurement method~\citep{kurniawan2013stress}. SCR part of EDA increases when encountered with an emotionally arousing situation~\citep{dawson2007electrodermal}. NS.SCR part of the EDA corresponds to the cognitive processes and the psycho-physiological states~\citep{nikula1991psychological}.

The skin conductance of a person is increased when the person is stressed, whereas skin conductance gets reduced when the person is relaxed~\citep{liao2005real}. GSR has been used for cognitive load measurement in the literature~\citep{shi2007galvanic}. The index and the middle finger of the hand of the subject are commonly used as a location for the placement of GSR electrodes because of the existence of a sufficient number of sweat glands to measure the skin response changes. The use of GSR sensors for stress measurement has been the focus of study in~\citep{healey2000wearable}. Whenever the person is under stress, the moisture in the human skin increases resulting in an increase in the SCL~\citep{giakoumis2012using,blechert2006identifying,ritz2000emotions,reinhardt2012salivary,hoehn1989somatic} and SCR~\citep{setz2009discriminating,ren2012affective,lee2004development,blechert2006identifying,hoehn1989somatic,nomikos1968surprise,lanzetta1976effects} part of the electrodermal activity. SCR peaks commonly appear between 1.5 and 6.5 seconds after the start of the stimulus. In another study, SCL was found to be the most effective marker for measuring stress in comparison to HRV and EMG. Some of the commonly extracted features from the electrodermal activity for human stress detection include SCR frequency, SCR amplitude, SCR latency, SCR rise time, SCR half recovery, SCR recovery time, SCR response onset, and SCR half recovery. Another interesting fact observed in a study conducted in~\citep{nomikos1968surprise} is that even the expectation of a stressful situation that has yet not occurred can cause an increase in the EDA similar to the situation if the event has occurred. Many other factors that can affect GSR measurement include the temperature and humidity of the environment. In summary, it can be concluded that the SCR and SCL part of the EDA consistently increases under stress conditions. A chronic stress measurement mechanism using GSR signals is presented in~\citep{panigrahy2017study}. Data from the participants is recorded in three different states i.e., sitting, standing, and sleeping. Stressed and relaxed conditions are discriminated against with a classification accuracy of 76.5\%. Another human stress measurement scheme for the office environment using physiological signals of GSR is proposed in~\citep{hernandez2011call}. An analysis of the self-reported measure obtained from the employees and call logs checked to determine the number of stressed and non-stressed calls were performed. The SVM classifier was used to classify the stressed and non-stressed individuals and a classification accuracy of 73\% was achieved. Another framework for the measurement of stress at the workplace using EDA signal is proposed in~\citep{kocielnik2013smart}. Self-Assessment Manikin Questionnaire was used as a subjective measure and pre-processing of data were performed by the removal of data from the first 15 seconds and the last 10 seconds of the recorded signal. The authors did not report any classification accuracy but they stated that the results obtained from the analysis are meaningful and provide useful information. A likelihood ratios-based stress detection framework using EDA sensors is presented in \cite{lee2021pen}. An accuracy of 87.5\% is achieved for stressed vs relaxed state classification. \Tab{tab3} presents a summary of human stress classification schemes using GSR signals.

\begin{table}[H] 
\caption{Summary of Human Stress Detection Studies using GSR signals.\label{tab:tab3}}
\scalebox{0.9}{
\newcolumntype{C}{>{\centering\arraybackslash}X}
\begin{tabularx}{\textwidth}{CCCCCCCC}
\toprule
\textbf{Method}	& \textbf{Type of Stress}	& \textbf{Number of Subjects (M/F)} & \textbf{Age} & \textbf{Stimulus}& \textbf{Features Domain}& \textbf{Classifier}& \textbf{Accuracy (Classes)}\\
\midrule
~\citep{healey2000wearable} & Acute & 9 & -- &  Car driving & Time & LDA & 96.00\% (4) \\
~\citep{blechert2006identifying} & Acute & 42 (14/28) & 42.2$\pm$9.9 & Pictures & Time & DFA & 83.30\% (2) \\
~\citep{setz2009discriminating} & Acute & 33 (33/0) & 24.06 & MIST & Time & LDA & 82.80\% (2) \\
~\citep{ren2012affective} & Acute & 30 (14/16) & 26.8$\pm$2.56 & SCWT & Time & NB & 85.5\% (2) \\
~\citep{lee2004development} & Acute & 80 & -- & SCWT & Time & MLP, GRNN, ANFIS & 96.67\% (2) \\
~\citep{panigrahy2017study} & Acute & 10 & -- & Computer game & Time & J48 & 76.50\% (2) \\
~\citep{hernandez2011call} & Acute & 9 (4/5) & -- & Call center & Time & SVM & 73.41\% (2) \\

\bottomrule
\end{tabularx}
}

   \noindent{\footnotesize{MIST: Montreal Imaging Stress Task, SCWT: Stroop color-word test, NB: Naive Bayes, MLP: Multilayer Perceptron, J48: Decision Tree, LDA: Linear Discriminant Analysis, DFA: Discriminant function analysis, GRNN: Generalized regression neural network, ANFIS: Adaptive network-based fuzzy inference system}}
\end{table}

\subsubsection{Electrocardiography based Stress Detection}

ECG is one of the most commonly used techniques for monitoring the functionality of the heart. ECG is a non-invasive modality used for the assessment of the electrical activity of the heart in real time. The activity of the heart is correlated to the human central system. Apart from the monitoring of heart functionality, it is also useful for human stress measurement~\citep{ahn2019novel}. The most commonly used method for the measurement of ECG is a 12-lead ECG technique. In this technique, nine sensors are placed on the human body at specified locations. Three main sensors are placed on the right arm, left arm, and left leg of the person. The sensor placed on the right leg act as a reference electrode for the ECG acquisition system. Even though the complete picture of the heart cannot be obtained by using only these three sensors, but a physician can use these schemes for quick analysis in case of emergency treatment. For higher resolution results, six sensors are placed on the chest of the individual. Using these nine sensors along with the leads (Lead I, Lead II, Lead III) interconnecting their results in a total of twelve leads. One of the most important advantages of this twelve lead system is that it gives detailed information about the heart activity of the subject thus leading to a better diagnosis and cure, whereas, on the other hand, the largest disadvantage of this 12 lead system is that it produces a huge amount of data especially when recording is done for many hours. ECG signals are characterized by peaks that include  P, Q, R, S, T, and U. Each of these peaks has its characteristics and gives a piece of specific information about the heart activity of the individual~\citep{al2007hardware}. Commonly used parameters for the assessment of ECG signals include P, PR, QRS complex, and QT. For medical purposes, all these four parameters are evaluated. For other applications, there may be some peaks that are more important than others.

A wide range of studies for human stress measurement using ECG signals has been presented in literature~\citep{karthikeyan2012study,karthikeyan2011ecg}. A human stress classification scheme using ECG signal and mental arithmetic task as a stimulus is presented in~\citep{karthikeyan2012study}. Statistical features are extracted using discrete wavelet transform and low and high-frequency bands of the ECG signals are analyzed separately. Three-level human stress classification i.e., low stress, medium stress, and high stress is performed using a kNN classifier. Maximum classification accuracy of 96.3\% and 75.9\% is achieved for low-frequency and high-frequency bands, respectively using the covariance feature. Another human stress recognition framework using ECG and discrete wavelet transform is presented~\citep{karthikeyan2011ecg}. The Stroop color-word test is used as a stimulus and heart rate variability is extracted as a feature from the recorded ECG signal. An accuracy of 96.41\% is achieved for stress vs relaxed state classification using the kNN classifier. Another human stress assessment scheme based on a mono-fuzzy index extracted from ECG or GSR signals is presented in~\citep{charbonnier2018multi}. Four different stress tasks are used in the experiment which includes a mental arithmetic stress task, a mental arithmetic control task, a trier social stress test, and a trier social control test, and a classification accuracy of 72\% was achieved for stress and no-stress classes. In~\citep{liu2014listen}, the authors presented a stress classification scheme using ECG signals. The dataset used in the experiment was adopted from PhysioNet~\citep{goldberger2000physiobank} and it consists of physiological signals of GSR and ECG recorded while drivers are in a rest state or were experiencing stressful events. Time and frequency domain features of the heart rate variability and spectral power of the ECG signal are used. An F-measure of 0.85 is achieved for stress classification using an SVM classifier. Acute stress classification using ECG signals is presented in~\citep{tanev2014classification}. Four different stimuli which include images, audio, mental tasks, and rest state are used in the experiment. Linear, as well as non-linear features, are extracted from the HRV data obtained from the ECG signals and classification accuracy of 80\% is achieved for acute stress classification. ECG signal has been analyzed for human stress classification in~\citep{bong2012analysis}. Time-domain features which include heart rate, mean R peak amplitude and mean R-R intervals are extracted from the ECG signal. kNN and SVM were used for classification and mean classification accuracy of 77.69\% and 66.49\% is achieved for two and three classes, respectively. Short-term ECG and heart rate variability signals are used for human stress classification in a study conducted in~\citep{karthikeyan2013detection}. The Stroop color-word test is used as a stimulus, and pre-processing of the acquired data is performed using a wavelet de-noising algorithm. Frequency domain features are extracted from the HRV signal which is obtained from the recorded ECG signals. Classification is performed using a probabilistic neural network and kNN classifiers with an average achieved an accuracy of 91.66\%. A driver stress recognition framework using ECG is proposed in~\citep{keshan2015machine}. The study aimed at the classification of stress at three different levels i.e., low, medium, and highly stressed. Seven different classifiers which include the Naive Bayes, Logistic Regression, Multilayer Perceptron, SVM, J48, kNN, and random forest classifiers are used for the classification purpose. The decision tree algorithm gave the best classification results with achieved accuracy of 88\% for three classes. A stress measurement mechanism among students during an oral exam using ECG signals is presented in~\citep{castaldo2016detection}. ECG data of the student is recorded during the oral exam as well as after the vacations which acted as a baseline recording. Time and frequency domain features are extracted from the recorded data and subjected to classification using Naive Bayes, Decision Tree, SVM, and Multilayer Perceptron classifiers. The best classification accuracy of 80\% is achieved using a decision tree classifier. \Tab{tab4} presents a summary of human stress classification schemes using ECG signals.

\begin{table}[H] 
\caption{Summary of Human Stress Detection Studies using ECG signals.\label{tab:tab4}}
\scalebox{0.9}{
\newcolumntype{C}{>{\centering\arraybackslash}X}
\begin{tabularx}{\textwidth}{CCCCCCCC}
\toprule
\textbf{Method}	& \textbf{Type of Stress}	& \textbf{Number of Subjects (M/F)} & \textbf{Age} & \textbf{Stimulus}& \textbf{Features Domain}& \textbf{Classifier}& \textbf{Accuracy (Classes)}\\
\midrule
~\citep{karthikeyan2012study} & Acute & 10 (0/10) & 20-25 & MAT & Wavelet & kNN & 96.30\% (4) \\
~\citep{karthikeyan2011ecg} & Acute & 10 (0/10) & 20-25 & SCWT & Wavelet & kNN & 96.41\% (2) \\
~\citep{charbonnier2018multi} & Acute & 20 & 19-30 & SCWT & Frequency & mono-feature fuzzy index & 72.00\% (4) \\
~\citep{tanev2014classification} & Acute & 10 (8/2) & 22-26 & IAPS, IADS & Time and Frequency & NB & 90.00\% (2) \\
~\citep{bong2012analysis} & Acute & 5 & -- & Audio-visual & Time & SVM,kNN & 77.69\% (2)  66.49\% (3) \\
~\citep{karthikeyan2013detection} & Acute & 60 (30/30) & 21-25 & SCWT & Time and Frequency & kNN, PNN & 91.66\% (2) \\
~\citep{keshan2015machine} & Acute & 17 & -- & Driving & Time & NB, LR, MLP, SVM, DT, kNN, RF & 88.00\% (3) \\
~\citep{castaldo2016detection} & Acute & 42 & -- & Oral Examination & Time and Frequency & NB, DT, SVM, MLP & 80.00\% (2) \\

\bottomrule
\end{tabularx}
}

   \noindent{\footnotesize{MAT: Mental Arithmetic Task, SCWT: Stroop color-word test, NB: Naive Bayes, MLP: Multilayer Perceptron, LR: Logistic Regression, DT: Decision Tree, SVM: Support Vector Machine, kNN: k- Nearest Neighbors, RF: Random Forest, IAPS: International Affective Picture System, IADS: International Affective Digital Sounds, PASAT: Paced Auditory Serial Addition Task, FDA: Fisher discriminant algorithm, PNN:  Probabilistic neural network }}
\end{table}

\subsubsection{Heart Rate based Stress Detection}
HR is one of the most widely used measures of human stress available in the literature. Heart rate is defined as the number of heartbeats in one minute (measured in beats per minute (bpm)). The RR interval of the ECG signal, which is defined as the interval between consecutive heartbeats has an inverse relationship with the heart rate of a person. In literature, there exist a large number of studies that report a significant increase in heart rate when facing a stressful situation~\citep{giannakakis2017stress,engert2014exploring,vinkers2013effect,lundberg1994psychophysiological,krantz2004consistency,finsen2001muscle,reinhardt2012salivary,acerbi2016wearable,moriguchi1992spectral,steptoe2001acute,ring2002shifting,tugade2004resilient,vuksanovic2007heart,schubert2009effects,clays2011perception,lackner2011phase,van2015ambulatory}. A human stress detection framework based on facial cues using features of eye movement, mouth activity, head movement, and heart rate acquired via PPG signal is presented in~\citep{giannakakis2017stress}. Four different stressors which include social exposure, emotional recall, stressful images, and stressful videos are used in the experiment. Feature selection is applied to select the optimum set of features for the discrimination of a stress state from a neutral state. Five different classifiers which include kNN, Generalized Likelihood Ratio, SVM, the Naïve Bayes, and the AdaBoost classifier are employed for classification purposes. Maximum classification accuracy of 91.68\% is achieved by the AdaBoost classifier using social exposure stressors. Another study presenting a comparison of the use of thermal infrared imaging for measurement of human stress with other stress biomarkers of heart rate, heart rate variability, alpha-amylase, cortisol, and finger temperature is presented in~\citep{engert2014exploring}. Two different stressors which include the cold pressor test and the trier social stressor test are used in the experiment. The study reported the fact that under stressful situations heart rate of the subjects is increased whereas it decreased in the recovery phase. In another study conducted in~\citep{lundberg1994psychophysiological}, the authors reported an increase in the heart rate of the individuals under stress conditions as compared to baseline rest state condition. Women are found to have an increased heart rate as compared to men when facing stressors, whereas men have higher blood pressure as compared to women under stress conditions in a study presented in~\citep{krantz2004consistency}. The cardiovascular response of individuals in response to computer mouse work with and without memory demands is discussed in a study presented in~\citep{finsen2001muscle}. The study found that with the increasing memory demands, the heart rate of the individuals is increased. A new stress induction protocol named Mannheim Multi-component Stress Test (MMST) is designed in a study conducted in~\citep{reinhardt2012salivary}. The MMST protocol included mental arithmetic tasks, effective images, sounds, and motivational stressors. The heart rate of the subjects is found to have an increasing pattern when facing the stressor. Another human stress measurement study based on wearable sensors is presented in~\citep{acerbi2016wearable}. The mean of the heart rate feature is extracted from the recorded heart rate variability signal. The study concluded that stressed and non-stressed individuals have significant differences in the heart rate of the subjects. The influence of acute mental stress on the cardiovascular response and concentrations of inflammatory cytokines are examined in a study conducted in~\citep{steptoe2001acute}. The study reported that participants have an increased heart rate and blood pressure when facing stressors as compared to the baseline condition. Another human stress measurement study conducted in~\citep{vuksanovic2007heart} proved the fact that an increase in the heart rate of the subject in mental stress aloud condition is due to the changes in autonomous modulation of the spectral power of high-frequency bands of the ECG signal. A study to examine the effects of chronic and short-term stress on heart rate and heart rate variability is presented in~\citep{schubert2009effects}. The speech task has been used as a stressor for the experiment and time, frequency, and phase domain measures were examined. The study reported the finding that the heart rate of the subjects is significantly increased when performing the public speaking task as compared to the rest state. A study to correlate the perception of work stressors with the measure of heart rate variability is presented in~\citep{clays2011perception}. The mean of correlation, multiple linear regression, and ANOVA is used to analyze the HRV signal. The mean of the heart rate is extracted as a feature from the HRV signal and it is found that the mean HR was raised in the high work stressor group as compared to the low stressor group.

On the contrary, few studies exist in the literature which report no change in the heart rate under stress~\citep{mcduff2014remote,blechert2006identifying,hynynen2011incidence,cinaz2013monitoring,mcduff2016cogcam}. Remote measurement of cognitive stress using heart rate, heart rate variability, and breathing rate have been performed in a study conducted in~\citep{mcduff2014remote}. Physiological data is acquired from the participants in the rest state and stress condition i.e., while performing a mental arithmetic task. The study concluded that there is a significant difference in the breathing rate and the heart rate variability of the subjects in stress vs rest condition, whereas the heart rate of the subject did not show any significant difference in stressed vs relaxed state. Identifying the difference between anxiety and rest state using a variety of physiological signals which include EDA, breathing rate, and cardiovascular measure of heart rate variability and heart rate is presented in~\citep{blechert2006identifying}. Physiological data were acquired in the rest state as well as while facing electric shocks. The study concludes that EDA showed a significant difference in the rest vs stressed state whereas cardiovascular measures show very little difference between the two groups. A study to correlate the self-reported questionnaire with the cardiac autonomic modulation in real-life scenarios is discussed in~\citep{hynynen2011incidence}. The PSS questionnaire is filled out by the participants and the participants were grouped into low and high-stress groups based on PSS scores. R-R interval data is recorded while participants were sleeping at night and during the orthostatic test after awakening in the morning. R-R interval data is used to extract HRV and HR features in time as well as frequency domain. The study concluded that a high score on the stress questionnaire is correlated with lower HRV in the orthostatic test. Moreover, there is no difference observed in the heart rate and heart rate variability of the low and high-stress participants. A new stress recognition framework using the using contact-free camera as an apparatus and computer-based tasks as a stimulus is proposed in~\citep{mcduff2016cogcam}. PPG signals are recorded and the heart rate, heart rate variability, and breathing rate features are extracted and used for the identification of stress during the tasks. The study identified the fact that heart rate variability has significant changes during the stressor where on the other hand there is no difference in the heart rate and breathing rate of the two groups. It can be observed from the above studies that heart rate has been widely used as a marker for human stress measurement because it is a reliable indicator of arousal due to stress. \Tab{tab5} presents a summary of human stress classification schemes using heart rate.

\begin{table}[H] 
\caption{Summary of Human Stress Detection Studies using Heart Rate Measure.\label{tab:tab5}}
\scalebox{0.9}{
\newcolumntype{C}{>{\centering\arraybackslash}X}
\begin{tabularx}{\textwidth}{CCCCCCCC}
\toprule
\textbf{Method}	& \textbf{Type of Stress}	& \textbf{Number of Subjects (M/F)} & \textbf{Age} & \textbf{Stimulus}& \textbf{Features Domain}& \textbf{Classifier}& \textbf{Accuracy (Classes)}\\
\midrule
~\citep{blechert2006identifying} & Acute & 42 (14/28) & 42.2$\pm$9.9 & Pictures & Time & DFA & 83.30\% (2) \\
~\citep{giannakakis2017stress} & Acute & 23 (16/7) & 45.1$\pm$10.6 & SCWT, IAPS, videos & Time & Adaboost & 91.68\% (2) \\
~\citep{mcduff2014remote} & Acute & 10 (3/7) & 18-30 & MAT & Frequency & SVM & 85.00\% (2) \\
~\citep{mcduff2016cogcam} & Acute & 10 (5/5) & 18-28 & BCST & Frequency & NB & 86.00\% (2) \\

\bottomrule
\end{tabularx}
}

    \noindent{\footnotesize{MAT: Mental Arithmetic Task, SCWT: Stroop color-word test, NB: Naive Bayes, SVM: Support Vector Machine, IAPS: International Affective Picture System, DFA: Discriminant function analysis, BCST: Berg Card Sorting Task }}
\end{table}

\subsubsection{Skin Temperature based Stress Detection}
Skin temperature is the temperature of the outermost surface of the body. The normal skin temperature of the outer surface of the skin lies between $33.5$ and $36.9^oC$. Our sense of hot and cold depends on the amount of energy to and from the skin. Skin temperature depends on the temperature of the air and time spent in that environment. Human skin temperature is strongly correlated to the heart activity and sweat reaction of an individual. Changes in skin temperature are connected to stressful and anxious conditions~\citep{mcfarland1985relationship}. Skin temperature has been measured at a variety of locations on the human body like a finger, arm, face, and armpits. Measurement of different positions gives different results under stress because the temperature on some parts of the body increases whereas on some other parts of the body temperature decreases.

In~\citep{zhai2006stress}, a skin temperature-based human stress measurement method has been developed. Skin temperature has a negative correlation to human stress, i.e., a decrease in stress level corresponds to an increase in ST and vice versa~\citep{reisman1997measurement}. A patch-based human stress monitoring using skin temperature and skin conductance is proposed in~\citep{yoon2016flexible}. Skin temperature has shown a negative correlation with the level of chronic stress~\citep{lee2010wearable,torii1992fall}. Changes in skin temperature to identify different levels of stress are studied in~\citep{karthikeyan2012descriptive}. The Stroop color-word test is used as a stimulus and the probabilistic neural network as a classifier to achieve an accuracy of 88\% for four levels of stress. The effect of the core and peripheral body temperature on human stress is discussed in~\citep{vinkers2013effect}. The study reported a decrease in the temperature of the fingertips and palm whereas, on the contrary, the temperature of the upper arm increased. An acute human stress measurement system using skin temperature is presented in~\citep{herborn2015skin}. Skin temperature when measured with an axillary thermometer tends to increase under stressful situations~\citep{marazziti1992psychological}. Some other studies analyzing skin temperature on the surface of the finger under human stress report a decrease in temperature~\citep{lee2004development,rimm1996psychological,vinkers2013effect,karthikeyan2012descriptive,engert2014exploring}. Slope of skin temperature has been used in some stress studies instead of temperature means value~\citep{barreto2007significance}. A study reporting different temperature changes in different parts of the body under a stressful stimulus of an interview~\citep{rimm1996psychological}. The temperature on the hands of the person is decreased whereas the cheeks and eyes of the person tend to show an increase in temperature. Moreover, there also exists a temperature difference in the left and right cheeks of the participants. \Tab{tab6} presents a summary of human stress detection schemes using skin conductance.

\begin{table}[H] 
\caption{Summary of Human Stress Detection Studies using Skin Temperature Measure.\label{tab:tab6}}
\scalebox{0.9}{
\newcolumntype{C}{>{\centering\arraybackslash}X}
\begin{tabularx}{\textwidth}{CCCCCCCC}
\toprule
\textbf{Method}	& \textbf{Type of Stress}	& \textbf{Number of Subjects (M/F)} & \textbf{Age} & \textbf{Stimulus}& \textbf{Features Domain}& \textbf{Classifier}& \textbf{Accuracy (Classes)}\\
\midrule
~\citep{lee2004development} & Acute & 80 & -- & SCWT & Time & MLP, GRNN,ANFIS & 96.67\% (2) \\
~\citep{zhai2006stress} & Acute & 32 & 21-42 & SCWT, Emotional pictures & Time and Frequency & SVM & 90.10\% (2) \\
~\citep{karthikeyan2012descriptive} & Acute & 60 (30/30) & 22.5$\pm$2.5 & SCWT & Time & PNN & 88.00\% (4) \\

\bottomrule
\end{tabularx}
}

       \noindent{\footnotesize{SCWT: Stroop color-word test, MLP: Multilayer Perceptron, PNN: Probabilistic Neural Network, GRNN: Generalized regression neural network, ANFIS: Adaptive network-based fuzzy inference system. SVM: Support Vector Machine}}
\end{table}

\subsubsection{Respiratory Rate-based Stress Detection}

The respiration rate of a person can be defined as the number of breaths a person takes in a duration of one minute. Two of the most common measures of respiration are breath rate and breath amplitude or depth~\citep{simoes1991respiratory}. The breath rate of a person is increased under stressful conditions, whereas on the contrary it is decreased in calm situation~\citep{vinkers2013effect,mcduff2014remote,grossman1983respiration}. Stress is associated with the irregularities of the respiratory rate~\citep{singh2013stress}, the shift from abdominal to thoracic breathing~\citep{ahmed2015rebreathe}, and faster and shallower breathing~\citep{kreibig2010autonomic}.  The breath rate sensor has been reported to be an accurate estimation of the respiratory rate. Breathing activity monitoring using chest cavity expansion has been reported in~\citep{stern2001psychophysiological}. For the detection of human stress, respiratory signals have been acquired using an elastic hall effect sensor placed at the lower part of the chest~\citep{healey2005detecting} and by use of thermistors placed in the nasal passage of the subject~\citep{shin1998estimation}. Respiratory rate signal has also been used in combination with other biomedical sensors for the assessment of human stress~\citep{hosseini2011classification}. Oxygen consumption rate has been extracted from the respiratory rate of the person and is considered a considerably reliable measure of human stress because the oxygen demand is increased under stress~\citep{seematter2002metabolic}. In~\citep{fernandez2018mental}, the authors proposed another stress recognition study using respiratory signals. Another study for human stress assessment using a respiration sensor is presented in~\citep{shan2020respiratory}. \Tab{tab7} presents a summary of human stress classification schemes using respiration rate.

\begin{table}[H] 
\caption{Summary of Human Stress Detection Studies using Respiratory Rate Measure.\label{tab:tab7}}
\scalebox{0.9}{
\newcolumntype{C}{>{\centering\arraybackslash}X}
\begin{tabularx}{\textwidth}{CCCCCCCC}
\toprule
\textbf{Method}	& \textbf{Type of Stress}	& \textbf{Number of Subjects (M/F)} & \textbf{Age} & \textbf{Stimulus}& \textbf{Features Domain}& \textbf{Classifier}& \textbf{Accuracy (Classes)}\\
\midrule
~\citep{mcduff2014remote} & Acute & 10 (3/7) & 18-30 & MAT & Frequency & SVM &85.00\% (2) \\
~\citep{ahmed2015rebreathe} & Acute & 25 (15/10) & 18-35 & SCWT and Public speaking & Time and Frequency & GEE & 88.00\% (2) \\
~\citep{hosseini2011classification} & Acute & 15 (15/0) & 20-24 & Picture presentation test &  Time and Frequency & SVM & 76.95\% (2) \\
~\citep{wijsman2013wearable} & Acute & 30 (25/5) & 19-53 & Calculation and memory task & Time and Frequency & GEE & 74.50\% (2) \\
~\citep{rigas2011real} & Acute & 13 (10/3) & 22-41 & Car driving & Time and Frequency & BN & 96.00\% (2) \\
~\citep{singh2013novel} & Acute & 10 & -- & Car driving & Time & ANN & 80.00\% (2) \\
~\citep{wijsman2011towards} & Acute & 30 (25/5) & 19-53 & Calculation, puzzle and memory task & Time and Frequency & ANN, LBN, QBN, FLS & 80.00\% (3) \\
~\citep{shan2020respiratory} & Acute & 89 (47/42) & 18-23 & SCWT & Time & SVM & 93.90\% (2) 93.40\% (2) 89.05\% (2) \\
~\citep{fernandez2018mental} & Acute & 43 (26/17) & 18-22 & Maths problem & Time and Frequency & MLP & 94.44\% (2) \\

\bottomrule
\end{tabularx}
}

    \noindent{\footnotesize{MAT: Mental Arithmetic Task, SCWT: Stroop color-word test, FDA: Fisher Discriminant Analysis, GEE: generalized estimating equation, SVM: Support Vector Machine, BN: Bayesian networks, ANN: Artificial Neural Network, LBN: Linear Bayes Normal, QBN: Quadratic Bayes Normal, FLS: Fisher’s Least Square, MLP: Multilayer Perceptron}}
\end{table}

\subsubsection{Heart Rate Variability based Stress Detection}

HRV is the measure of the variation in the time interval between consecutive heartbeats of an individual. ANS activities can be reliably measured by using the HRV parameter and is a strong tool for human stress assessment~\citep{pflanzer2013galvanic}. HRV shows distinct changes in response to changes in individuals~\citep{acharya2006heart}. HRV can be obtained using ECG as well as PPG sensor data.  HRV measurement methods based on ECG signals have been developed in~\citep{clifford2002signal}. An ECG recordings-based study is conducted in~\citep{sloan1994effect} to analyze the relationship of RR intervals (time between consecutive heartbeats), HRV, and human stress. The study concludes that the increase in heart rate is correlated with the decrease in RR interval. Another study for human stress assessment using physiological signals of ECG and HRV is introduced in~\citep{karthikeyan2013detection}. Classification accuracy of 94.66\% for normal vs stressed class using a fusion of ECG and HRV signals is achieved. In~\citep{jobbagy2017hrv}, authors have proposed a stress recognition system to characterize human stress using HRV signals. The influence of HR and HRV on human stress is discussed in~\citep{taelman2009influence}. The HR and HRV of the subjects are recorded in the rest state as well as while performing a mental stressor. The study concluded that the HR and HRV of the subject change when facing a mental stressor and hence can be used as a potential biomarker for the assessment of human stress. Another study about the association of mental stress with HRV is presented in~\citep{salahuddin2007dependence}. Correlation between the perceived stress face by college students and the HRV is explored in~\citep{lombardo2019relationship}. Another cognitive stress measurement model using HRV with an accuracy of 85\% is proposed in~\citep{mcduff2014remote}. A deep learning model for the identification of mental stress in firefighters using HRV data is presented in~\citep{oskooei2019destress}. A study about the correlation of mental stress and HRV of the students during the university final examination is presented in~\citep{hammoud2019stress}. The study reports that HRV in female students is significantly lower as compared to their male counterparts before and after taking the exam. A study to find the correlation between the perceived mental stress and the HRV parameters of the subjects is discussed in~\citep{orsila2008perceived}. A strong correlation between the perceived stress and the values of triangular interpolation of rhythm-to-rhythm (RR) interval histogram (TINN) and the root mean square of differences of successive RR intervals (RMSSD) of the HRV data obtained in the morning and during the workday is found in the study.

Some studies claim that the HRV data needs to be of around five minutes for some reasonable analysis~\citep{malik1996heart}, whereas, on the other hand, some studies negate this conclusion by claiming that an even smaller amount of data could be used as a reliable marker of human stress~\citep{hall2004acute,salahuddin2007ultra}. The standard deviation of the NN interval (SDNN) is found to be reduced under stressful condition~\citep{blechert2006identifying,acerbi2016wearable,schubert2009effects,clays2011perception,hynynen2011incidence,cinaz2013monitoring,bernardi2000effects,taelman2011instantaneous,tharion2009short,visnovcova2014complexity,madden1995effects}. A study to monitor human stress using physiological signals of electrodermal activity and heart rate variability recorded via wearable sensors is presented in~\citep{acerbi2016wearable}. A new stress-inducing protocol named TransSafe (The Ambient Response to Avoid Negative Stress and enhance SAFEty) is developed in this study. Subjective questionnaires of the State-Trait Anxiety Inventory (STAI) and Shortened State Stress Questionnaire (SSSQ) are filled by the participants before and after facing the stressor. Time and frequency domain features are extracted from the HRV signal. A statistical test is applied to the extracted features and a significant difference is found in some of the features of the EDA and HRV signal. A study to examine the effect of short-term and chronic stress on the heart rate variability of the subject is conducted in~\citep{schubert2009effects}. The speech task has been used as a stressor in the experiment and it is found that the standard deviation of the R-R interval got reduced when facing the stressor. Perception of work stressor relationship with heart rate variability is examined in a study conducted in~\citep{clays2011perception}. Perception of the work stressors is measured using a 27-item job stress questionnaire. An association between the percentage of differences between adjacent normal RR intervals (pNN50), lower high-frequency power, and a higher ratio of the low frequency over high-frequency power and the worker's stress is found. Moreover, no significant correlation between low-frequency power and worker stress is found. An investigation of the relationship of self-reported measures with heart rate variability in real-life situations is explored in~\citep{hynynen2011incidence}. SDNN features extracted from the HRV signal got reduced in a high-stress condition when compared to a low-stress condition. Monitoring of mental workload in office work scenarios using heart rate variability feature is proposed in~\citep{cinaz2013monitoring}. The NASA Task Load Index questionnaire is used to obtain the subjective mental workload of the participants. Time and frequency domain features are extracted from the HRV signal and the pNN50 feature was found to decrease significantly and the SDNN feature is found to give a consistent decrease under stress conditions. Another study to access that whether talking or reading aloud or silently affects heart rate variability is presented in~\citep{bernardi2000effects}. An increase in the speed of breathing and a decrease in the mean and variance of the RR interval as compared to normal breathing are observed when reading silently as compared to reading aloud. A study to monitor instantaneous changes in heart rate activity due to mental workload in an office environment is proposed in~\citep{taelman2011instantaneous}. The participants are asked to perform a low mental workload task and a high mental workload task twice where each of these tasks is followed by a rest state condition. A significant difference in the heart rate and heart rate variability is observed under mental workload conditions as compared to baseline rest conditions. A study to explore the heart rate variability in students during examination time is presented in~\citep{tharion2009short}. The mean of the RR interval is reported to be significantly lower whereas mean arterial pressure and SDNN are found to be higher during the examination time. A study to understand the relation of acute mental stress and complexity and time asymmetry of HRV signal is proposed in~\citep{visnovcova2014complexity}. Two different stimuli which include SWT and mental arithmetic tasks are used. The study reveals the fact that SDNN was found to be significantly lower when facing a stressful situation as compared to the recovery period. The effect of mental state on the heart rate and blood pressure variability in both males and females is examined in a study conducted in~\citep{madden1995effects}. The mental arithmetic task is used as a stimulus for the experiment. As compared to the control condition, the stressor condition causes a decrease in SDNN, log standard deviation of systolic blood pressure, log total power, and log fractal powers. Root Mean Square of the Successive Differences (RMSSD) is another HRV feature that has been explored in the literature and it has been established to decrease under stress~\citep{acerbi2016wearable,ring2002shifting,hynynen2011incidence,cinaz2013monitoring,taelman2011instantaneous,tharion2009short}. Authors in~\citep{li2009longitudinal} reported that RMSSD, a time-domain feature of HRV and high frequency (HF) power, a frequency domain feature of HRV get decreased in stress condition. Another important frequency-domain feature of HRV discussed in the literature is the ratio of Low-frequency power (LF) to HF power which is increased under stressful situation~\citep{mcduff2014remote,blechert2006identifying,acerbi2016wearable,moriguchi1992spectral,vuksanovic2007heart,schubert2009effects,clays2011perception,cinaz2013monitoring,mcduff2016cogcam,lucini2002hemodynamic,taelman2011instantaneous,tharion2009short,taelman2009influence,hjortskov2004effect,hall2004acute}. Very low-frequency (VLF) band of the HRV signal is found to be increased in some studies~\citep{acerbi2016wearable,moriguchi1992spectral}. Another HRV feature named correlation dimension D2 is reduced under stress in a study conducted on university students during their examination time in~\citep{melillo2011nonlinear}. A new framework for the remote measurement of heart rate variability using a webcam for the detection of stress is proposed in~\citep{bousefsaf2013remote}. HRV is investigated for different stress factors which include stressed, tensed, concentrated, and stimulated conditions. The study concluded that remote measurement of HRV can be used as a reliable marker of human stress assessment. Another stress recognition study using HRV signals is discussed in~\citep{kim2008detection}. A self-reporting questionnaire was used to label the participants into low and high-stress groups. HRV data were recorded for three different periods during the day and it was concluded that the highly stressed participants showed a decrease in the HRV patterns as compared to the low-stress group. Moreover, using logistic regression as a classifier accuracy of 63.2\% is achieved for the low vs high stressed group. Another HRV-based stress measurement scheme using time and frequency domain features is presented in~\citep{boonnithi2011comparison}. Stress detection using ECG and HRV features is presented in~\citep{melillo2013classification}. Variation in the heart rate variability of the students in rest conditions and during the examination phase was examined using a non-parametric classifier called Classification and Regression Tree (CART). Sensitivity and specificity of 83.33\% and 90.48\% for the stress vs rest state classification is achieved. Another study for four-level stress classification i.e., no stress, low stress, medium stress, and high stress using HRV features is discussed in~\citep{vanitha2014hierarchical}. The database used for the experiment in this study was the MIT-BIH multi-parameter database where different driving tasks are used as a stimulus. Hierarchical SVM is used as a classifier to classify the four stress state with a classification accuracy of 92\%. Driver stress level recognition using HRV features along with support vector machine classifier is performed in~\citep{munla2015driver}. The database used in this experiment is stress recognition in automobile driver database and a classification accuracy of 83\% is reported. A stress recognition measurement scheme for driver stress monitoring using HRV data is proposed in~\citep{wang2013k}. DriveDB database is used in this study for measuring driver stress and features are extracted using parameter-based methods. Kernel-based class separability (KBCS) method for feature selection is used to select the optimum feature set. LDA and PCA algorithms are used for dimensionality reduction. Next, the classification of driver stress is performed using a kNN classifier with an achieved accuracy of 97\%. Taking into consideration the studies presented in the literature, it is evident that the relationship between HRV to stress is not very straightforward. However many of the studies presented in the literature present a consistent relationship with certain HRV features and hence can help draw some useful conclusions. \Tab{tab8} presents a summary of human stress detection schemes using HRV.

\begin{table}[H] 
\caption{Summary of Human Stress Detection Studies using Heart Rate Variability Measure.\label{tab:tab8}}
\scalebox{0.9}{
\newcolumntype{C}{>{\centering\arraybackslash}X}
\begin{tabularx}{\textwidth}{CCCCCCCC}
\toprule
\textbf{Method}	& \textbf{Type of Stress}	& \textbf{Number of Subjects (M/F)} & \textbf{Age} & \textbf{Stimulus}& \textbf{Features Domain}& \textbf{Classifier}& \textbf{Accuracy (Classes)}\\
\midrule
~\citep{blechert2006identifying} & Acute & 42 (14/28) & 42.2$\pm$9.9 & Pictures & Time & DFA & 83.30\% (2) \\
~\citep{karthikeyan2013detection} & Acute & 60 (30/30) & 21-25 & SCWT & Time and Frequency & kNN, PNN & 91.66\% (2) \\
~\citep{mcduff2014remote} & Acute & 10 (3/7) & 18-30 & MAT & Frequency & SVM &85.00\% (2) \\
~\citep{mcduff2016cogcam} & Acute & 10 (5/5) & 18-28 & BCST & Frequency & NB & 86.00\% (2) \\
~\citep{melillo2011nonlinear} & Acute & 42 & -- & University Examination & Time & LDA & 90.00\% (2) \\
~\citep{melillo2013classification} & Acute & 42 (19/23) & 20-28 & University Examination & Time and Frequency & CART & 87.00\% (2) \\
~\citep{vanitha2014hierarchical} & Acute & 16 & -- & Car driving & Time and Frequency & hierarchical SVM & 92.00\% (4) \\
~\citep{munla2015driver} & Acute & 16 & -- & Car driving & Time and Frequency & SVM-RBF & 83.00\% (2) \\
~\citep{wang2013k} & Acute & 27 & -- & Car driving & Time and Frequency & kNN & 97.00\% (2) \\
~\citep{kim2008detection} & Chronic & 68 & 10-30 & Baseline & Time and Frequency & LR & 66.1\% (2) \\

\bottomrule
\end{tabularx}
}

    \noindent{\footnotesize{MAT: Mental Arithmetic Task, SCWT: Stroop color-word test, DFA: Discriminant function analysis, kNN: k- Nearest Neighbors, PNN: Probabilistic Neural Network, SVM: Support Vector Machine, NB: Naive Bayes, LDA: Linear Discriminant Analysis, CART: Classification and Regression Tree, RBF: radial basis function, LR: Logistic Regression, BCST: Berg Card Sorting Task}}
\end{table}

\subsubsection{Blood Volume Pressure based Stress Detection}

BVP is a method to measure the amount of pressure exerted on the blood vessels. When measuring blood pressure, we get two values, the first value is systolic blood pressure (SBP) and the second number is called diastolic blood pressure (DBP). The human body releases a large number of stress hormones under stressful conditions which increases blood pressure~\citep{gasperin2009effect} thus making blood pressure a good indicator for stress measurement~\citep{pickering1996environmental}. A 3-year duration study conducted revealed the fact that individuals who have stress at the workplace tend to have higher SBP and DBP, whereas during sleep they have increased SBP~\citep{schnall1998longitudinal}. SBP and DBP have been reported to be higher during a mental stressor task in~\citep{ring2002shifting,lundberg1994psychophysiological,carroll2003blood,carroll2011blood}. A study presented in~\citep{ring2002shifting} discussed the hemodynamics due to which the increase in blood pressure occurs during exposure to stress for a longer time duration. Mean arterial pressure (MAP), cardiac output (CO), and total peripheral resistance (TPR) parameters were measured during three phases of the experiment i.e., rest phase, mental arithmetic task phase, and recovery phase. MAP increased at a constant rate during the stressor, CO increased during the first half of the stressor and decreased back to rest state condition toward the end of the task, whereas on the other hand TPR kept on increasing as the mental arithmetic task progressed. In a study conducted in~\citep{lundberg1994psychophysiological}, it is found that the blood pressure of the subjects i.e., both systolic and diastolic increased in case of stress session as compared to the baseline condition. Authors in~\citep{carroll2003blood} present a study to examine the relationship of human stress to the future value of blood pressure and how it is affected by the gender, age, and socioeconomic condition of the subject. The blood pressure of the subjects was recorded in rest condition and under mental stressors. Moreover, five years of follow-up resting-state blood pressure data of the participants were also available. The findings of the study are that the systolic blood pressure reaction to human stress is found to be positively correlated with the follow-up systolic blood pressure where no correlation could be found with diastolic blood pressure. Another conclusion of the study is that the magnitude of the predicted blood pressure has an association with the gender and socioeconomic position of a person. Another study to correlate the reaction of blood pressure to acute stress and future blood pressure condition is proposed in~\citep{carroll2011blood}. Blood pressure readings in a rest state and with facing stressors are recorded. Moreover, after twelve years, resting-state blood pressure reading was again taken. The study concluded that the systolic blood pressure positively made a prediction about future systolic pressure and there is an increasing pattern of blood pressure over the span of 12 years. However, these findings are not observed in the case of diastolic blood pressure. However, in another study, the author claims that the mental stress of a person does not affect the recorded BVP~\citep{hjortskov2004effect}. There exists a wide of studies in which SBP and DBP is reported to increase under stressful conditions~\citep{finsen2001muscle,vinkers2013effect,lundberg1994psychophysiological,krantz2004consistency,moriguchi1992spectral,steptoe2001acute,ring2002shifting,bernardi2000effects,hjortskov2004effect,schnall1998longitudinal,carroll2003blood,carroll2011blood}. Men have been found to have higher blood pressure under stress as compared to women in a study conducted in~\citep{krantz2004consistency}. In a study conducted in~\citep{vinkers2013effect}, authors reported that when participants are facing a standard stressor there is an increase in their blood pressure i.e., systolic and diastolic as compared to a baseline recording. Authors in~\citep{finsen2001muscle} reported that an increase in the blood pressure of the participants is observed when increasing memory demands. An increase in the blood pressure of the participants is observed when facing a stressor in an experimental study conducted in~\citep{moriguchi1992spectral}. The correlation of mental stress with the cardiovascular response is examined in~\citep{steptoe2001acute}. The blood pressure of the participants is recorded during the rest state as well as while performing stressful tasks. The stressed group has significantly higher blood pressure as compared to the control group. The effect of reading aloud or silently on the blood pressure of the participants is examined in a stress measurement study performed in~\citep{bernardi2000effects}. The effect of mental stress on the HRV and blood pressure of the participants performing computer work is examined in~\citep{hjortskov2004effect}. The study concludes that reading silently causes an increase in the blood pressure of the participant. Hence, looking at these trends, BVP can be considered a reliable marker of stress.

\subsubsection{Photoplethysmography based Stress Detection}
PPG is a technique to measure the blood volumetric changes in the vessels~\citep{challoner1979photoelectric}. PPG is a widely accepted technique being used in clinical applications as well as commercial devices, e.g., a pulse oximeter. PPG sensor is quite simple consisting of an infrared light source that is used to illuminate tissue of the skin and a photodetector is used to measure the changes in the light illumination due to blood flow.  Many commercial devices that measure blood pressure, oxygen saturation, and cardiac output are based on a PPG sensor. For the acquisition of PPG signals, currently, a large variety of devices are available in the market. The component of a PPG data acquisition system includes a light source, an LED for lightening up the tissue, and a photodetector to receive the light and measure the variations of the light. Light sources commonly used in PPG sensors include red, light green, or infrared color. Green color has a shorter wavelength and thus it produces a larger variation of light intensity to cardiac changes~\citep{maeda2011advantages}. Heart rate calculation has also been performed using the PPG sensors~\citep{kageyama2007wavelet}. Many algorithms have been developed to measure the heart rate and heart rate variability parameters, which can be used to measure different physiological responses including human stress~\citep{vstula2003evaluation}. PPG signal can also be used to calculate the value of pulse rate, pulse rate variability, blood volume pressure, blood oxygen saturation level, and blood pressure~\citep{giannakakis2019review}. \Tab{tab9} presents a summary of human stress detection schemes using PPG signals.

\begin{table}[H] 
\caption{Summary of Human Stress Detection Studies using PPG Signal.\label{tab:tab9}}
\scalebox{0.9}{
\newcolumntype{C}{>{\centering\arraybackslash}X}
\begin{tabularx}{\textwidth}{CCCCCCCC}
\toprule
\textbf{Method}	& \textbf{Type of Stress}	& \textbf{Number of Subjects (M/F)} & \textbf{Age} & \textbf{Stimulus}& \textbf{Features Domain}& \textbf{Classifier}& \textbf{Accuracy (Classes)}\\
\midrule
~\citep{mcduff2014remote} & Acute & 10 (3/7) & 18-30 & MAT & Frequency & SVM &85.00\% (2) \\
~\citep{mcduff2016cogcam} & Acute & 10 (5/5) & 18-28 & BCST & Frequency & NB & 86.00\% (2) \\
~\citep{chauhan2018real} & Acute & 10 & 30-58 & PASAT & Frequency and Wavelet & Adaboost & 93.00\% (2) \\
~\citep{cho2019instant} & Acute & 17 (8/9) & 29.82$\pm$12.02 & MAT & Frequency & ANN & 78.33\% (2) \\
~\citep{li2018photoplethysmography} & Acute & 178 (85/93) & 16-36 & MAT & Frequency & elastic net & 86-91\% (2) \\
~\citep{zangroniz2018estimation} & Acute & 50 (28/22) & 20-28 & IAPS & Time and Frequency & DT & 82.35\% \\

\bottomrule
\end{tabularx}
}

   \noindent{\footnotesize{MAT: Mental Arithmetic Task, ANN: Artificial Neural Network, SVM: Support Vector Machine, NB: Naive Bayes, DT: Decision Tree, BCST: Berg Card Sorting Task, PASAT: Effects of the Paced Auditory Serial Addition Task, IAPS: International Affective Picture System}}
\end{table}

\subsubsection{Salivary Cortisol based Stress Detection}
Cortisol is a well-known biomarker for measuring psychological stress. Salivary cortisol has been used by physicians as a diagnostic tool for the measurement of psychological stress and many other diseases for over two decades and has been reported to be a very good measure of human stress~\citep{kirschbaum1989salivary}. As it is established that stress is a phenomenon that is affected by a wide range of factors, so a reliable measure is required for the accurate estimation of stress. It has been reported that when the person is under an acute stressor, the level of cortisol release increases~\citep{fink2000encyclopedia}. In~\citep{hellhammer2009salivary}, the author has presented a correlation between stress and cortisol level. Human HPA activity is affected by stress and is reflected in cortisol thus making it a very practical tool for stress detection. An acute stress measurement scheme using cortisol secretion is presented in~\citep{boucher2019acute}. A study measuring the response of sweating cortisol to human stress is presented in~\citep{tu2019sweat}. The study deduced the fact that there exists a strong association between the diet and cortisol of a person and therefore adjusting the diet may help in lowering the cortisol level and thus preventing stress-related health issues. The relationship between the changes in the cortisol level of a subject due to mental stress is discussed in~\citep{luo2012relationship}. The study concluded that the cortisol level of a person is increased due to mental stress. Another stress detection study using cortisol as a biomarker is presented in~\citep{nath2020validating}. The model achieved a classification accuracy of 92\%. Acute stress measurement using cortisol as a biomarker is discussed in~\citep{selvaraj2015psychological}. Emotional stress measurement using cortisol is introduced in~\citep{rey2014towards}. The study reveals that the cortisol level of men is found to increase under stress. Salivary cortisol has been used as a measure of assessing the mental workload in~\citep{nomura2009salivary}.

\subsection{Non-Wearable Sensor-based Human Stress Detection}

Non-wearable sensors are the type of sensors in which no physical device needs to be connected to the human body rather the data can be acquired at a considerable distance from the subject. Non-wearable sensors for human stress measurement can be subdivided into physical measures, behavioral measures, and computer vision-based measures. Physical measures are one in which some observable parameters of the human body like human pupil dilation, human speech, human eye activity, and body postures are recorded, whereas, on the other hand, behavioral measures are the ones in which human stress is measured based on the interaction of the subject with some device like keyboard, mouse, and smartphones. The third and last type of non-wearable sensors used for human stress measurement is computer vision-based sensors like video cameras and thermal imaging.

\subsubsection{Physical Measures}
A physical property is defined as a property that is observable by humans with the naked eye. To acquire physical measures, sophisticated equipment, and sensors are required. Physical measures of stress can be subdivided into four main categories, which include pupil dilation, speech-based measures, eye movement, and body postures. Literature corresponding to each of these categories is given below.

\begin{enumerate}
\item \textit{Pupil Dilation:}
The eye pupil is a hole located at the center of the iris, which allows light to enter the retina. The color of the eye pupil is black because the light entering the pupil is either absorbed directly by the eye tissues or absorbed after diffusion. The pupil of the eye may appear to open i.e., dilate, or close i.e., constrict but it is the iris that governs its movement. Under bright lighting conditions, the pupil constricts to allow less light to enter the eye, whereas under dark conditions the pupil dilates to allow more light to enter the eye. The size of the pupil is controlled by two muscles i.e., constrictor and dilator pupillae, which are in turn controlled by the sympathetic (SNS) and parasympathetic (PNS) part of the ANS~\citep{beatty2000pupillary}. Just like the physiological responses of an individual, pupil dilation is not under subject control and is strongly associated with cognitive and emotional arousal~\citep{bradley2008pupil}.

The relationship between the affective states and pupil dilation has been discussed in a number of studies~\citep{onorati2013reconstruction,partala2003pupil,al2013using,bradley2008pupil,ren2012affective,pedrotti2014automatic}. Moreover, pupil dilation has been used as a marker of stress and anxiety assessment~\citep{honma2013hyper,simpson1971effects,baltaci2016stress,zhai2006stress}. Another human stress measurement study based on pupil diameter along with physiological signals of Galvanic Skin Response (GSR), Blood Volume Pulse (BVP), and Skin Temperature (ST) is proposed in~\citep{zhai2006stress}. SVM is used for the classification of stress and relaxed state with an achieved accuracy of 90.10\%. The study concluded that in comparison to physiological signals pupil diameter was a more effective indicator of stress. In a laboratory environment, when the subject is presented with a stressful stimulus the pupil diameter is increased~\citep{de2016acute}. Stress measurement based on pupil dilation has been the subject of study in~\citep{barreto2007non}. An increase in pupil dilation diameter suggests that the pupil dilates at a higher frequency and it means the person is in a stressed state. Experimental studies have shown that due to negative and positive arousing sounds, the diameter of the human pupil is increased quite significantly~\citep{partala2003pupil}. The mean value of pupil diameter has also been used as a parameter for stress detection. Increasing mean values over a time period shows the increasing level of stress. Images having negative valance tend to have a stronger effect on the eye pupil of the subject who feels more stressed~\citep{kimble2010eye}. Moreover, positive as well as negative sounds cause an increase in the pupil diameter of the person~\citep{partala2003pupil}. Public speaking anxiety also affects pupil size~\citep{simpson1971effects}. The size of the pupil is directly proportional to the anxiety level of a person i.e, the more the anxiety level more the pupil diameter and vice versa~\citep{wang2011attention}. Another study about the widening of a pupil under stress is presented in~\citep{liao2005real}. In~\citep{torres2015pupil}, authors presented a study for human stress classification scheme using pupil diameter. Mental arithmetic task questions were asked of the participants in front of a camera and the changes in pupil diameter were observed. The authors concluded that pupil diameter can be a good indicator of mental stress but for better classification results it needs to be combined with physiological signals.

The use of pupil dilation for stress analysis has some limitations, which need to be addressed. The size of the human pupil is not constant throughout life i.e., the size of the pupil decreases with increasing age~\citep{winn1994factors}. Ambiguity exists on the effect of gender in the pupil size because there exist some studies which show no correlation~\citep{winn1994factors}. On the other hand, some studies show gender affects pupil size when faced with a painful~\citep{ellermeier1995gender} and audio stimulus~\citep{partala2003pupil}. Lightening conditions affect pupil size due to the in-built light reflexes of the human being~\citep{pedrotti2014automatic,reeves1920response}. Thus, to use pupil dilation for human stress assessment it is necessary to take into consideration these limitations and precautionary measures need to be taken beforehand. \Tab{tab10} presents a summary of human stress classification schemes using pupil dilation.

\begin{table}[H] 
\caption{Summary of Human Stress Detection Studies using Pupil Dilation.\label{tab:tab10}}
\scalebox{0.9}{
\newcolumntype{C}{>{\centering\arraybackslash}X}
\begin{tabularx}{\textwidth}{CCCCCCCC}
\toprule
\textbf{Method}	& \textbf{Type of Stress}	& \textbf{Number of Subjects (M/F)} & \textbf{Age} & \textbf{Stimulus}& \textbf{Features Domain}& \textbf{Classifier}& \textbf{Accuracy (Classes)}\\
\midrule
~\citep{ren2012affective} & Acute & 30 (14/16) & 26.8$\pm$2.56 & SCWT & Time & NB & 85.5\% (2) \\
~\citep{zhai2006stress} & Acute & 32 & 21-42 & SCWT and Emotional pictures & Time and Frequency & SVM & 90.10\% (2) \\
~\citep{pedrotti2014automatic} & Acute & 33 (16/17) & 23-54 & Driving task & Frequency & ANN & 79.20\% (2) \\
~\citep{baltaci2016stress} & Acute & 11 (9/2) & 29-40 & IAPS &  Time & ABRF & 65-83.8\% (2) \\

\bottomrule
\end{tabularx}
}

      \noindent{\footnotesize{SCWT: Stroop color-word test, ANN: Artificial Neural Network, SVM: Support Vector Machine, NB: Naive Bayes, ABRF: Adaboost with Random Forest, IAPS: International Affective Picture System}}
\end{table}

\item \textit{Speech Based Measures:}
Stress measurement using speech features has been one of the focuses of the research community. Stress in the voice is defined as “observable variability in certain speech features due to a response to stressors~\citep{murray1996towards}. Stress measurement from a human voice is a dynamic process. Stress is measured by the nonverbal component of voice. Speech components have variations while facing a stressful stimulus~\citep{womack1999n}. Vocal stress analysis has been performed to identify the discriminating voice features of stressed and neutral conditions~\citep{lefter2015recognizing}. The pitch of the speech signal is found as a distinct feature under emotional stress and is increased when an individual is feeling stressful~\citep{williams1972emotions,hansen1988analysis,cairns1994nonlinear,junqua1996influence,protopapas1997fundamental,hansen2007speech,gharavian2012statistical,lu2012stresssense,kurniawan2013stress,sondhi2015vocal,hansen2011robust}.  In~\citep{nwe2003speech}, the author suggests that the change in speed of fundamental frequency of voice is the most important feature for stress measurement. Speech signals-based human stress measurement has also been discussed in~\citep{fernandez2003modeling,healey2005detecting,lefter2011automatic}. Articulatory, excitation, and cepstral-based features have been used in~\citep{womak1996improved} to identify stress in speech signals, and a classification accuracy of 91\% is achieved. A study conducted in~\citep{cairns1994nonlinear} distinguished between the loud, angry, neutral, clear speech, and Lombard effect by using speech features of intensity, spectral tilt, and energy. Another speech-based stress classification scheme using spatial and spectral domain features is presented in~\citep{devillers2006real}. The frequency of the speech signal in an angry, Lombard effect, and loud state is different from each other~\citep{hollien2002forensic}. In~\citep{simantiraki2016stress}, the author extracted the spectral tilt feature from the speech signal and found it to be less negative under stressful conditions. The energy of the equivalent rectangular bandwidth band obtained from the spectrogram and Gabor filters are used in ~\citep{he2009stress} to identify stress recognition with an accuracy of 79\%. Stress classification in a noisy environment is performed using the Teager Energy Operator (TEO) features in~\citep{hanson1993finding}. Speech signal has been analyzed for stress using multi-resolution wavelet analysis (MWA) in~\citep{sarikaya1998subband} and a fusion of TEO and MWA in~\citep{fernandez2003modeling}. Hidden Markov Model (HMM) and TEO have been used for the analysis of stress in speech signals~\citep{hansen2011robust}. Physical features of the vocal cords of a person are examined in~\citep{yao2012physical} to identify stress in speech. The performance of automatic speech recognition could be improved if the speaker stress is accurately identified~\citep{kadambe2007study}. Another speech-based human stress measurement scheme is proposed in~\citep{soury2013stress}. TSST was used as a stimulus and an SVM classifier were used to classify the stress with an achieved recall rate of 72\%. \Tab{tab11} presents a summary of human stress classification schemes using speech-based measures.

\begin{table}[H] 
\caption{Summary of Human Stress Detection Studies using Speech based Measures.\label{tab:tab11}}
\scalebox{0.9}{
\newcolumntype{C}{>{\centering\arraybackslash}X}
\begin{tabularx}{\textwidth}{CCCCCCCC}
\toprule
\textbf{Method}	& \textbf{Type of Stress}	& \textbf{Number of Subjects (M/F)} & \textbf{Age} & \textbf{Stimulus}& \textbf{Features Domain}& \textbf{Classifier}& \textbf{Accuracy (Classes)}\\
\midrule
~\citep{kurniawan2013stress} & Acute & 10 & -- & SCWT, MAT & Time and Frequency & SVM & 92.00\% (2) \\
~\citep{womack1999n} & Acute & 44 (30/14) & 22-76 & Speech & Time and Frequency & HMM & 92.41\% (2) \\
~\citep{lefter2015recognizing} & Acute & 16 & -- & Speech & Time and Frequency & BN & 73.00\% (2)  \\
~\citep{cairns1994nonlinear} & Acute & 32 (19/13) & 22-76  & SUSAS & Frequency & HMM & 99.10\% (2) \\
~\citep{hansen2007speech}& Acute & 32 (19/13) & 22-76  & SUSAS & Frequency & HMM & 73.80\% (2) \\
~\citep{lu2012stresssense}& Acute & 14 (4/10) & 22.86  & Job Interview & Frequency & GMM & 81.00\% (2) \\
~\citep{fernandez2003modeling} & Acute & 4 & -- & driver speech & Frequency & SVM  & 51.20\% (4) \\
~\citep{womak1996improved} & Acute & 32 (19/13) & 22-76 & SUSAS & Frequency & HMM & 80.64\% (3) \\
~\citep{simantiraki2016stress} & Acute & 32 (19/13) & 22-76 & SUSAS & Frequency & RF & 92.06\% (2) \\
~\citep{he2009stress} & Acute & 32 (19/13) & 22-76 & SUSAS & Frequency & GMM & 81.00\% (2) \\
~\citep{sarikaya1998subband} & Acute & 32 (19/13) & 22-76 & SUSAS & Frequency & MLP & 70.00\% (2) \\
~\citep{soury2013stress} & Acute & 29 (12/17) & -- & TSST & Time and Frequency & SVM & 72.00\% (2) \\

\bottomrule
\end{tabularx}
}
   \noindent{\footnotesize{MAT MAT: Mental Arithmetic Task, SCWT: Stroop color-word test, SVM: Support Vector Machine, FDA: Fisher discriminant algorithm, HMM: Hidden Markov Model, BN: Baysian network, GMM: Gaussian Mixture Model, RF: Random Forest, MLP: Multilayer Perceptron}}
\end{table}

\item \textit{Eye Activity:}
Functioning and behavior of the eye are affected by stress and this fact is supported by different studies. A study for human stress detection using eye blink rate and brain activity of the subject was proposed in~\citep{haak2009detecting}. The stimulus used in the experiment was driving a car in a simulator on a road that included steep and sharp curves and were having many attention-seeking advertising boards. While driving on this road, stressful emotions were elicited in the drivers resulting in a change in eye blink rate and brain activity. A correlation between the eye blinks of the participants and the experienced stress level was established and a higher frequency of eye blinks was observed as an indication of the individual experiencing a stressful condition. The stressful stimulus causes an increase in the eye blink rate of the person as reported in the studies conducted in~\citep{haak2009detecting,giannakakis2017stress}. A biometric identification application of human stress detection is proposed in~\citep{pavlidis2000thermal}. Images are used to detect the facial expressions and eye movements corresponding to anxiety, alertness, and fearfulness, and rapid eye movement under a stress state was reported. The biometric application was based on the idea of "what you are doing" instead of the traditional approach of "who are you". The study reported encouraging results of the proposed scheme. A human stress detection framework based on facial cues using features of eye movement, mouth activity, head movement, and heart rate acquired via PPG signal is presented in~\citep{giannakakis2017stress}. Four different stressors which include social exposure, emotional recall, stressful images, and stressful videos were used in the experiment. Social exposure stressors included a task of the text reading and self-description speech. Eye blink rate was reported to be increased in response to stressful images and the Stroop color-word test, whereas reading a difficult text causes the eye blink rate to get reduced. Moreover, an increase in the eye aperture was observed in a stressful situation. Stress affects the eye gaze behavior of a person. In~\citep{laretzaki2011threat}, authors presented a study to determine if and how threat and trait anxiety interact to affect the stability of gaze fixation. Video oculography was used to estimate the gaze position with and without gaze fixation stimulus in a safe and verbal threat condition in subjects characterized for their trait anxiety. Trait anxiety significantly showed that there is a gaze fixation instability under threat conditions. Some stress detection studies have employed different gaze features like gaze direction, congruence, and size of gaze-cue for the assessment of stress~\citep{fox2007anxiety,staab2014influence}. A study to investigate the role of neutral, angry, happy, and fearful facial expressions in enhancing orienting to the direction of eye gaze is presented in~\citep{fox2007anxiety}. Photographs of faces with either direct or averted gaze are used as a stimulus in the experiment. A target letter appeared unpredictably to the left and right sides of the face. approximately 300 ms or 700 ms after the eye gaze direction changed. It is observed from the results that the response time of the participant is less when the eyes in the image gazed toward the subject as compared to the conditions when the eye gazed away from the subject. An enhanced orientation to the eye-gaze of faces with fearful expressions is reported in participants with high trait anxiety scores as compared to participants with low trait anxiety scores. A review to analyze the effect of anxiety on the ocular motor control and the gaze of the subject is presented in~\citep{staab2014influence}. Another human stress measurement scheme using the eye-tracking feature is proposed in~\citep{mokhayeri2011mental}. SWT is used to induce stress in the participants of the experiment. A genetic algorithm is employed to detect the human eye and the noise is removed by using a fuzzy filter. Fuzzy-SVM classifier is used to classify human stress into two classes with an accuracy of 70\%. \Tab{tab12} presents a summary of human stress classification schemes using eye activity measures.

\begin{table}[H] 
\caption{Summary of Human Stress Detection Studies using Eye Activity Measure.\label{tab:tab12}}
\scalebox{0.9}{
\newcolumntype{C}{>{\centering\arraybackslash}X}
\begin{tabularx}{\textwidth}{CCCCCCCC}
\toprule
\textbf{Method}	& \textbf{Type of Stress}	& \textbf{Number of Subjects (M/F)} & \textbf{Age} & \textbf{Stimulus}& \textbf{Features Domain}& \textbf{Classifier}& \textbf{Accuracy (Classes)}\\
\midrule
~\citep{giannakakis2017stress} & Acute & 23 (16/7) & 45.1$\pm$10.6 & SCWT, IAPS, videos & Time & Adaboost & 91.68\% (2) \\
~\citep{mokhayeri2011mental} & Acute & 60 & 20-28 & SCWT & Time & Fuzzy-SVM & 70.00\% (2) \\

\bottomrule
\end{tabularx}
}

   \noindent{\footnotesize{SCWT: Stroop color-word test, SVM: Support Vector Machine, IAPS: International Affective Picture System}}
\end{table}

\item \textit{Body Postures:}
Body language is a non-verbal type of communication in which physical behavior instead of words is used to convey information. Visual cues-based behavioral features for human stress measurement are presented in~\citep{aigrain2015person}. Behavioral body language features used in the study are visual cues that are extracted from the data acquired by Kinect and an HD camera. The stimulus used for eliciting stress in the participant was a mental arithmetic task. Classification accuracy of 77\% is achieved using a support vector machine classifier. Another human stress measurement scheme based on activity-related behavioral features is proposed in~\citep{giakoumis2012using}. Accelerometer, video-based camera, ECG, and GSR sensors are used to record the behavioral features of the subject. The Stroop color-word test is used as a stress-inducing stimulus for the experiment. The study concluded that the behavioral features extracted correlated with the self-reported response. Behavioral features proved to be better as compared to other physiological signals for the measurement of stress. Classification is performed using linear discriminant analysis and maximum classification accuracy of 96.30\% is achieved. A human stress recognition framework by use of behavioral features extracted by their interaction with technological devices is proposed in~\citep{carneiro2012multimodal}. Eight different behavioral, cognitive, and physical features are examined for the analysis of the effect of different levels of acute stress. A statistical test is applied to measure the difference between different levels of stress. The study revealed the fact that the mean and maximum intensity of the touch is the feature that correlated strongly to human stress. It is also observed that if the stress level of an individual is high then there is less movement in the upper part of the human body. Emotional states including anxiety of an individual are examined using facial cues and gestures from the upper part of the body in~\citep{gunes2007bi}. Stress monitoring by the movement of head and mouth muscles is done~\citep{liao2005real,bevilacqua2018automated}. A study to analyze the facial cues for the estimation of the difference between boredom and stress of a computer game player is presented in~\citep{bevilacqua2018automated}. Seven different facial features are extracted from the players playing the game and it is concluded that 5 out of these 7 features showed a significant difference in boredom and stress state. The head moves under stressful conditions have been reported to be more often~\citep{liao2005real}, more quick~\citep{dinges2005optical,giannakakis2018evaluation}, and the overall head movement is also large~\citep{hadar1983head,giannakakis2018head}. In a study conducted in~\citep{dinges2005optical}, the authors proposed a scheme to detect facial changes during a performance by using optical character recognition in response to low and high stress. Workload and social feedback are used as stress-inducing stimuli in the experiment. The study concluded that the OCR algorithm when applied using mouth and eyebrow region features was able to identify around 75-88\% of the stressed and non-stressed individuals. A study conducted to find the association of the head poses with different kinds of stressors IS proposed in~\citep{giannakakis2018evaluation}. Four different stressors which included, social exposure, emotional recall, stress images or mental tasks, and stressful videos are used to induce stress in the participant. Video recording of the subject is performed when facing each stressor. Head movement and pose features are extracted from the recorded videos. The study reports the fact that more quick head movement is observed in participants when facing stressful situations. The Pitch feature showed a significant difference in the stress and neutral state. The highest classification accuracy of 98.6\% for neutral vs stress state classification is achieved using a kNN classifier with K=3. A study to analyze the head movement in the context of speech during neutral and stress conditions is presented in~\citep{giannakakis2018head}. The tasks involved in stimulus presented to the participants included neutral tasks, interviews, reading text, anxious and stressful event recall, and stressful images and videos. Translational and rotational head movements are used as a feature to assess the stress and neutral state. The study reveals the fact that facing a stressful situation makes the head movement pattern swift and fast. Emotional situations have been identified using head shakes and nods in~\citep{adams2015decoupling}. Another study about the relationship of body postures with human stress level is presented in~\citep{arnrich2009does}. The study is aimed at finding whether stress-related information can be obtained from the posture data in an office work scenario. MIST is used as a stimulus and features are extracted from the pressure distribution on the chair and given to a self-organizing map classifier to classify the stress response. A classification accuracy of 73.75\% is achieved for the proposed scheme. \Tab{tab13} presents a summary of human stress classification schemes using body postures.

\begin{table}[H] 
\caption{Summary of Human Stress Detection Studies using Body Postures.\label{tab:tab13}}
\scalebox{0.9}{
\newcolumntype{C}{>{\centering\arraybackslash}X}
\begin{tabularx}{\textwidth}{CCCCCCCC}
\toprule
\textbf{Method}	& \textbf{Type of Stress}	& \textbf{Number of Subjects (M/F)} & \textbf{Age} & \textbf{Stimulus}& \textbf{Features Domain}& \textbf{Classifier}& \textbf{Accuracy (Classes)}\\
\midrule
~\citep{carneiro2012multimodal} & Acute & 19 & -- & Computer game & Time & DT & 78\% (2) \\
~\citep{giannakakis2018evaluation} & Acute & 24 (17/7) & 47.3$\pm$9.3 & MAT, stressful images & Time & GLR & 97.90\% (2) \\
~\citep{dinges2005optical} & Acute & 60 (29/31) & 30 & Workload and social feedback & Time & OCR & 75-88\% (2) \\
~\citep{arnrich2009does} & Acute & 33 (33/0) & 24.06 & MIST & Frequency & SOM & 73.75\% (2) \\

\bottomrule
\end{tabularx}
}

    \noindent{\footnotesize{MIST: Montreal Imaging Stress Task, MAT: Mental Arithmetic Task, OCR: Optical Character Recognition, SOM: Self-Organizing Map, DT: Decision Tree, GLR: Generalized Likelihood Ratio}}
\end{table}

    \end{enumerate}

\subsubsection{Behavioral Measures}
      Behavior measures correspond to the type of behavior a person adopts when interacting with a certain thing. Behavioral measures have been used in literature for the detection of human stress. Behavioral measures of stress can be sub-divided into the following types i.e., interaction with a computer mouse, interaction with a computer keyboard, and interaction with smartphones. Literature corresponding to each of these types is presented as follows.

      \begin{enumerate}
        \item \textit{Interaction with Computer Mouse:}
        Different mouse interaction-based stress measurement studies have been developed in the literature. An approach to measuring human stress by embedding sensors in the computer mouse to measure the physiological signals while the user is using the mouse is presented in~\citep{liao2005real}. Cameras, pressure sensors, temperature sensors, and GSR sensors are integrated within the mouse to measure the physiological signals to correlate them to the stress of an individual. A capacitive mouse that measures the amount of interaction and the pressure exerted on the computer mouse is developed in~\citep{hernandez2014under}. The authors concluded that the stressed individuals have a significantly higher contact time with the mouse as compared to non-stressed individuals. Another model of how the user moves the mouse under stress is developed in~\citep{sun2014moustress}. The authors proposed an arm-hand dynamics model to measure the muscle stiffness of the subject while moving the mouse. Mouse speed, click rate, and mouse inactivity have been correlated to stress in~\citep{lim2014detecting}. Different features extracted from the movement of the mouse are associated with stress during examination~\citep{carneiro2015using}. \Tab{tab14} presents a summary of human stress classification schemes using computer-mouse interaction.

\begin{table}[H] 
\caption{Summary of Human Stress Detection Studies using Computer Mouse Interaction.\label{tab:tab14}}
\scalebox{0.9}{
\newcolumntype{C}{>{\centering\arraybackslash}X}
\begin{tabularx}{\textwidth}{CCCCCCCC}
\toprule
\textbf{Method}	& \textbf{Type of Stress}	& \textbf{Number of Subjects (M/F)} & \textbf{Age} & \textbf{Stimulus}& \textbf{Features Domain}& \textbf{Classifier}& \textbf{Accuracy (Classes)}\\
\midrule
~\citep{sun2014moustress} & Acute & 49 (23/26) & 20 & Computer task & Frequency & SVM & 70.00\% (2) \\
~\citep{carneiro2015using} & Acute & 53 & -- & Online Exam & Time & NB, DT & 86.40\% (2) \\

\bottomrule
\end{tabularx}
}

   \noindent{\footnotesize{SVM: Support Vector Machine, DT: Decision Tree, NB: Naive Bayes}}
\end{table}

        \item \textit{Interaction with Computer Keyboard:}
        Interaction with a computer keyboard has also been used as a measure of human stress in the literature. A study presented in~\citep{rodrigues2013keystrokes} developed a keyboard dynamics-based approach to measure stress in university students by recording their key latency and writing speed on the keyboard. Another study considered average key latency, average typing speed on the keyboard, and the error rate as a feature to measure the human stress~\citep{lim2014detecting}. Case-based reasoning systems along with multiple features extracted from the interaction with the keyboard have been used for human stress classification~\citep{andren2005case}. A pressure-sensitive keyboard has been developed in~\citep{hernandez2014under} that gives a pressure value between 0 and 255 for each keystroke on the keyboard. Using the pressure values obtained from the keyboard, the stress of the user is measured. Keystroke dynamics is an important measure of human stress because stress causes an effect on different muscles of the human body like around arms, hands, and shoulder muscles. Author in~\citep{gunawardhane2013non} developed a stress measurement scheme using three features of keystroke dynamics which include durations between key pressers of specific digraphs, trigraphs, and an error rate of backspace and delete. A statistical test was applied to the data to show that there exists a significant difference in the keystroke dynamics of the stressed and non-stressed individuals.

        \item \textit{Interaction with Smartphones:}
        Smartphones have completely been revolutionized in the last decade and evolved into a mini-computer with far more functionalities and power as compared to traditional mobile phones. Due to this technological revolution, smartphones have facilitated physicians to get real-time data of patients in just no time. Moreover, applications for smartphones have been built, which enable the users to monitor their health and get advice or alerts on a particular situation~\citep{kailas2010mobile}. Mobile phones are being embedded with sensors like heart rate and body temperature to analyze the health state in a very cost-effective manner. Mobile applications make use of these in-built sensors for health monitoring. Even though these applications are being built with a lack of any kind of scientific validity, but due to their low or even no cost can reach hundreds of thousands of users in no time. Azumio’s Stress Check and StressViewer~\citep{carneiro2017new} apps utilize the light and the camera of the smartphone to monitor the heart rate of the user. Different apps are also available that not only measure the stress but also provide breathing and some other exercises for relieving the stress. DeStressify~\citep{lee2018evaluation} is another app, which is developed to relieve stress based on music. EDA sensor is used by stress-relieving apps named PIP Relax and Race~\citep{dillon2016smartphone}. In these two apps, the user has to participate in a race and the participant which is more relaxed wins the race. DroidJacket~\citep{colunas2011droid} is another app integrated with VitalJacket-a shirt with an integrated ECG sensor to continuously monitor the health of a person~\citep{colunas2011droid}. A specific sensor platform named a Personal Biomonitoring System is used along with smartphones to measure the stress~\citep{gaggioli2012system}. Another smartphone-based approach for the measurement of stress using the features of the speech signal of the user is discussed in~\citep{lu2012stresssense}. A stress classification accuracy of 82.9\% is achieved for indoor scenarios and 77.9\% for outdoor scenarios. Stress-related changes in human behavior using GPS, WiFi, Bluetooth, phone calls, and SMS logs features of a smartphone are explored in~\citep{bauer2012can}. Human stress measurement at the workplace is very important to measure for the better mental and physical health of the employees. A human stress measurement scheme to measure the stress of workers during the workday and sleep at night is presented in~\citep{muaremi2013towards}. Features extracted from the audio, physical activity, and communication data recorded during daytime and heart rate variability data recorded during night sleep are used to build logistic regression models. A leave-one-out cross-validation scheme is used for the classification purpose and a classification accuracy of 61\% is achieved for three-level stress classification. A new stress recognition framework named AMMON (\textbf{A}ffective and \textbf{M}ental health \textbf{MON}itor) which is a speech analysis library for analyzing effect and stress directly on mobile phones is proposed in~\citep{chang2011s}. Stress classification is performed and a classification accuracy of 84\% is achieved for two-class stress classification. A stress recognition framework using the accelerometer sensor of the smartphone is proposed in~\citep{garcia2015automatic}. Oldenburg burnout inventory (OLBI) questionnaire is used to acquire the subjective response from the users and features are extracted in the time and frequency domain and fed to a Naive Bayes and decision tree classifier. A classification accuracy of 71\% is achieved for user-specific models. A driver stress monitoring system using inertial sensors is proposed in~\citep{lee2016wearable}. For comparison of the results, GSR, self-reporting questionnaire, and facial expressions are employed. Forty-six features are extracted from the data which were subjected to feature selection resulting in 22 features. SVM classifier is used to discriminate the low-stressed participants from the high-stressed participants with achieved accuracy of 94\%. A student stress measurement mechanism using smartphone sensors is proposed in~\citep{gjoreski2015automatic}. The authors used an accelerometer, GPS, WiFi, time and duration of calls, light sensor data, and a self-reported questionnaire for the classification of stress. Forty-seven features are extracted from the acquired data and a classification accuracy of 60\% is achieved for three classes. Classification in this study is performed in such a manner that the data of each student is divided into two parts ie., some features of each student are used for training, and some features are used for testing. An automatic stress measurement system for graduate students using smartphone data is proposed in~\citep{bogomolov2014pervasive,bogomolov2014daily}. Smartphone data which included mobile phone activity of call and SMS logs and Bluetooth proximity hits, weather conditions, and personality traits are recorded from 227 students for a duration of one year. Weather conditions are divided into mean temperature, pressure, total precipitation, humidity, visibility, and wind speed whereas personality traits are obtained by using the big five personality trait questionnaire and are labeled into extraversion, neuroticism, agreeableness, conscientiousness, and openness to experience. Subjective labeling of stress for the recorded data is obtained by using a self-reported questionnaire. A wide range of classification algorithms is applied but the random forest algorithm proved to be the best with an achieved classification accuracy of 72\% and a feature reduction from 500 features to 32 features. Another human stress mechanism using mobile phone activity both in the laboratory environment and out of the lab environment is proposed in~\citep{ciman2016individuals}. For the controlled lab environment part of the experiment, an android application that contained search and writing tasks is developed. The user activities which are monitored during the performance of these tasks include users' tap, scroll, swipe, and text input gestures. Stressors are used to induce stress in the participants and the Experience Sampling Method is used to obtain the subject stress scores. kNN, SVM, decision trees, and neural networks classifiers are used with achieved accuracy of 80\%. For the out-of-lab environment, the activities of the subjects including the type of applications used, user physical activity, and light values of the screen are recorded using the daily life usage of mobile phones. The achieved classification accuracy for this part of the experiment is 70\%. Another smart sensor-based and context-aware stress measurement scheme for daily life stress monitoring is proposed in~\citep{gjoreski2017monitoring}. Real-life data of 55 days is recorded and precision and recall of 95\% and 70\% are achieved, respectively. Smartphone data has been used in the stress recognition framework proposed in~\citep{gimpel2015mystress}. The authors developed an android application and used 36 software and hardware sensors in their study. The authors did not report any classification accuracy but they found that high smartphone usage, average battery temperature, the maximum number of running applications, and the frequency of switching the display on are the features that are strongly correlated to stress. Another daily life stress monitoring system using smartphone sensors is proposed in~\citep{sysoev2015noninvasive}. Audio, gyroscope, accelerometer, ambient light sensor data, screen mode changing frequency, self-assessment, and activity type are used as a feature, and NASA-TLX is used as a subjective stress questionnaire. Activity recognizer is used along with the stress recognition system to achieve a classification accuracy of 77\%. Another stress recognition framework called as StayActive is developed for android phones in~\citep{kostopoulos2017stress}. Social interaction, physical activity, and sleeping patterns are used for the measurement of stress. A fusion of offline mathematical modeling and online machine learning models is used to identify stress and give some relaxation therapy when the stress is identified. The Circumplex Model of Effect with some modifications about stress measurement is used as a questionnaire to obtain the subjective score. The number of sleeping hours is used as a feature of sleep patterns, the number of touches on the screen, the number of calls, and SMSs are used as a feature of social interaction. The author did not report any accuracy but they intended to use physiological signals along with a smartphone in the future to further improve the results of the proposed stress measurement scheme. Another unsupervised stress classification scheme using smartphone data is proposed in~\citep{vildjiounaite2018unobtrusive}. The hidden Markov model is used for stress classification and an accuracy of 68\% is achieved. \Tab{tab15} presents a summary of human stress classification schemes using smartphone interaction.

\begin{table}[H] 
\caption{Summary of Human Stress Detection Studies using Smartphone Interaction.\label{tab:tab15}}
\scalebox{0.9}{
\newcolumntype{C}{>{\centering\arraybackslash}X}
\begin{tabularx}{\textwidth}{CCCCCCCC}
\toprule
\textbf{Method}	& \textbf{Type of Stress}	& \textbf{Number of Subjects (M/F)} & \textbf{Age} & \textbf{Stimulus}& \textbf{Features Domain}& \textbf{Classifier}& \textbf{Accuracy (Classes)}\\
\midrule
~\citep{lu2012stresssense} & Acute & 14 (4/10) & 22.86 & Job interviews,Marketing jobs & Time and Frequency & GMM & 81.00\% (2) \\
~\citep{muaremi2013towards} & Acute & 35 (24/11) & 25-62 & Day long recording & Time and Frequency & LR & 61.00\% (3)\\
~\citep{garcia2015automatic} & Chronic & 30 (18/12) & 37.46$\pm$7.26 & Day long recording & Time and Frequency & DT, NB,ONB & 71.00\% (2)\\
~\citep{lee2016wearable} & Acute & 8 (6/2) & 30$\pm$5 & Car Driving & Time and Frequency & SVM & 94.78\% (2) \\
~\citep{gjoreski2015automatic} & Chronic & 48 & -- & Day long recording & Time & SVM, DT, RF & 60.00\% (3)\\
~\citep{ciman2016individuals} & Acute & 13 (7/6) & 22-32 & MAT & Time & kNN, SVM,DT, NN & 80.00\% (5) \\
~\citep{gjoreski2017monitoring} & Chronic & 21 & 28$\pm$4 & Day long recording & Time and Frequency &  DT, NB,kNN, SVM, RF, ES & 73.00\% (3) \\
~\citep{sysoev2015noninvasive} & Chronic & -- & -- & Real life activities & Time & RF, SL & 77.00\% (2) \\
~\citep{vildjiounaite2018unobtrusive} & Chronic & 30 & -- & Real life activities & Time and Frequency  & HMM & 68.00\% (2)\\ 

\bottomrule
\end{tabularx}
}
    \noindent{\footnotesize{MAT MAT: Mental Arithmetic Task, SVM: Support Vector Machine, DT: Decision Tree, kNN: K- nearest neighbors, NN: Neural Network, NB: Naive Bayes, ONB: Ordinal Naive Bayes, ES: Ensemble Selection, SL: Simple Logic, GMM: Gaussian Mixture Models, LR: Logistic Regression, RF: Random Forest, HMM: Hidden Markov Model}}
\end{table}

      \end{enumerate}

    \subsubsection{Vision based Measures}
     Vision-based measures have also been used for human stress detection and used some kind of imaging modality for measuring the response of the user. Vision-based techniques can be sub-divided into thermal infrared (IR) and computer vision-based techniques. Literature for each of these sub-divisions is given below.

     \begin{enumerate}
       \item \textit{Thermal Infrared Imaging:}
       Thermal IR imaging is a non-invasive and contactless technique used to measure the temperature of the human skin. In this technique, a thermal infrared camera is used to record the oxygen tissue saturation level. The benefit of this technique is that it is not affected by skin color and lighting conditions. When a person is feeling stressed, the flow of blood in the vessels get increases, and hence the temperature in the adjacent regions is raised. Human affective states like fear~\citep{levine2001face}, arousal~\citep{nozawa2009correlation}, and stress~\citep{ioannou2014thermal} have been recognized using thermal imaging~\citep{nhan2009classifying}. In most of the affect recognition studies, facial skin temperature is measured using thermal IR imaging to extract some useful information~\citep{nhan2009classifying,shastri2008imaging} Skin temperature of the nose, chin, and corrugator is affected due to stressors~\citep{hong2016real}. Even though no conclusive remarks can be given of the effect of stress on a specific region of the face. However, in studies conducted in~\citep{puri2005stresscam,chen2014detection}, forehead temperature is raised under stressful conditions. Periorbital areas also show signs of increasing temperature under anxious states~\citep{pavlidis2002thermal,pavlidis2000thermal}. Skin temperature is also reported to increase in supraorbital and periorbital areas under stress~\citep{shastri2008imaging}. Studies conducted in~\citep{engert2014exploring,vinkers2013effect,kang2006determining} showed an increase in nose temperature when an unknown task is presented to the participants. A decrease in the temperature of the perinasal area is found under stressful condition~\citep{pavlidis2012fast,shastri2012perinasal}. Another thermal camera-based stress recognition framework is proposed in studies conducted in~\citep{cho2017deepbreath,cho2017thermsense}. A thermal camera is used to detect breathing and a respiratory spectrogram was used to extract features. SWT and the mental arithmetic task are used as a stimulus to induce stress in the lab environment. Convolutional Neural Network (CNN) classifier is used for two and three-level stress classification. Two-level stress classification accuracy of 84.59\% and three-level stress classification of 56.52\% is achieved for the proposed scheme. \Tab{tab16} presents a summary of human stress classification schemes using thermal imaging.

\begin{table}[H] 
\caption{Summary of Human Stress Detection Studies using Thermal Imaging.\label{tab:tab16}}
\scalebox{0.9}{
\newcolumntype{C}{>{\centering\arraybackslash}X}
\begin{tabularx}{\textwidth}{CCCCCCCC}
\toprule
\textbf{Method}	& \textbf{Type of Stress}	& \textbf{Number of Subjects (M/F)} & \textbf{Age} & \textbf{Stimulus}& \textbf{Features Domain}& \textbf{Classifier}& \textbf{Accuracy (Classes)}\\
\midrule
~\citep{nhan2009classifying} & Acute & 12 (3/9) & 24$\pm$2.9 & Visual stimulus & Time and Frequency & LDA & 70-80\% (6) \\
~\citep{hong2016real} & Acute & 41 & 20-65 & TSST & Frequency & DEFP & 90.00\% (2) \\
~\citep{chen2014detection} & Acute & 21 (19/2) & 25 & TSST & Frequency & Binary classifier & 88.10\% (2) \\
~\citep{cho2017deepbreath} & Acute & 8 (5/3) & 18-53 & SCWT, MAT & Time & CNN & 84.59\% (2), 56.52\% (3) \\

\bottomrule
\end{tabularx}
}

   \noindent{\footnotesize{TSST: Trier Social Stressor Test, MAT: Mental Arithmetic Task, SCWT: Stroop color-word test, LDA: Linear Discriminant Analysis, DEFP: Differential Energy between Philtrum and Forehead, CNN: Convolutional Neural Network}}
\end{table}

       \item \textit{Computer Vision:}
       In computer vision-based, human stress assessment, many different organs, and locations of the human body have been used, but the face is the most commonly used location for monitoring stress. Authors proposed a computer recognition algorithm for use by astronauts to identify the facial changes occurring in response to low and high stressor~\citep{dinges2005optical}. Another study to identify the stress of the drivers using facial expressions is presented~\citep{gao2014detecting}. Thermal imaging can be used to measure blood perfusion which is correlated to human stress~\citep{derakhshan2014preliminary}. Another study analyzed the recorded video by use of both temporal thermal spectrum and visible spectrum video features to measure the stress~\citep{sharma2014thermal}. Another human stress measurement scheme using Kinect sensors is proposed in~\citep{aigrain2015person}. The participants of the experiment are asked to give answers to the time-constrained arithmetic questions in front of a video camera. Facial features along with body postures are extracted and fed to an SVM classifier for stress classification. An accuracy of 77\% is achieved for stress vs non-stressed class classification. Changes in the facial blood flow using thermal and visible cameras are used for stress detection in the study conducted in~\citep{mohd2015mental}. It was reported in the study that facial thermal features are difficult to attain due to low contrast so they applied a Nostril mask to focus on the nostril area. Noise smoothing is performed using graph cut algorithms and feature extraction is performed using Scale Invariant Feature Transform resulting in a classification accuracy of 88.6\% for two classes. Facial hyperspectral imaging (HSI) technique and tissue oxygen saturation (StO2) data are used to identify stress in a study conducted in~\citep{chen2014detection}. Results obtained from thermal imaging and HSI are compared in the proposed scheme. TSST is applied to induce stress in the participants of the experiment. It is reported that the StO2 in the eye and forehead of the subject are discriminative features for the identification of stress. An accuracy of 76\% with automatic thresholding and 88\% for manual thresholding is achieved for two-level stress using computer vision-based techniques.

\begin{table}[H] 
\caption{Summary of Human Stress Detection Studies using Computer Vision Based Techniques.\label{tab:tab17}}
\scalebox{0.9}{
\newcolumntype{C}{>{\centering\arraybackslash}X}
\begin{tabularx}{\textwidth}{CCCCCCCC}
\toprule
\textbf{Method}	& \textbf{Type of Stress}	& \textbf{Number of Subjects (M/F)} & \textbf{Age} & \textbf{Stimulus}& \textbf{Features Domain}& \textbf{Classifier}& \textbf{Accuracy (Classes)}\\
\midrule
~\citep{gao2014detecting} & Acute  & 21 &  -- & Car driving & Time & SVM & 90.50\% (2) \\
~\citep{derakhshan2014preliminary} & Acute & 12 &  -- & peak of tension (POT) test & Time & SVM & 96.00\% (2) \\
~\citep{sharma2014thermal} & Acute & 35 (22/13) & -- & Video clips & Time and Frequency & SVM & 86.00\% (2) \\
~\citep{aigrain2015person} & Acute & 14 (11/3) & 24.8$\pm$2.8 & Public speaking, MAT &  Time and Frequency & SVM & 77.00\% (2) \\

\bottomrule
\end{tabularx}
}

    \noindent{\footnotesize{MAT: Mental Arithmetic Task, SVM: Support Vector Machine}}
\end{table}

     \end{enumerate}

\section{Multimodal Stress Detection}
\label{sec:msa}
Multimodal human stress detection has been focused on in literature in a wide range of studies. The primary aim of the multimodal stress detection framework is to increase the system accuracy as compared to single-modality stress measurement systems. Multimodal stress detection schemes available in the literature can be sub-divided into (i) fusion of data recorded from different physiological modalities, (ii) fusion of data obtained from motion and physiological sensors (iii) fusion of data obtained from imaging modalities and physiological sensors, (iv) fusion of data obtained from smartphones and physical, behavioral and physiological sensors. In this section, we will discuss available literature for stress measurement using all kinds of multimodal fusion approaches.  
  
A multimodal human stress measurement is proposed in~\citep{al2016mental}, where EEG and functional near-infrared spectroscopy (fNIRS) are used to classify acute stress. Classification accuracy of 96.6\% is achieved with EEG and fNIRS data. Another human stress measurement system using GSR and skin temperature is carried out in~\citep{kyriakou2019detecting} with an achieved accuracy of 84\%. Another real-time stress detection scheme using heart rate, skin conductance, and accelerometer sensors is proposed in~\citep{can2019continuous}. Using a fusion of features from these sensors an accuracy of 92.15\% is achieved for three-level stress classification. A multimodal stress classification framework for drivers using the physiological signal of PPG and inertial sensors of accelerometer, gyroscope, and magnetometer is proposed in~\citep{lee2016stress}. Driving was performed in a simulator environment and a driving behavior survey questionnaire is used to obtain the subjective response from the subjects. Time and frequency domain features are extracted from the acquired data and stress classification is performed using an SVM classifier with a radial basis function (RBF) with an achieved accuracy of 95\% for two classes. A stress recognition framework for driver stress monitoring using physiological signals of GSR, ECG, and respiration is proposed in~\citep{chen2017detecting}. Features are extracted in time, frequency, and wavelet domain, and PCA and Sparse Bayesian Learning (SBL) algorithms are applied for feature selection and SVM for classification resulting in an accuracy of 99\% for three classes. Another multimodal driver stress recognition framework using DRIVE DB from the PHYSIONET database and physiological signals of GSR, EMG, and ECG is proposed in~\citep{ghaderi2015machine}. Three-level stress classification is performed using kNN and SVM classifier and a classification accuracy of 98\% is achieved. Another multimodal stress classification scheme using physiological signals of BVP, HR, ST, GSR, and RR is proposed in~\citep{gjoreski2016continuous}. The mental arithmetic task is used for the induction of stress and 63 features are extracted from the recorded data. For the in-lab experiment, two (non-stressed, stressed) and three (no stress, low stress, and high stress) level stress classification is performed. For two-level stress classification, an accuracy of 83\% is achieved whereas for a three-class problem, an accuracy of 72\% is achieved. For the out-of-lab environment, experimental activities of walking, sitting, running, and cycling are recorded. All these activities are numbered according to their intensities i.e., 1 for lying and 5 for running. To determine the stress-inducing interval, the average intensity of the interval was calculated. The importance of recording the activities along with their intensity was to provide context to the situation to create a distinction between a strong physical activity and a stressful situation. The day-long activity is subdivided into one-hour episodes. An accuracy of 76\% and 92\% is achieved with no-context and with context scenarios, respectively. Another multimodal stress recognition frame framework using ECG signals along with activity and contextual data is proposed in a study conducted in~\citep{maier2014mobile}. The application they designed measured the stress and when the stress exceeded a certain threshold the user is allowed to quit a stressful situation or is given some relaxation therapy to reduce the level of stress. Accelerometer and GPS data are added to the HRV data obtained from ECG data to achieve the accuracy of the system. The authors did not report any classification accuracy for the proposed scheme. Another stress measurement scheme using accelerometer and EDA data obtained from the wrist sensors, call and SMS data, location and screen on/off features obtained from mobile phones, and subjective stress score using questionnaires is proposed in~\citep{sano2013stress}. Sequential Forward Floating Selection (SFFS) is used to select the optimum set of features which selected screen on/off, mobility, call acceleration, and EDA data features, and an accuracy of 75\% is achieved for class stress classification using SVM and kNN as a classifier. Moreover, high-stress behavior is found to be associated with acceleration during sleep in the second half of the day, a small amount of SMS, and screen time. Another multimodal stress classification scheme using respiratory, ECG, and accelerometer sensors is proposed in~\citep{hovsepian2015cstress}. The experimental setup included baseline recording, public speaking task, mental arithmetic task, and cold pressure test. This data is used to train the model using laboratory settings and data from 23 participants is recorded in the out-of-lab environment for testing purposes. SVM is used as a classifier and a recall of 89\% is achieved on the test data from the lab environment and a classification accuracy of 72\% is achieved on testing data from the out-of-lab environment. Another multimodal stress measurement study using an accelerometer, EDA, and Bluetooth sensor is presented in~\citep{zubair2015smart}. The experiment is performed in a controlled environment and logistic regression is used as a classifier to achieve an accuracy of 91\% for two-class stress classification. A real-time stress recognition methodology using HR and EDA signals is proposed in~\citep{de2010two}. The public speaking task is used as a stimulus and the kNN classifier is used to detect stress and relaxed state with an accuracy of 95\%. PPG and EDA signals are used to identify stress in subjects in a study conducted in~\citep{sandulescu2015stress}. TSST is used to induce stress and an SVM classifier is used to classify stress into two levels with an accuracy of 80\%. Another study using a combination of EDA and HRV signals for measurement of stress is proposed in~\citep{martinez2017real}. Puzzles are used as a stress-inducing stimulus and an F-measure value of 0.984, 0.970, and 0.943 is achieved for the high, medium, and low-stress groups, respectively. Another stress detection scheme using physiological and sociometric sensors is proposed in~\citep{mozos2017stress}. Physiological sensors include EDA and PPG whereas sociometric sensors include microphone and accelerometer sensors. Public speaking is used as a stressor to induce stress in the participants and kNN, SVM, and AdaBoost are used as a classifier. AdaBoost classifier produces the highest accuracy of 94\% for discriminating stress and neutral condition. In another study presented in~\citep{kurniawan2013stress} author made use of GSR and speech signals for the measurement of stress. SWT, TMST (Trier Mental Stress Test), and TSST are used to induce stress in the participants. Time and frequency domain features are extracted from the speech signals. K-means, GMM, SVM, and decisions are used for classification purposes, SVM produced the best results for each type of stressor. Speech signals resulted in better classification accuracy as compared to EDA data. The fusion of speech data with the GSR signal did not increase the classification accuracy for stress classification. Another multimodal stress classification scheme using cardiac features along with EMG, GSR, and respiratory sensors and a Kinect-based video camera is proposed in~\citep{aigrain2016multimodal}. Moreover, in addition to the self-reported questionnaire feedback from the psychology expert is also added to the proposed system. SVM is used as a classifier and a classification accuracy of 85\% for two classes is achieved. Pupil dilation and periorbital temperature data are used for stress classification in a study proposed in~\citep{baltaci2014role}. IAPS is used as a stress-inducing stimulus for the experiment. The decision tree classifier resulted in a classification accuracy of 90\% for two classes. The study proposed in~\citep{baltaci2014role} is improved by authors in their study conducted in~\citep{baltaci2016stress} by the addition of entropy feature for the physiological signals. The study resulted in better classification accuracy as compared to the earlier studies by using the AdaBoost classifier with the Random Forest classifier instead of the decision tree. Human gaze and mouse click behavior are used for stress classification in~\citep{huang2016stressclick}. The stress stimulus used in the experiment is a mental arithmetic task. Random forest classifier achieved the highest accuracy of 60\% for two classes for the generalized stress model. Another multimodal stress recognition framework for computer users using physiological signals of EEG, ECG, EMG, and EOG is proposed in~\citep{akhonda2014stress}. EEG data and the subjective questionnaire score is obtained only once at the start of the experimental procedure, whereas ECG, EOG, and EMG data are acquired continuously. A neural network is used as a classifier and three class stress recognition is performed with an achieved accuracy of 80\%.

A stress detection system using physiological signals of ECG, respiration rate, skin temperature, and GSR is presented in~\citep{shi2010personalized} with a recall of 80\%. Another human stress detection system using wearable EEG and ECG sensors is presented in~\citep{ahn2019novel}. Stroop color-word and the mental arithmetic test is used as a stimulus and an accuracy of 87.5\% is achieved for stress classification. The fusion of ECG, EMG, GSR, and respiration signals are used to measure the driver's stress in~\citep{healey2005detecting}. Classification accuracy of 97\% is achieved for three levels of stress. Another study to analyze the impact of human stress on the sleep pattern using wearable sensors of ECG, GSR, body temperature, and respiration is presented in~\citep{muaremi2014monitoring}. SVM, kNN, NN, RF (Random Forest), and logistic regression classifiers are used to classify the stress and SVM produced the best classification accuracy of 73\% for three classes of stress. A cluster-based technique for the detection of perceived stress using EEG, ECG, GSR, and EMG signals is presented in~\citep{xu2014cluster}. Real-time human stress classification using ECG and thoracic electrical bioimpedance (TEB) signals is presented in~\citep{mohino2015assessment} with an error rate of 21\%. Another stress detection study focusing the working people using ECG and GSR signals is presented in~\citep{sriramprakash2017stress}. The database used in the study was SWELL-KW and a classification accuracy of 72.82\% is achieved for stressed vs non-stressed classes.

Another stress recognition system using respiratory rate along with GSR, EEG, and blood volume pressure is presented in~\citep{hosseini2010emotional} with a classification accuracy of 82.7\%. A stress assessment study for the office environment using HR, HRV, GSR, EMG, and respiratory rate is discussed in~\citep{wijsman2013wearable} with 74.5\% as a reported accuracy. Driver stress has been monitored using ECG, GSR, and respiration rate in~\citep{rigas2011real}. GSR, EMG, respiration, and HR combination have also explored driver stress level~\citep{singh2013novel}. A study to classify the emotional states of stress and anger using GSR, EMG, BVP, and respiratory rate signals is discussed in~\citep{picard2001toward}. ECG, GSR, EMG, and respiration rate have also been explored for human stress assessment in~\citep{wijsman2011towards}. HRV, respiration, GSR, EMG, and geographical locations have been used for the detection of mental stress in~\citep{choi2011minimally}. Stress level estimation using GSR, EMG, HR, and respiration sensors is presented in~\citep{gjoreski2016continuous}. A human perceived stress measurement system using smartphone PPG and thermal imaging is presented in~\citep{cho2019instant} with a reported average classification accuracy of 78.3\%.

Another study to correlate the results of the human stress detection system based on ECG, EDA, and EEG sensors with the changes in the cortisol level of the subject is discussed in~\citep{betti2017evaluation}. The study reveals that the changes in cortisol levels were strongly in line with the physiological signal response. Classification accuracy of 86\% is achieved for the stress classification. Another stress classification scheme using HRV features and cortisol as a reference is proposed in~\citep{liew2015classifying}. Classification accuracy of 80\% is achieved using a cortisol bio-marker. ECG and salivary cortisol are used for the measurement of different levels of psycho-social stress in~\citep{nater2005human}. Bad childhood experience has been reported to cause a change in the physiological processes and can determine the magnitude of the stress response. HRV, BVP, ECG, and salivary cortisol are used in~\citep{aimie2018stress} to propose a stress response index that could be used as a future biomarker for such individuals. A fusion of keyboard strokes along with the linguistic features of the written text has been used in~\citep{vizer2009automated} to measure the cognitive and physical stress of the user. The fusion of features from pupil diameter and physiological signals is used in~\citep{barreto2007non,zhai2008stress} to measure the stress. The stress of a computer user is measured by a fusion of features from physical appearance, physiological signals, and behavioral data in a study conducted in~\citep{liao2005real}. In another study, accelerometer data and recorded videos are analyzed to monitor the stress response~\citep{giakoumis2012using}. A novel stress detection framework based on three different state-of-the-art technologies which include Deep Learning (DL), blockchain, and the Internet of Medical Things (IoMT) has been proposed in \cite{qi2023blockchain}. The authors claim that the blend of these three technologies can help in mitigating the effects occurring due to long-term stress. The deep learning part of the system was evaluated on a publicly available WESAD dataset with a reported average accuracy of 99\%. A stress detection framework for individuals suffering from communication diseases like COVID-19 is proposed in \cite{khan2022secure}. Wearable sensors which include ECG, oximeter, temperature, accelerometer, GPS, and pressure sensors are embedded in the clothes of the subject. The data from these sensors is transferred to a cloud-based service to execute the machine-learning-based prediction of human stress. A study focusing on human stress measurement of construction workers using heart rate, galvanic skin response, and the camera is presented in \cite{rescio2023ambient}. An accuracy of 75.1\% is reported for binary stress classification. A multimodal human stress assessment scheme using temperature, respiration, ECG, and EDA signals is presented in \cite{mohammadi2022integrated}. A classification accuracy of 96\% is reported using the KNN classifier. A deep learning-based multimodal worker's stress recognition framework using electrocardiogram (ECG), respiration (RESP), and video data is presented in \cite{seo2022deep}. An average accuracy of 73.3\% and 54.4\% is reported for two and three-level stress classification. A CNN-based stress detection framework using physiological signals of EDA and ECG is presented in \cite{kuttala2023multimodal}. An accuracy of 97.6\% is achieved for three-level stress classification. A human stress measurement scheme using a fusion of data from EDA, PPG, and ST sensors is presented in \cite{nath2021smart}. An accuracy of 94\% is reported for two-level stress classification. A stress level detection framework using a fusion of features from ECG, EDA and EMG while playing a virtual reality video game is presented in \cite{orozco2022stress} with an accuracy of 99\% for four-level stress classification. A driver stress recognition scheme using a fusion of data from ECG, GSR and eye-tracking is presented in  \cite{vaitheeshwari2022stress} with a reported accuracy of 95.1\%. A multimodal human stress assessment scheme using a fusion of data from GSR, PPG, and EEG signals in response to public speaking activity is presented in \cite{arsalan2021human}. An accuracy of 96.25\% is reported for two-level stress classification. A study for the assessment of mental fatigue in construction workers using EEG signals and facial features is presented in \cite{mehmood2022validity}. A strong correlation was found between the recorded EEG signals  and the facial features of the construction workers.     
\Tab{tab18} presents a summary of multimodal human stress classification schemes available in the literature.


\begin{table}[H] 
\caption{Summary of Multimodal Human Stress Detection Studies.}\label{tab:tab18}
\scalebox{0.7}{

\newcolumntype{C}{>{\centering\arraybackslash}X}

\begin{tabularx}{1\fulllength}{CCCCCCCCC}
\toprule
\textbf{Method}	& \textbf{Type of Stress}	&  Modalities & \textbf{Number of Subjects (M/F)} & \textbf{Age} & \textbf{Stimulus}& \textbf{Features Domain}& \textbf{Classifier}& \textbf{Accuracy (Classes)}\\
\midrule
~\citep{al2016mental} & Acute & EEG, fNIRS & (22 (22/0)) & 22-30 & MIST & Frequency & SVM & 95.10\% (2) \\
~\citep{kyriakou2019detecting} & Acute & GSR, ST & 19 (8/11) & 25-45 & Audio & Time & -- & 84.00\% (2) \\
~\citep{can2019continuous} & Acute & HR, GSR, Acc &  21 (18/3) & 20 & Computer programming & Frequency & MLP & 92.15\% (3) \\
~\citep{lee2016stress} & Acute & Acc, Gyro, Mag & 28 (18/10) & 35$\pm$16 & Car driving & Time and Frequency & SVM & 95.00\% (2) \\
~\citep{chen2017detecting} & Acute & ECG, GSR, RR & 9 & -- & Car driving & Time and Frequency & SVM & 99.00\% (3) \\
~\citep{ghaderi2015machine} & Acute & RR, GSR, HR, EMG & 17 & -- & Car driving & Time and Frequency & SVM, kNN & 98.00\% (3) \\
~\citep{gjoreski2016continuous} & Acute & BVP, HR, ST,GSR and RR & 21 & 28$\pm$4.1 & MAT & Time and Frequency & RF & 92.00\% (2) \\
~\citep{sano2013stress} & Acute & SC, Acc & 18 (15/3) &28$\pm$ 7.8 & Daily Life Activity & Time and Frequency & SVM, kNN & 75.00\% (2) \\
~\citep{hovsepian2015cstress} & Chronic & RR, ECG,Acc & 23 & -- & Day long recording & Time & SVM & 95.30\% (2) \\
~\citep{zubair2015smart} & Acute & EDA, Acc, Bluetooth & 12 & -- & MAT, Emotional pictures & Time & LR & 91.00\% (2)\\
~\citep{de2010two} & Acute & HR, EDA & 80 (0/80) & 19-32 & Public Speaking & Time & kNN & 95.00\% (2) \\
~\citep{sandulescu2015stress} & Acute & PPG, EDA & 5 & 18-39 & TSST & Time and Frequency & SVM & 80.00\% (2) \\
~\citep{mozos2017stress} & Acute & EDA, PPG, Acc, microphone & 18 & -- & Public Speaking & Time and Frequency & kNN, SVM, AdaBoost & 94.00\% (2) \\
~\citep{kurniawan2013stress} & Acute & GSR, Speech & 10 & -- & SCWT, TMST, TSST & Time and Frequency & k-mean, SVM, GMM & 92.00\% (2) \\
~\citep{aigrain2016multimodal} & Acute & EMG, GSR, RR, Kinect & 21 (6/15) & 26.3$\pm$4.6 & MAT & Time and Frequency & SVM & 85.00\% (2) \\
~\citep{baltaci2016stress} & Acute & pupil dilation, periorbital temperature & 11 (9/2) & 29-40 & IAPS & Time & ABRF & 65\%-84\% (2) \\
~\citep{huang2016stressclick} & Acute & Eye gaze, mouse click behaviour & 20 (13/7) & 20-33 & MAT & Time & RF & 60.00\% (2) \\
~\citep{akhonda2014stress} & Acute & EEG, ECG, EMG, EOG & 12 & -- & Computer Work & Time and Frequency & NN & 80.00\% (3) \\
~\citep{ahn2019novel} & Acute & EEG, ECG & 7 & 29.3$\pm$2.4 & MAT, SWT & Time and Frequency & SVM & 87.50\% (2) \\
~\citep{healey2005detecting} & Acute & ECG, EMG, GSR & 24 & -- & Car driving & Time and Frequency & FDA & 97.00\% (3) \\
~\citep{muaremi2014monitoring} & Acute & ECG, GSR, ST, RR & 10 (7/3) & 41 & Sleep data & Time and Frequency & SVM, kNN, NN, RF, LR & 73.00\% (3) \\
~\citep{xu2014cluster} & Chronic & EEG, ECG, GSR, EMG & 44 (44/0) & 28.6$\pm$7.2 & PASAT & Time and Frequency & k-Mean & 85.20\% (3) \\
~\citep{sriramprakash2017stress} & Chronic & ECG, GSR & 25 & -- & Office work & Time and Frequency & SVM, kNN & 72.82\% (2) \\
~\citep{hosseini2010emotional} & Acute & GSR, EEG, BVP & 15 (15/0) & 20-24 & IAPS & Time and Frequency & SVM & 84.10\% (2) \\
~\citep{wijsman2013wearable} & Acute & HR, HRV, GSR, EMG, RR & 30 (25/5) & 19-53 & calculation, puzzle and memory task & Time and Frequency & GEE & 74.50\% (2) \\
~\citep{rigas2011real} & Acute & ECG, GSR, RR & 13 (10/3) & 22-41 & Car driving & Time and Frequency & BN & 96.00\% (2) \\
~\citep{wijsman2011towards} & Acute & ECG, RR, GSR, EMG & 30 (25/5) & 19-53 & calculation, puzzle and memory task & Time and Frequency & LBN & 80.00\% (2) \\
~\citep{gjoreski2016continuous} & Acute & GSR, EMG, HR, RR & 26 & -- & Daily life activity & Time and Frequency & RF & 92.00\% (2) \\
~\citep{cho2019instant} & Acute & PPG, Thermal imaging & 17 (8/9) & 29.82 & Mental workload &
Time and Frequency & NN & 78.33\% (2) \\
~\citep{betti2017evaluation} & Acute & ECG, EDA, EEG & 26 (8/7) & 60.8$pm$9.5 & MAST & Time and Frequency & SVM & 86.00\% (2) \\
~\citep{liew2015classifying} & Acute & HRV, Cortisol & 22 (17/5) & 21 & TSST & Time and Frequency & FAM & 80.00\% (2) \\
~\citep{vizer2009automated} & Acute & keystroke, linguistic features & 24 (10/14) & 18-56 & Cognitive task & Time & DT, kNN, SVM, ANN & 75.00\% (2) \\
~\citep{barreto2007non} & Acute & BVP, GSR, ST, PD & 32 & 21-42 & SCWT & Time and Frequency & SVM & 90.10\% (2) \\
~\citep{zhai2008stress} & Acute & BVP, GSR, ST, PD & 32 & 21-42 & SCWT & Time & SVM & 90.10\% (2) \\
~\citep{giakoumis2012using} & Acute & ECG, GSR, Acc, Video & 21 (17/4) & 30.4$\pm$3.7 & SCWT & Time & LDA & 100\% (2) \\
\bottomrule
\end{tabularx}
}

   \noindent{\footnotesize{TSST TSST: Trier Social Stressor Test, TMST: Trier Mental Stress Test, PASAT: Effects of the Paced Auditory Serial Addition Task, LDA: Linear Discriminant Analysis, LBN: Linear Bayes Normal, MAST:  Maastricht Acute Stress Test, FAM: Fuzzy ARTMAP, GMM: Gaussian Mixture Model, ABRF: Adaptive Boosting Random Forest, FDA: Fisher Discriminant Analysis, GEE: Generalized Estimating Equations}}
\end{table}

\section{Future Directions}
\label{sec:fd}
In this section, future directions which could be adopted for the improvement of existing human stress measurement methods and the open challenges that still need to be addressed in the literature are discussed.

One of the most important limitations of most of the existing stress measurement schemes is that they are performed in a well-controlled lab environment whereas measurement of stress in real-life poses a different set of challenges as compared to the lab environment. The challenges posed by real-life stress measurement scenarios include the occurrence of multiple stressors at the same time (unlike lab environment where a particular stressor is given to the participant at one time), incomplete contextual information like lack of information about changing room temperature, and the effect of outdoor environmental conditions on the acquired physiological signals.
The addition of contextual information in real life, as well as laboratory settings, has been found useful toward efficient measurement of human stress in a study conducted in~\citep{gjoreski2016continuous} thus supporting the fact that contextual information is important and has not been widely considered in the available literature. In future stress measurement studies, challenges occurring due to the out-of-lab environment should be explored to make the available stress measurement method more practical in a real-life scenario.

The availability of the ground truth for labeling data used for training the stress detection model is another open research issue. Either in the lab environment or real-life scenario, we don't have the ground truth and the majority of the studies in the literature have to rely on subjective scoring which can vary from person to person. It has been observed that physiological data from two participants may have the same pattern but one of the participant's labeled as "stressed" whereas the other is labeled as "non-stressed" in subjective labeling~\citep{liapis2015stress}. This type of labeling ambiguity degrades the performance of the system and thus needs to be rectified by developing a more robust mechanism for labeling the data.

Power consumption of data acquisition is another important factor that needs to be considered while recording the data in an out-of-lab environment. In lab stress measurement system can use medical-grade and other data acquisition devices with constant power supply and there is no power outage, whereas in real-life scenarios all the devices whether they are EEG headsets, skin conductance, and heart rate measurement modules or smartwatches all have limited battery life and can last approximately around 4-5 hours. But to record the activities of a user for a complete day, the power consumption of the devices needs to be kept at a minimum level so that the battery can work for more time. Existing literature has not examined this power consumption factor and needs to be addressed to be able to manufacture efficient and long-lasting data acquisition devices for monitoring stress, especially for real-life environments.

Noise affects the data acquisition part of the stress recognition measurement in the lab as well as the out-of-lab environment. Physiological signals which include EEG, EMG, ECG, PPG, and RR are most commonly affected by the individual body movement parts and severely degrade the signal quality. Artifact removal techniques include low, high, and bandpass filtering, least mean square, notch filtering, recursive least mean square, principal
component analysis (PCA), independent component analysis (ICA), and wavelet denoising. Even though these techniques have been found quite useful in the removal of noise from the bio-signals in the lab environment, but still the data acquired out-of-lab presents a wider range of challenges like environmental conditions i.e., temperature, humidity, and lightning affecting these recorded physiological signals and other body functions. An example of the effect of these environmental conditions includes pupil diameter changes in response to light stimulation and changes in skin conductance of the human body due to temperature and physical activity. Thus in an outdoor environment, these conditions are very difficult to remain constant which can somehow be maintained in-lab environment thus posing a greater challenge and affecting the stress measurement study results. 

Most of the human stress measurement studies available in the literature have focused on acute stress and very little attention has been given to chronic stress detection techniques. Even though it is a fact that chronic stress can be fatal in comparison to acute stress and has given billions of dollars lost to many companies and is affecting their worker's health severely~\citep{gottlieb2013coping}. Chronic stress if poorly managed and prolonged for a longer time duration turns into a traumatic disorder that can permanently damage an individual's life. Thus in future stress measurement work chronic stress measurement should be given its due importance so that it can be properly diagnosed and treated before it becomes a permanent part of one's life. Most of the stress measurement studies in the literature have used their dataset for the detection of stress and thus it is difficult to compare the results of different schemes directly because of the lack of standard publicly available stress measurement datasets, hence the curation of data for measurement of stress like DEAP database for emotions~\citep{koelstra2011deap} using globally accepted practices is the need of the hour. 

The deep learning-based method has been proven a powerful mechanism for classification in a variety of applications including handwriting and image recognition but the use of deep learning for human stress detection has still not been explored to a great extent in the literature. It is because deep learning-based algorithms require a large dataset to train the model and acquiring a large dataset for stress is a cumbersome task and it is still an open research problem to acquire a large dataset for the deep learning models to be implemented. Only a few studies focusing on deep learning-based stress classification have been presented in the literature~\citep{song2017development} and thus this area needs further exploration. Keeping in view all the above mentioned limitations of the existing methods, new work on human stress detection should address these challenges so that a more robust system that is practically possible to be implemented for real-life stress monitoring solutions in the future.    

\section{Conclusion}
\label{sec:conc}
Human stress is one of the most significant challenges faced by modern civilization with a long-lasting impact on both society and the social life of an individual. This also impacts the economic condition of those individuals that are facing stress-related situations. Therefore, automatic recognition of stress is of utmost importance to bring positive change in society. Some of the important benefits of automated stress monitoring could include a better attitude toward the workplace, an increase in productivity, and a decrease in the number of road accidents to name a few. In general, better stress management for individuals would have far-reaching benefits for both individuals and society. Towards this, we have surveyed available literature for human stress measurement including both subjective and objective measures. Commonly used stressors used for inducing stress followed by publicly available stress measurement databases are also examined. Multimodal approaches for human stress detection along with a discussion about the limitations of the existing work and the open research issues which need to be addressed in future stress measurement studies are also explored. In particular, we have focused on methods that use or are suitable for employing artificial intelligence techniques for automated stress detection. The information presented here will provide a platform for designing future studies for stress detection, particularly for employing AI with domain knowledge. With this, we also provide a better context for methods that can be used for daily life stress detection (acute) and for situations that are more demanding (chronic). Towards, this we have comprehensively highlighted the current challenges and future directions.

\vspace{6pt}

\reftitle{References}

\PublishersNote{}
\end{document}